\documentclass{ar-1col}

\usepackage[utf8]{inputenc}
\usepackage{graphicx}
\usepackage{epsfig}
\usepackage[comma]{natbib}
\usepackage[]{color}
\usepackage{CJK}
\usepackage[printwatermark]{xwatermark}
\usepackage{longtable}

\newcommand{\nodata}{$...$}
\newcommand{\apj}{Ap. J.}
\newcommand{\apjs}{Ap. J. S.}
\newcommand{\apjl}{Ap. J.}
\newcommand{\aj}{A. J.}
\newcommand{\baas}{Bulletin of the American Astronomical Society}
\newcommand{\mnras}{MNRAS}
\newcommand{\nat}{Nature}
\newcommand{\na}{New Astronomy}
\newcommand{\prd}{Phys. Rev. D}
\newcommand{\prl}{Phys. Rev. Lett.}
\newcommand{\aap}{Astron. Astrophys.}
\newcommand{\aapr}{Astron. Astrophys. Rv.}
\newcommand{\araa}{ARAA}
\newcommand{\apss}{Astrophysics and Space Science}
\newcommand{\pasa}{PASA}
\newcommand{\pasp}{PASP}
\newcommand{\pasj}{PASJ}
\newcommand{\ssr}{Space Science Reviews}

\def\gtrsim{\mathrel{\hbox{\rlap{\hbox{\lower4pt\hbox{$\sim$}}}\hbox{\raise2pt\hbox{$>$}}}}}

\newcommand{\kms}{km~s\ensuremath{^{-1}}}

\newcommand{\mbh}{\ensuremath{M_\mathrm{BH}}}

\newcommand{\msigma}{\ensuremath{M_{\mathrm{BH}}-\sigmastar}}
\newcommand{\msun}{\ensuremath{M_{\odot}}}

\newcommand{\sigmastar}{\ensuremath{\sigma_{\ast}}}

\def\lesssim{\mathrel{\hbox{\rlap{\hbox{\lower4pt\hbox{$\sim$}}}\hbox{$<$}}}}
\def\gtrsim{\mathrel{\hbox{\rlap{\hbox{\lower4pt\hbox{$\sim$}}}\hbox{$>$}}}}

\def\lax{{$\mathrel{\hbox{\rlap{\hbox{\lower4pt\hbox{$\sim$}}}\hbox{$<$}}}$}}
\def\gax{{$\mathrel{\hbox{\rlap{\hbox{\lower4pt\hbox{$\sim$}}}\hbox{$>$}}}$}}

\begin{document}

\markboth{Greene et al.}{Intermediate-Mass Black Holes}

\title{Intermediate-Mass Black Holes}

\author{Jenny E. Greene,$^1$ Jay Strader,$^2$ and Luis C. Ho$^3$
\affil{$^1$Department of Astrophysical Sciences, Princeton University, Princeton, NJ 08544, USA; email: jgreene@astro.princeton.edu}
\affil{$^2$Center for Data Intensive and Time Domain Astronomy, Department of Physics and Astronomy, Michigan State University, East Lansing, MI 48824, USA}
\affil{$^3$Kavli Institute for Astronomy and Astrophysics, Peking University, Beijing 100871, China; Department of Astronomy, School of Physics, Peking University, Beijing 100871, China}
}

\begin{abstract}
We describe ongoing searches for intermediate-mass black holes with \mbh$\approx 100-10^5$~\msun. We review a range of search mechanisms, both dynamical and those that rely on accretion signatures. We find: \\
$\bullet$ Dynamical and accretion signatures alike point to a high fraction of $10^9-10^{10}$~\msun\ galaxies hosting black holes with $\mbh \,\sim 10^5$~\msun. In contrast, there are no solid detections of black holes in globular clusters. \\
$\bullet$ There are few observational constraints on black holes in any environment with $\mbh \approx 100-10^4$~\msun. \\
$\bullet$ Considering low-mass galaxies with dynamical black hole masses and constraining limits, we find that the \mbh-\sigmastar\ relation continues unbroken to \mbh$\sim 10^5$~\msun, albeit with large scatter. We believe the scatter is at least partially driven by a broad range in black hole mass, since the occupation fraction appears to be relatively high in these galaxies. \\
$\bullet$ We fold the observed scaling relations with our empirical limits on occupation fraction and the galaxy mass function to put observational bounds on the black hole mass function in galaxy nuclei.\\
$\bullet$ We are pessimistic that local demographic observations of galaxy nuclei alone could constrain seeding mechanisms, although either high-redshift luminosity functions or robust measurements of off-nuclear black holes could begin to discriminate models. 

\end{abstract}

\maketitle

\tableofcontents

\section{Introduction}

\subsection{Motivation}

``Intermediate-mass'' black holes (IMBHs) are often introduced as what they are not. They are not stellar-mass black holes, which are formed in the deaths of massive stars and are historically thought to be $\sim 10$~\msun\ \citep{remillardmcclintock2006}. They are not supermassive black holes, which are historically considered to have masses of $10^6-10^{10}$~\msun. The question is often framed: are there black holes with masses between these two classes?

We frame the central question differently. At some point in cosmic time, black holes between 10 and $10^6 M_{\odot}$ had to exist, in order to make the $10^9 M_{\odot}$ black holes that are seen only hundreds of millions of years after the Big Bang \citep[e.g.,][]{banadosetal2018}. The central question of this review is whether we can find evidence for these ``intermediate-mass’’ black holes. The corollary is whether, based on the mass distributions and environments of the black holes, we can determine how these IMBHs were formed. Currently we have no concrete evidence for black holes with masses $ \gtrsim 100-10^5$~\msun, although there are some important candidates in this mass range that we will discuss. Finding objects, and characterizing the black hole mass function in this range, is of interest for many reasons. 

Extending scaling relations to this regime may provide unique insight into the evolution of black holes \citep[e.g.,][]{kormendyho2013,shankaretal2016,pacuccietal2018}, along with the possible importance of feedback for dwarf galaxies \citep{silk2017,bradfordetal2018,pennyetal2018,dickeyetal2019}. Demographics of black holes at lower masses will help elucidate the dynamical evolution of dense stellar systems \citep[e.g.,][]{milleretal2002, portegieszwartetal2002,gurkanetal2004,antoninirasio2016}.

Intermediate-mass black holes will also be prime sources of gravitational radiation for upcoming gravitational wave detectors in space \citep[\emph{Laser Interferometer Space Antenna [LISA]}, e.g.,][]{amaro-seoaneetal2015}. To determine the rates and interpret the gravitational wave signals, we need independent measurements of the black hole number densities \citep[e.g.,][]{stonemetzger2016,macleodetal2016,eracleousetal2019}. Furthermore, ongoing and future time-domain surveys are detecting more and more tidal disruption events. In principle, lower-mass black holes should be significant contributors to the detected tidal disruption events, and they are starting to be detected \citep{maksymetal2013,weversetal2017,vanvelzen2018}. 

Finally, we do not know exactly how emission from accretion flows onto black holes will evolve with black hole mass: at lower masses the accretion disk gets hotter and the bolometric luminosity drops, perhaps leading the phenomenology to look more like X-ray binaries and less like accreting supermassive black holes \citep[e.g.,][]{cannetal2019}. The emergent spectrum of such sources is of interest not only to understand accretion, but also in thinking about the impact of ``mini-quasars'' on the formation of the first galaxies and the reionization of the Universe \citep[e.g.,][]{madauhaardt2015}. 

For all of these reasons, the time is right to review the current state of our knowledge, and to prepare for the rich new upcoming data sets that will bear on these questions.  

\subsection{Definition}

Observationally, we have compelling evidence for black holes with \mbh$\lesssim \, 100$~\msun\ and with \mbh$\gtrsim \, 10^5$~\msun. Work over the past 25 years has established clearly that galaxy centers harbor ``normal'' active galactic nuclei (AGN) with masses \mbh$\approx 10^5$~\msun\ and higher (\S \ref{sec:searchnuclei}). Existential discovery space lies in the mass range of \mbh$\approx 100-10^5$~\msun. Nevertheless, from the perspective of understanding the growth and demographics of black holes, we will show that black holes in the range \mbh$\approx 10^5-10^6$~\msun\ also encode important (and poorly quantified) information. Thus we cover them under the purview of this review as well. 


\begin{figure}
\vbox{ 
\vskip 0mm
\hskip +5mm
\includegraphics[width=0.85\textwidth]{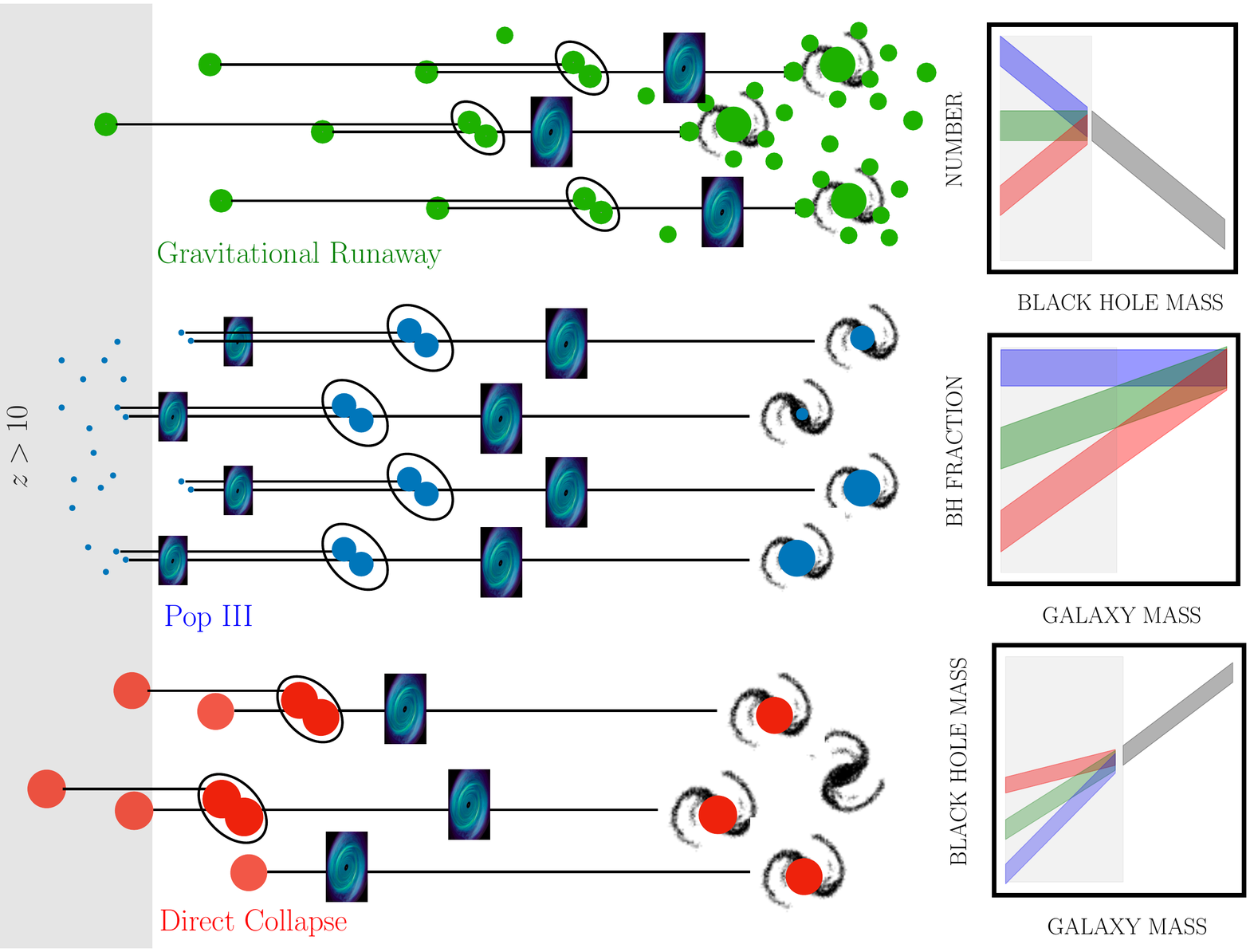}
}
\vskip -0mm
\caption{Possible observable differences between different seeding scenarios. Early formation through direct collapse (red) or Population III stars (blue) occur at $z>10$, while gravitational runaway (green) can happen throughout cosmic time. As cosmic structures evolve, the seed black holes will suffer mergers (black ovals) leading to the emission of gravitational waves, as well as accretion episodes (blue disks) that could be observed as active galactic nuclei. At the present day, differences in black hole mass functions, occupation fractions, and black hole-galaxy scaling relations may ensue from different seeding channels, for simplicity here shown only for nuclear black holes. Grey bars in these relations show where we do not yet have observational constraints.}
\label{fig:formcartoon}
\end{figure}

\subsection{Historical Context}

The growth of supermassive black hole seeds, and the existence of IMBHs, have both long been discussed in the literature, even before we were completely sure of the existence of astrophysical black holes \citep[e.g.,][]{eardleypress1975}. Quickly thereafter, the community developed theoretical ideas about how one might form non-stellar--mass black holes; the famous diagram from \citet{rees1978} includes all the formation mechanisms we consider today, including gravitational runaway \citep{bahcallostriker1975,begelmanrees1978,quinlanshapiro1990,lee1993}, the collapse of Population III stars \citep[e.g.,][]{bondetal1984,madaurees2001}, and the idea of a ``direct collapse'' into a black hole \citep[e.g.,][]{haehneltrees1993,loebrasio1994,koushiappasetal2004}. There was also early interest in intermediate-mass black holes as a possible explanation for dark matter \citep[e.g.,][]{laceyostriker1985,carr1994}, although there are now many limits on black holes as dark matter \citep[e.g.,][]{tisserandetal2007,alihamoudetal2017,zumalacseljak2018}. 

As observations of young quasars push to earlier and earlier times \citep[e.g.,][]{fanetal2006,mortlocketal2011,banadosetal2018}, the community has recognized the significant challenge of creating such massive black holes so quickly \citep[e.g.,][]{haiman2013}, leading to a desire to make relatively massive seed black holes, further motivating searches for IMBHs. Accordingly, theoretical work identified that local observations of BHs in dwarf galaxies, which suffer fewer mergers and less accretion, may provide an additional key to understanding seeding processes \citep[e.g.,][]{volonterietal2008}. 

In parallel, the \emph{Einstein} telescope enabled the discovery of the first ultra-luminous X-ray sources \citep[ULXs; e.g.,][]{longetal1981,fabbiano1989}, whose luminosities are nominally above the Eddington limit for compact stellar objects. One natural explanation for the high X-ray luminosities of these sources is that they are powered by more massive compact objects, namely IMBHs \citep[e.g.,][ and references therein; \S \ref{sec:searchnuclei}]{kaaretetal2017}. The ultra-luminous X-ray sources show association with young star clusters and this spurred continuing interest in the formation of IMBHs in dense stellar clusters through gravitational runaway \citep[e.g.,][]{ebisuzakietal2001,portegieszwartetal2002,milleretal2002}. Puzzling suggestions of abnormal numbers of young stars in the Galactic Center also sparked interest in a possible IMBH to carry in the stars \citep[e.g.,][]{ghezetal2003,hansenmilosavljevic2003}.

Thanks to the successful detection of supermassive black holes in galaxy nuclei \citep[e.g.,][]{kormendyrichstone1995}, dynamical searches for IMBHs at the centers of globular clusters like M15 were also underway, yielding conflicting results \citep[e.g., ][]{gebhardtetal2000,gebhardtetal2002,gerssenetal2002,mcnamaraetal2003}. Early dynamical searches for black holes in the nuclei of low-mass galaxies also reported non-detections \citep{gebhardtetal2001,vallurietal2004}. Comprehensive searches for AGN signatures in nearby galaxies likewise pointed to a low fraction of AGN in late-type and low-mass galaxies \citep[e.g.,][]{hfs1997,kauffmannetal2003}. These facts together led the community to conclude that black holes are rarely found in the nuclei of $<10^{10}$~\msun\ galaxies (i.e. galaxies less massive than the Milky Way). Even today, there is a severe dearth of information about black hole demographics in sub-$L_*$ galaxies. 

Real progress in the last decade, based both on dynamical studies (\S \ref{sec:dynamics}) and accretion-based ones (\S \ref{sec:searchnuclei}), suggests instead that a high fraction ($>50$\%) of $10^9-10^{10}$~\msun\ galaxies do harbor central black holes, likely with \mbh$>10^5$~\msun\ \citep[][ and \S \ref{sec:demographics}]{milleretal2015,nguyenetal2019}. We use this recent progress as a motivation to revisit both scaling relations (\S \ref{sec:scaling}) and the allowed range of black hole mass functions for \mbh$<10^6$~\msun\ (\S \ref{sec:demographics}).

We are not the first to attempt a review of the linked subjects of IMBH searches and seeding mechanisms  \citep{volonteri2010,greene2012,reinescomastri2016,mezcua2017}. There is a related review about seeding mechanisms in Inayoshi et al.\ this volume.

\section{Formation Paths: Where to Find IMBHs}
\label{sec:formation}

We briefly review the primary theoretical channels for seeding supermassive black holes (Figure \ref{fig:formcartoon}). We then describe the predictions for local observations in terms of number densities and environments of IMBHs.

\subsection{Seeding Models}
\label{sec:seedmodels}

We know that black holes are made in the death of massive stars. This process happens at the present day, but at early times we expect theoretically that the first (so-called Population {\small III}) stars were quite massive, due to the inability of molecular hydrogen gas to cool \citep[][and references therein]{brommlarson2004,karlssonetal2013}. Such massive stars would likely end their lives as black holes with \mbh\,$\sim 100$~\msun\ \citep[e.g.][]{fryeretal2001}, excepting first stars in the mass range of $140-260$~\msun\ that would explode as pair-instability supernovae and leave no remnant \citep{hegeretal2003}. 

As shown by \citet{madaurees2001}, if these first massive stars are rare, such that only $\sim$ one is made in each high density peak at very high redshift, then the number density of black holes would be well-matched to the number of supermassive black holes today. However, growing from $\sim 100$~\msun\ to a billion suns in a fraction of a Gyr requires dramatic fueling \citep{haimanloeb2001}. Specifically, for a typical $e-$folding time of 45 million years, 14 such $e-$foldings are available at $z=7$, but 17 are needed if the black hole grows at the Eddington limit \citep[and references therein]{madauetal2014}. Arranging this uninterrupted growth seems challenging even for these rare quasars \citep[e.g.,][]{johnsonbromm2007,milosavljevicetal2009}. On the other hand, a few episodes of super-Eddington accretion could largely alleviate the tension, and super-Eddington accretion now appears to arise naturally in numerical experiments \citep[e.g.,][]{jiangetal2019}. 

A different channel, with no local analogs, is that of collapsing gas clouds forming a massive seed black hole without passing through all the phases of stellar evolution. These ``direct collapse'' models form seed black holes with \mbh~$\sim 10^4-10^6$~\msun\  \citep[e.g.,][]{brommloeb2003,loebrasio1994,lodatonatarajan2006,begelmanetal2006}, perhaps passing through a ``quasi-star'' phase \citep[e.g.,][]{begelman2010}. This channel can only operate at very high redshift, since pristine gas is required to suppress cooling and fragmentation.  Recent literature has discussed in detail how rare the conditions may be to ensure that the gas does not cool and fragment into smaller units \citep[e.g.,][]{visbaletal2014,habouzitetal2016}, typically assuming that an elevated Lyman-Werner background is needed \citep[e.g.,][]{omukai2001}. Some works instead argue that additional photons from star formation in nearby halos \citep[e.g.,][]{dijkstraetal2008,visbaletal2014}, additional gas heating through mergers \citep{yoshidaetal2003,wiseetal2019}, or even star formation within individual halos \citep[][]{dunnetal2018}, may provide sufficient heating or ionizing photons to make this channel viable. We emphasize that the number density of halos that suffer direct collapse is very uncertain, with predictions differing on the halo gas accretion rates as a function of mass, the required elevation of the ionizing background, and the possible sources of ionizing photons. See a comprehensive review in Inayoshi et al., this volume.

A final class of models makes $\sim 10^3-10^4$~\msun\ black holes in a gravitational runaway event within a dense stellar cluster. Modern discussion divides globular cluster IMBH formation scenarios into ``slow" and ``fast" versions, depending on whether the formation occurs over $\sim 100$ Myr to 1 Gyr or $\lesssim$ a few Myr, respectively. Here we focus specifically on the {\it formation} of the seed at the centers of dense clusters. In any scenario where an IMBH does form, the present-day mass may well be higher due to accretion of gas and/or stars (e.g., \citealt{rosswogetal2009,macleodetal2016,sakuraietal2017}; \S \ref{sec:demographics}). We also note that while normal single stars are not expected to produce black holes $> 100 M_{\odot}$, it is plausible that repeated mergers in star clusters could occasionally produce IMBHs larger than 100~\msun\ by this channel \citep{rodriguezetal2019}.  

The modern incarnation of the ``slow" scenario \citep{milleretal2002} envisions a black hole of mass $\gtrsim 50 M_{\odot}$ forming as the remnant of a massive star, then growing though mergers of mass-segregated black holes, leading to $\sim 10^3 M_{\odot}$ IMBHs in a few $\times 10\%$ of massive globular clusters. In this scenario the occupation fraction is set by considering the central densities observed in Galactic globular clusters. Lower-mass ($\sim 10$--$20 M_{\odot}$) black hole seeds will not lead to a runaway since early stellar interactions will tend to eject these lighter black holes from the cluster. A governing uncertainty in this scenario is whether even $50 M_{\odot}$ seeds can form, as such remnants might be rare or absent in single star evolution \citep{woosley2017}.

The ``fast" scenario was re-introduced in a contemporary form by \citet{portegieszwartetal2002}, who used a combination of simulations and analytic calculations to argue that sufficiently dense star clusters will undergo a collisional runaway of massive stars, the product of which could eventually lead to the formation of an IMBH. Such supermassive stars could potentially also produce enough ejected material with unusual abundances to explain the multiple stellar populations seen in globular clusters \citep{gielesetal2018}. Both this scenario and the direct collapse scenario may be sensitive to metallicity: at high metallicity, stellar winds may remove so much mass that collapse to a black hole is uncertain \citep[e.g.,][]{mapelli2016}. In the context of very low metallicity and very high redshift, some models predict that very dense and massive clusters may form black holes at their centers \citep[e.g.,][]{devecchivolonteri2009,katzetal2015,sakuraietal2017}. While clusters this dense are not found at lower redshift, hypothetical Pop III star clusters could optimistically be detectable with the \emph{James Webb Space Telescope} (\emph{JWST}).

\subsection{From Seeds to Local Predictions: Population III and Direct Collapse Channels}

Currently it is challenging to compare predictions of local demographics based on the seeding mechanism. To that end, we gather approximate predicted number densities and mass distributions of present day IMBH populations as a function of channel and modeling approach. Even with perfect knowledge of the number density and halo locations of black hole seeds at early times, there are many unknown factors at play as the seeds evolve towards the present day \citep[e.g.,][]{volonterietal2008,volonteri2010,mezcua2017,buchneretal2019,mezcuaetal2019}. We outline some of these uncertainties and then compare the predictions from different types of models.

When galaxies merge, the central black holes may also merge, and any differences in the mass and angular momentum of the two black holes can translate into anisotropic gravitational wave emission during the merger, which in turn can impart a linear momentum ``kick'' to the merged remnant \citep[e.g.,][]{redmountrees1989}. A major uncertainty in going from initial seed models to local observations of black holes in galaxy nuclei comes from the unknown fraction of black holes that are ejected via gravitational recoil. These kicks may be more than sufficient to exceed the escape velocity of low-mass halos \citep[e.g.,][]{haiman2004,campanellietal2007}, but the most common mergers with much lower-mass black hole secondaries likely never lead to recoil \citep{volonterirees2006}. In practice, the distribution of kick velocities will depend on the mass ratios and relative spins of the incoming black holes, for which we do not have complete models. There are other dynamical uncertainties to consider. Such low-mass black holes in such shallow potentials may never settle at the centers of their galaxies \citep{bellovaryetal2019}. Away from the center of the potential, the black holes may not grow efficiently, and would likely be missed with current searches \citep[although see][]{reinesetal2019}.

Finally, tracking accretion in a physically motivated way requires prohibitively high spatial resolution for cosmological simulations. In particular, if super-Eddington accretion is possible, then the differences between the Population III channel and the direct collapse channel may be washed out at early times \citep[e.g.,][]{volonterignedin2009,alexandernatarajan2014}. Thus, it is possible that the high redshift luminosity function and number of gravitational wave detections from growing black holes may be more sensitive to seeding \citep[e.g.,][]{sesanaetal2007,tanakahaiman2009}. Furthermore, late-time growth contributes in an unconstrained way to the present-day black hole mass distribution. Nuclear star clusters, the hosts of at least some candidate IMBHs (\S \ref{sec:dynamics}), experience multiple bursts of star formation \citep[e.g.,][]{walcheretal2006,carsonetal2015,kacharovetal2018}. Such bursts may well feed the black hole, and additional accretion via stellar interactions may also occur \citep[e.g.,][]{stoneetal2017,alexanderbaror2017}. 

\subsubsection{Number Densities}

The Population III and direct collapse channels are distinguished from the dynamical runaway channels in that they happen exclusively at very high redshift and rely on the presence of metal-free gas, so we will refer to them collectively as early seeding models. As described above, the direct collapse scenario requires finely tuned conditions. For those theoretical works that consider direct collapse viable, then both channels are generally thought to make at most one IMBH per halo (although as emphasized above, direct collapse black hole may be considerably rarer), and thus yield very similar number density predictions (Table \ref{tab:predict}).  

The hope for observable differences between these two seeding mechanisms has hinged on the idea that ''heavy'' seeding would leave lower numbers of more massive seeds while Population III remnants would be more common but of lower mass \citep[e.g.,][]{vanwassenhoveetal2010,greene2012}. Two approaches have been taken to predict the observable signatures of seeding mechanisms on the black hole population today, semi-analytic modeling and hydrodynamic simulations. 

Semi-analytic modeling \citep{volonterietal2008} can quickly track the evolution of black holes using merger trees, and explore the dependence on all the uncertain accretion and merging prescriptions \citep[others include][]{volonterinatarajan2009,vanwassenhoveetal2010}. Most of the semi-analytic models in the literature that compare Population III and heavy seeds adopt the heavy seeding model of \citet{lodatonatarajan2006}, which predicts a high number density of heavy seeds $\sim 0.1$~Mpc$^{-3}$ \citep{volonteri2010}, as compared to $\sim 10^{-4}$~Mpc$^{-3}$ quoted by Inayoshi et al.\ this volume. Thus, these are optimistic scenarios for heavy seeds, and likely depend in detail on the assumed ionizing background \citep[e.g.,][]{valianteetal2016}. Recently, \citet{bellovaryetal2019} also used zoom-in hydrodynamical simulations to track seeds to the present day as formed in high-redshift via a direct collapse scenario \citep{dunnetal2018}. We will take the recent work by \citet{ricartenatarajan2018} as representative of the semi-analytic approach, because they consider a range of fueling mechanisms. The two have comparable predicted occupation fractions in the context of direct collapse. However, Ricarte \& Natarajan find a very wide range of acceptable occupation fractions for the Population III channel (see also \S \ref{sec:demographics}). We note that according to predictions from both models, the occupation fraction starts to change dramatically below galaxy host masses of $10^9$~\msun, a regime that is not yet empirically constrained.

We translate these occupation fractions into number densities for the two channels (Table \ref{tab:predict}). To convert the occupation fractions into number-density predictions, we take a galaxy stellar mass function \citep{wrightetal2017} and integrate from $10^7-10^{10}$~\msun, incorporating the range of occupation fractions presented in Table \ref{tab:predict}. Note that we have observational constraints only down to $10^9$~\msun, and so these predictions represent a significant extrapolation, but are consistent with existing observations in all cases. These numbers are also roughly consistent with the numbers quoted by \citet{volonteri2010}, which again relies on the optimistic direct collapse channel from \citet{lodatonatarajan2006}. Although mass distributions may differ between the two, the full range of predicted occupation fractions for the direct collapse and Population III channels are quite similar. We emphasize that the accretion history will be tracked eventually with active galaxy luminosity functions \citep[e.g.,][]{civanoetal2019}, which will provide additional constraints on the fueling models and may allow us to distinguish between seeding models more effectively (\S \ref{sec:future}).

\subsection{From Seeds to Local Predictions: Gravitational Runaway}

We now attempt to translate theory on the different channels of gravitational runaway into predictions for the number density and mass distribution of IMBHs at present (Table \ref{tab:predict}). Unique to this channel is the expectation of $\gg 1$ massive black hole per galaxy, which will be quantifiable if we ever manage to detect or rule out non-nuclear black holes. 

Recent work has revisited the slow formation scenario described in \S \ref{sec:seedmodels} in the context of dynamical formation of black hole--black hole binaries observable as gravitational wave mergers (e.g., \citealt{rodriguezetal2018}). Indeed, there is some prospect for useful observational constraints for this channel using gravitational wave observations from the current generation of detectors \citep{kovetzetal2018}. The predictions for the slow scenario depend in essential ways on whether the dynamics of close encounters are treated in a relativistic manner. While there is no simple expression for the gravitational recoil velocity as a function of the black hole masses, mergers with more extreme mass ratios tend to have lower velocities. The predictions also depend strongly on the unknown initial spins of black holes, which can lead to enormous velocities that will eject nearly all systems from clusters \citep{campanellietal2007}. Even if the initial spins are low, post-merger products will have higher spins, suggesting that successive mergers are more likely to eject black holes. \cite{holley-bockelmannetal2008} found that in most cases the IMBH will be ejected over a wide range of seed masses, unless all of the encounters have mass ratios $M_{\rm low}/M_{\rm high} \lesssim 0.05$. To ensure that this condition is met would require that the initial seed be $\gtrsim 500 M_{\odot}$, and that it is surrounded by a sub-cluster of $\lesssim 25 M_{\odot}$ stellar-mass black holes. Seeds of lower mass would have a high likelihood of ejection if other black holes are present. 

In the very densest clusters, the escape velocity might be high enough to allow merger products to remain bound \citep{antoninietal2019}, which would limit IMBH retention to rare dense clusters $\gtrsim$ a few $\times 10^6 M_{\odot}$. Some other studies (e.g., \citealt{gierszetal2015}) that advocate for slow IMBH formation in star clusters sometimes miss the crucial contribution of recoil in the dynamics of IMBH formation through this channel. While the occupation fraction from the slow channel is uncertain, current simulations are consistent with a low or negligible value for typical globular clusters. 

For the the fast channel, \citet{portegieszwartetal2002} argue that fast runaway only applies to clusters with short initial relaxation times. A short relaxation time excludes most typical globular clusters, unless perhaps they are born with primordial mass segregation. \citet{gurkanetal2004} find that runaway collisions should be common in dense clusters, but the final product of the runaway is uncertain, since a ``supermassive" star formed by collisions will evolve on a timescale comparable to the collision rate \citep{freitagetal2006}. The metallicity of the stars and the possible presence of gas in the core add additional complications. Hence it is difficult to predict whether a supermassive star will form at all, its expected mass if it does form, and whether it would eventually collapse to an IMBH, or instead destroy itself entirely as a pair-instability supernova \citep{speraetal2017}. 

Other papers have discussed the possibility of fast seeds selectively forming only at high redshift in star clusters at the centers of dense and relatively massive halos (e.g., \citealt{sakuraietal2017}). If so, then most of these clusters will preferentially migrate to galaxy centers. However, there is voluminous evidence that most globular clusters of all metallicities formed in the same way that massive clusters still form today, rather than in a unique channel only operating in the early universe (e.g., \citealt{brodieetal2006,elbadryetal2019}). 

\citet{millerdavies2012} argue that all clusters above a central velocity dispersion of $\sim 40$ km s$^{-1}$ will necessarily form an IMBH through some mechanism, since neither primordial nor dynamically formed binaries of stars or stellar remnants can prevent core collapse for these velocity dispersions. This analytic argument rests on efficient ejection of stellar-mass black holes. \citet{breenetal2013} argue that black hole ejection is expected to be rather inefficient for typical globular clusters and hence core collapse will usually occur only after many relaxation times. This result is generally born out by numerical simulations (e.g., \citealt{kremeretal2019}), though few simulations of $\gtrsim 10^6 M_{\odot}$ clusters have yet been published.

Overall, there is little theoretical support for the idea that IMBH formation is favored in typical globular clusters. By contrast, many nuclear star clusters have high velocity dispersions and a range of relaxation times, and may be plausible sites for the formation of IMBHs. We return to this idea in \S \ref{sec:searchoff}. 

While old globular clusters are currently too dynamically evolved to suffer recent gravitational runaway, young star clusters or growing nuclear star clusters should still be in a position to grow massive BHs if this channel is robust. To date, no such trustworthy sources have been found (although see \S \ref{sec:searchoff} for some candidates). One way out of this constraint is to note that few sensitive optical, radio, or X-ray surveys for massive BHs are currently possible outside of the Milky Way, which itself has scarcely any young massive clusters (see \S \ref{sec:searchoff}). Another way out is that the channel may require low metallicities \citep{mapelli2016}, and then we would not expect it to operate in the local universe. However, metal-poor and metal-rich clusters have no observed substantial structural or dynamical differences, so if IMBHs form commonly but only in low-metallicity clusters, they have not yet been found to affect the clusters in any noticeable way. 

\subsubsection{Number Densities}
\label{sec:numdensgravrun}

To translate these predictions into number densities, we first consider galaxy nuclei, then the cluster population. If gravitational runaway happens, one would think that nuclear star clusters, the most massive and densest clusters, may be the most likely to form and retain IMBHs. However, not all nuclear star clusters may successfully form massive black holes by gravitational collapse \citep[][\S \ref{sec:searchoff}]{breenetal2013}. We therefore assume that under this channel, there will be a wide range of occupation fractions in galaxy nuclei, between 10-100\%. Adopting a measured range of nucleation fractions as a function of stellar mass from \citet[][see also Neumayer et al.\ in preparation]{sanchez-janssenetal2019}, we find a final number density of 0.02-0.25 Mpc$^{-3}$ (Table \ref{tab:predict}). This number could be higher than for early collapse models.

\subsection{From Seeds to Local Predictions: Wandering Black Holes}

In addition to black holes in galaxy centers, any seeding model will lead to a population of off-nuclear ``wandering'' black holes. We divide wandering black hole channels into those where the black hole is made via gravitational runaway in star clusters and those where the black hole is formed in the nucleus of a satellite galaxy. 

If gravitational runaway operates, \citet{holley-bockelmannetal2008} show that any IMBH formation scenario within globular clusters would lead to the majority of IMBHs with \mbh$<3000$~\msun\ being ejected. \citet{fragioneetal2018} use a semi-analytic model and explore a range of \mbh/$M_{\rm cluster}$ of 0.5-4\% to find that only a small percentage (3\%) of IMBHs would be retained in globular clusters today, while the majority of the $\sim 1000$ IMBHs in the Milky Way would either be ejected during black hole--black hole mergers (70-90\%) or have their cluster dissolve around them (4-30\%), leaving a large population of wandering black holes with a low-mass dense stellar cluster remaining bound. 

To find these wanderers, consider that at the present day $\sim 20$--40 stars are expected to remain surrounding the \mbh$\sim 10^4$~\msun\ IMBHs and perhaps $\sim 10^3$ stars surrounding rarer \mbh$\sim 10^5$~\msun\ \citep[see also ][]{komossamerritt2008}. The expected half-light radii of these remnant clusters are small, $\lesssim 1$ pc. Such clusters should be detectable as resolved stars in future wide-fields surveys such as the Large Synoptic Survey Telescope (LSST), since current observations are already discovering very low-mass globular clusters in the outer halo (e.g., HSC-1, with $\sim 250$ stars; \citealt{hommaetal2019}). Follow-up spectroscopy of even a few stars in such objects could reveal their provenance via an unexpectedly large velocity dispersion. Closer to the Sun, remnant recoil clusters containing only a handful of stars could be identified using \emph{Gaia} data as a population in a small patch of sky ($\lesssim$ a few $\times 10$") with a net proper motion and a substantial dispersion around that value. These IMBHs could also pass through molecular clouds in the disk, sparking into detectability from the additional fuel as a faint radio source with high proper motion \citep{fenderetal2013}.

Even if there are no black holes in globular clusters, disrupted infalling satellite galaxies that harbor IMBHs will deposit their black holes somewhere outside the nucleus of the larger halo. If some infalling satellites harbor a massive black hole (which is likely for at least some of the massive satellites; \S \ref{sec:demographics}) then we expect a population of wandering black holes \citep{governatoetal1994,schneideretal2002,volonterietal2003,islametal2004,bellovaryetal2010,micicetal2011,rashkovmadau2014}. A related pathway comes from very early accretion of satellites. Mergers between satellites could have lead to ejections of black holes that might still be wandering around the Milky Way. The recoiling black hole will likely take with it a tightly bound cluster of stars \citep{olearyloeb2009,olearyloeb2012}. O'Leary and Loeb estimate $\gtrsim 100$ black holes with \mbh$>10^3$~\msun\ (and $\gtrsim 10$ black holes with \mbh$>10^4$~\msun) in the Milky Way. That massive black holes have been detected in the centers of ultra-compact dwarf galaxies \citep{sethetal2014,ahnetal2018}, which are the dense remnants of tidally stripped galaxies, provides empirical evidence for this channel. 

\subsubsection{Number Densities}

We will make two different estimates for the number density of wanderers (non-nuclear IMBHs), first for those deposited by globular clusters, and then for those formed in satellites. We have decided to make our estimates as fully empirical as possible, by linking the number of wanderers from clusters directly to observed numbers of globular clusters per unit host stellar mass and number of wanderers from satellites to the number of ultra-compact dwarf galaxies per unit host stellar mass, assuming these are tidally stripped galaxies.

We start by calculating the number of globular clusters per galaxy using the stellar mass to globular cluster number relation from \citet{harris2016}. We assume a $V-$band mass-to-light ratio of two for galaxies with $M_* < 3 \times 10^{10}$~\msun, and three for more massive galaxies. Not all globular clusters host black holes. We assume that 10\% of massive globular clusters with $M_*>10^6$~\msun\ can form and retain massive black holes \citep{holley-bockelmannetal2008,fragioneetal2018}. To estimate the fraction of globular clusters with $M_*>10^6$~\msun, we integrate the globular cluster luminosity functions from \citet{jordanetal2007} for different galaxy masses. Integrating these luminosity functions, we find that $\sim 5\%$ of the globular clusters around a $\sim 2\times10^{10}$~\msun\ galaxy have $M_*>10^6$~\msun, while $\sim 10\%$ of the globular clusters around a $10^{11}$~\msun\ galaxy are of this mass or higher. Folding that in with the number of massive clusters per galaxy, we find a number density of 0.31 IMBH per cubic Mpc (Table \ref{tab:predict}). Again, we emphasize that this is a lower limit, since it does not include black holes in dissolved clusters or those that may have been ejected from their cluster after a merger.

We further assume that massive clusters ($M_* > 10^6$~\msun) in a galaxy today comprise a mixture of a tail of truly massive objects that formed as clusters ($\sim 80\%$) with the cores of stripped galaxies ($\sim 20\%$) that we refer to as ultra-compact dwarfs \citep[e.g.,][]{pfefferetal2016}. With the number of such clusters calculated in the prior paragraph, we then expect somewhere between zero and two ultra-compact dwarfs for the Milky Way (e.g., M54 and perhaps $\omega$ Cen), and somewhere between 75-150 in a cluster like Virgo, as observed \citep[][]{liuetal2015}. Our assumption that ultra-compact dwarfs do not dominate the globular cluster luminosity function even at high mass is also supported by the Jordan et al.\ luminosity functions, which do not change if all clusters with sizes greater than 5 pc (a proxy for stripped dwarf galaxies) are excluded.  

To calculate the number of wandering black holes due to disrupted satellites, we not only need the number of disrupted satellites, but also how often the satellite hosts a black hole. This is mass dependent. Based on work summarized in \S \ref{sec:demographics}, we take an occupation fraction between 10\% and 50\%, leading to at most one wandering black hole in a Milky Way-mass galaxy according to this scaling. Our numbers are somewhat lower than predictions from some theoretical works \citep{bellovaryetal2010,rashkovmadau2014} but slightly higher than what is presented by \citet{volonteriperna2005}. Integrating over the galaxy luminosity function \citep{wrightetal2017}, we find a comparable number density of IMBHs in wanderers in $L_*$ galaxies as there are in dwarf nuclei, $\sim 0.06-0.3$ per cubic Mpc (Table \ref{tab:predict}). 

A far more conservative number density is given by \citet{voggeletal2019}. They investigate the number density of black holes associated with ultra-compact dwarf galaxies found in cluster environments and likely to host black holes detectable by current techniques. That number density is $2 \times 10^{-3}$ supermassive black holes per cubic Mpc. They then extend their methodology to include ultra-compact dwarf galaxies around all elliptical galaxies to find $7-8 \times 10^{-3}$ supermassive black holes per cubic Mpc. Our much higher numbers, in contrast, are dominated by so-far unknown lower-mass black holes ($<10^5$~\msun) that we expect to be found in the halos of $\sim L_*$ galaxies due to stripping of low-mass satellites like Sagittarius and the LMC \citep[e.g.,][]{rashkovmadau2014}.

\subsection{Tests of Seeding Mechanisms}

Clearly number densities alone are too blunt a tool to distinguish different early seeding mechanisms. Other means have been discussed in the literature. The black hole mass function could encode seeding history, particularly if there is a mass cut-off corresponding to a heavy seeding mechanism (\S \ref{sec:demographics}). Occupation fractions and black hole scaling relations (\S \ref{sec:scaling}) might reflect differences in seeding mechanisms, or depend more on fueling than seeding \citep[e.g.,][]{volonterinatarajan2009,ricartenatarajan2018}. Transients (tidal disruptions and gravitational waves; \S \ref{sec:searchupcoming}) may provide our best hope to find wandering black holes, as well as to see direct evidence of early seeds from direct merger detections with \emph{LISA}.

In summary, theoretical challenges continue to impede our ability to make definitive predictions for local IMBH populations based on the channel. It is still unclear from the theoretical literature whether enough gas can be in a prime condition to facilitate enough direct collapse black holes. If yes, then the differences in local number density and mass distribution between direct collapse and Population III channels will likely be insufficient to distinguish between them. Likewise it is unclear whether Population III seeds can grow efficiently in low-mass halos at early times \citep[e.g.,][]{pelupessyetal2007,alvarezetal2009}. Additional information on the early growth history of these black holes may come from active galaxy luminosity functions \citep[e.g.,][]{civanoetal2019} and gravitational waves from black hole mergers at high redshift \citep{sesanaetal2007}.

\section{Stellar and Gas Dynamical Searches for IMBHs}
\label{sec:dynamics}

\subsection{Dynamics From Integrated Light Measurements}

\begin{figure}
\vbox{ 
\vskip 0mm
\hskip 0mm
\includegraphics[width=0.95\textwidth]{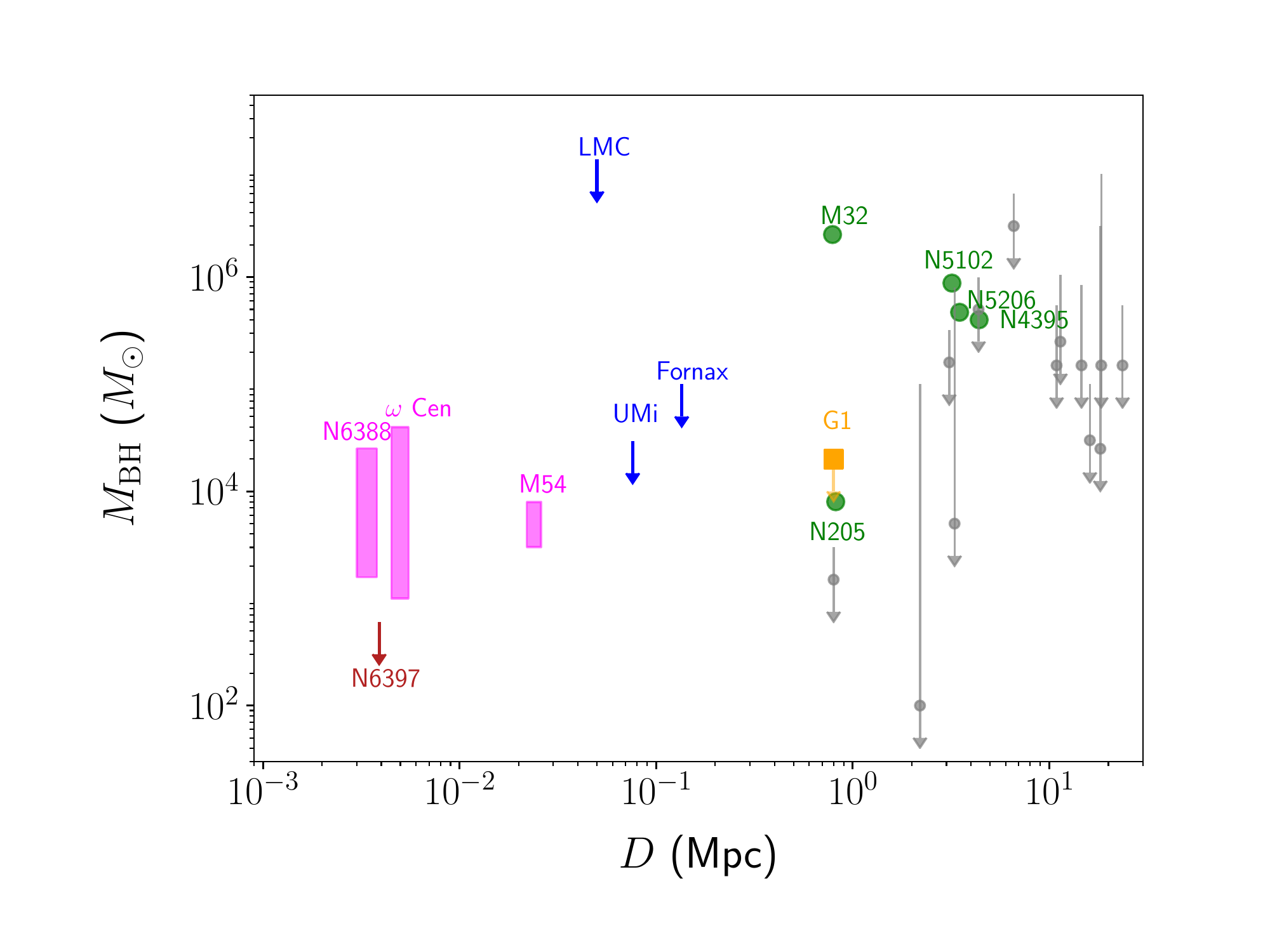}
}
\vskip -0mm
\caption{Summary of existing dynamical measurements or limits on black hole mass in low-mass stellar systems. Black hole masses in galactic nuclei (green circles) and limits (grey) are from various sources \citep[][ and references therein]{neumayerwalcher2012,nguyenetal2019}. For the limits from Neumayer et al., we show their ``best'' measurement (grey circles) and the ``maximum'' allowed black hole mass (arrow top). In pink are Galactic globular clusters with controversial black hole mass measurements; the bottom of the bars represent upper limits from radio observations \citep{tremouetal2018} while the top are dynamical measurements \citep{lutzgendorfetal2013}. The similarly contested measurement in G1 is shown in yellow \citep{gebhardtetal2005,vandermarelanderson2010}. The maroon upper limit for NGC 6397 is representative of the best limits possible in the nearest systems \citep{kamannetal2016}.
Blue arrows are upper limits on three Milky Way satellite dwarf galaxies, the LMC \citep{boyceetal2017}, Fornax \citep{jardelgebhardt2012}, and Ursa Minor \citep[][]{loraetal2009}. }
\label{fig:existinglimits}
\end{figure}

Dynamical modeling of supermassive black holes has a long history and has been reviewed elsewhere \citep[e.g.,][]{kormendyrichstone1995,kormendyho2013}. The kinematics can be measured from either stars or gas in the vicinity of the black hole. For stellar-dynamical modeling, state-of-the-art codes use Schwarzschild modeling \citep{schwarzschild1979} to jointly model the mass density of the central black hole, stars, and dark matter by orbit superposition \citep[e.g.,][]{rixetal1997,gebhardtetal2003}. The stellar mass profile is modeled from the light, which is then converted to a mass profile by solving for the stellar mass-to-light ratio ($M/L$), and if needed a dark matter halo component is also independently fitted \citep[e.g.,][]{gebhardtthomas2009}. The orbit families are weighted to match the spatially resolved stellar kinematical measurements. A major issue for all dynamical models is to distinguish between orbital anisotropy and mass (e.g., a black hole). This mass-anisotropy degeneracy is well-known \citep[e.g.,][]{binneymamon1982}. Other (related) issues include: (1) degeneracy between the stellar $M/L$ and the black hole mass (2) incomplete stellar orbit libraries and (3) the assumption of axisymmetry \citep[][]{vandenboschdezeeuw2010}.
The systematic uncertainties for black hole masses from gas dynamics are qualitatively different than for stellar dynamics, but given the large systematics for both kinds of analysis, we use them interchangeably in our discussion \citep[although see][for a more complete discussion]{walshetal2013}.

Low-mass stellar systems (galaxies, nuclear star clusters and globular clusters) present additional challenges. One is spatial resolution. To detect a black hole, the region within which the gravity of the black hole dominates the stellar motions must be resolved. This gravitational sphere of influence, roughly where the mass enclosed in stars equals that of the black hole, is given by: SOI (pc) $\approx$ 0.0043 ($M_{\rm BH}/\msun$) ($\sigma/\mathrm{km\,  s^{-1}})^{-2}$. If we consider black holes with $\mbh \, \sim10^5$~\msun, for typical \sigmastar$ \approx 10-30$~\kms, then at the limiting distance of current samples (4 Mpc) the angular size is not more than $\sim 0.2"$
\citep[e.g.,][]{nguyenetal2018}. Another challenge is $M/L$ determinations, which are more complicated in low-mass galaxies, due to ongoing star formation and dust. A third is non-axisymmetric structures such as disks and bars, and other irregular star-forming features \citep[e.g.,][]{kormendykennicutt2004}. A fourth is that individual bright stars can dominate the total light output, biasing the measured stellar dispersion \citep[e.g.,][]{lutzgendorfetal2015}. In galaxies with ongoing star formation, this issue is particularly important. In the Milky Way, \citet{feldmeieretal2014} have shown that one blue supergiant could erase the signature of the BH.

The issue of identifying a robust galaxy center is a fifth significant concern. For instance, center determinations for the Large Magellanic Cloud (LMC) have an uncertainty that is larger than a square degree \citep{vandermarelkallivayalil2014}. For more distant galaxies, the field has avoided issues of determining galactic centers by focusing on galaxies with known nuclear star clusters, which encompasses nearly all galaxies with $M_* \sim 10^9-10^{10}$~\msun\ \citep{georgievboker2014,sanchez-janssenetal2019}. At lower galaxy masses, as the nucleation fraction drops and galaxies appear more irregular, we will be fundamentally limited by centroid determinations.

Finally, one of the most challenging limitations to dynamical masses for $<10^4$~\msun\ black holes in dense stellar systems is confusion between a putative IMBH and a central cluster of stellar-mass remnants. As dense stellar clusters evolve, the compact remnants sink to the center of the cluster and segregate in mass. Thus, the detailed assumptions made about the presence of stellar remnants (especially stellar-mass black holes) have a crucial impact on the relative evidence for an IMBH \citep[e.g.,][and references therein]{mannetal2019}. A related issue is that because of mass segregation and the desire to constrain IMBHs of modest mass, there may be very few luminous stars within the sphere of influence to provide dynamical constraints (see next subsection). Studies that do not carefully consider a range of predictions for stellar remnants will not reach robust conclusions about IMBHs. Observations in galaxy nuclei are already approaching this limit for the nearest sources \citep{nguyenetal2019}. 

\subsection{Proper Motions}
\label{sec:pm}

When possible, stellar dynamics can be more precisely measured through proper motions of individual stars, which can break the mass-anisotropy degeneracy (e.g., \citealt{zocchietal2017}). Nevertheless, there are still challenges with these measurements, which generally are looking for less massive black holes as well.  

As a test case, consider the well-studied globular cluster 47 Tuc. 
The sphere of influence of a $\sim 1000 M_{\odot}$ black hole in 47 Tuc would only be $\sim 1"$, and few stars will be observable within this radius. For example, studies of the central $1"$ with the \emph{Hubble Space Telescope (HST)} have only produced precise proper motions for 11--12 stars \citep{mclaughlinetal2006,mannetal2019} down to a main-sequence mass of $\sim 0.65 M_{\odot}$. Since the central mass function is relatively depleted of low-mass stars due to mass segregation, the \emph{total} number of stars observable in the sphere of influence of such an IMBH, even down to the hydrogen burning limit with next-generation extremely large telescopes, might still only be $\lesssim 50$ stars. With this modest sample, it is not clear that the advantages in breaking the mass-anisotropy degeneracy gained from proper motions will be definitive in proving the presence of $\sim 1000~M_{\odot}$ IMBHs. On the other hand, a $\sim 3000$--$4000~M_{\odot}$ IMBH would have a factor of $\sim 10$--15 more stars within the sphere of influence, and hence be considerably easier to detect than a $1000~M_{\odot}$ IMBH. 

The concerns about a misleading population of dense stellar remnants still apply to proper motion measurements. The current state-of-the-art is to use N-body codes to model the possible signal from such a cluster of stellar-mass compact remnants \citep[e.g.,][]{baumgardt2017}, which still leaves ambiguous cases. A possible way to alleviate this degeneracy is to determine indirect tests of the presence of stellar-mass compact objects such as expected numbers of pulsars or X-ray binaries. Perhaps more promising is the finding of \citet{macleodetal2016} that IMBHs should typically acquire companions with orbital periods of years, corresponding to semi-major axes of $\sim$ 5--10 mas for typical globular cluster distances and $\sim 1000 M_{\odot}$ IMBHs. In the models, the companions undergo frequent exchanges, but about half of the time these companions would be main sequence stars or giants, most of which should be observable in data of sufficient depth with extremely large telescopes. The proper motion due to the orbit should be very high compared to other cluster members ($\gtrsim$ a few mas per year), making such stars readily identifiable even in short-baseline observations, and distinguishable from foreground stars by their precise central location. The theoretical predictions in \citet{macleodetal2016} cover a small range of parameter space and it would be worthwhile to see extensions to a wider set of IMBH masses and core densities. This method could not prove the absence of an IMBH in any particular cluster, since a visible companion cannot be guaranteed. However, if few hundred \msun\ black holes are common in clusters, companion studies could push down to lower masses than any other method. The timing of millisecond pulsars in even wider orbits could also reveal the presence of an IMBH, though the interpretation of these timing observations is not necessarily straightforward. We revisit some of these issues below in \S \ref{sec:globdynamics}.

\subsection{IMBH Demographics in Galaxy Nuclei From Dynamics}

The most definite existing dynamical measurements are for nucleated $10^9-10^{10}$~\msun\ galaxies within $\sim 4$ Mpc of the Sun. Ten such galaxies have published black hole masses or limits from stellar and gas dynamics (see Figure \ref{fig:existinglimits}). There are published detections in five of these galaxies: M32, NGC 5102, NGC 5206, NGC 205 \citep{nguyenetal2018,nguyenetal2019}, and NGC4395 \citep{denbroketal2015}. There are published upper limits for five additional galaxies: NGC 300 and NGC 7793 \citep{neumayerwalcher2012}, NGC 404 \citep{nguyenetal2018}, the LMC \citep{boyceetal2017} and M33 \citep{gebhardtetal2001}. Taking these ten objects and five detections at face value, we have a lower limit on the occupation fraction of $>50$\%. This is a lower limit because the measurements are really only sensitive to black holes with \mbh$>10^5$~\msun\ for galaxies outside of the Local Group.   

It is worth commenting specifically on the case of the LMC. Later in this section we discuss indirect evidence (from a hypervelocity star) for a massive black hole somewhere in the LMC. Because the LMC is so close to the Milky Way, we have high spatial resolution and thus sensitivity to even lower-mass black holes. However, the increased resolution is debilitating. No measurement of the center of the LMC is in good agreement with any other \citep{vandermarelkallivayalil2014}, and because there is an offset bar it is very challenging to know where to search for a putative massive black hole. The best limit we have thus far was made with great effort by \citet{boyceetal2017} using VLT/MUSE observations, but their limit of \mbh$<10^7$~\msun\ is not terribly constraining.

For galaxies with lower masses ($M_*<10^9$~\msun) within the Local Group, there are three other interesting published upper limits to mention. In order of decreasing stellar mass, we have limits in the dwarf galaxies Sagittarius, Fornax, and Ursa Minor. In the case of Sagittarius, the nucleus is the globular cluster also known as M54. Since the galaxy is actively being disrupted, indirect means are needed to estimate the original mass of the galaxy. Summing the light in the tidal features \citep{niedersteostholtetal2010}, modeling of the stream \citep{laporteetal2018}, and stellar abundances and abundance ratios \citep{deboeretal2014} all suggest that the galaxy was one of the more massive satellites, comparable to the Small Magellanic Cloud (SMC), with a stellar mass of a few $\times 10^8$~\msun. Dynamical measurements of the center of M54 are reviewed below, but there is no consensus on the presence of a black hole in this cluster \citep{ibataetal2009,baumgardt2017}.

Moving downward in mass to the Fornax dwarf galaxy, \citet{jardelgebhardt2012} present an orbit modeling limit of \mbh$<10^5$~\msun\ ($3\sigma$ limit) on any black hole. Finally, there is an interesting published limit on a massive black hole in the galaxy Ursa Minor. \citet{loraetal2009} argue that any centrally located black hole with \mbh$>3 \times 10^4$~\msun\ would dissolve observed stellar clumps. The major caveat here, also noted by the authors, is that the initial location of the black hole in the galaxy is unconstrained \citep{bellovaryetal2019}. To find any $10^3-10^4$~\msun\ black holes that may be lurking in Local Group dwarfs will likely require proper motions with an extremely large telescope \citep[e.g.,][]{greeneetal2019}.

Finally, at slightly larger distances out to 10 Mpc, \citet{neumayerwalcher2012} perform Jeans modeling for a sample of nine nucleated spiral galaxies in the $M_*=10^9-10^{10}$~\msun\ range. There are published upper limits for NGC 3621 \citep{barthetal2009} and NGC 4474 \citep{delorenzietal2013} as well. Together, these limits will prove crucial in measuring scaling relations in this mass range (\S \ref{sec:scaling}).

\subsubsection{Sample Completeness}

In \S \ref{sec:demographics} below, we will use the dynamical sample of ten galaxies with $10^9<M_*/\msun<10^{10}$ and $D<4$ Mpc as one constraint on the occupation fraction. Although there is not a truly volume-complete sample of galaxies with dynamical measurements, there is no obvious bias in the galaxies that have been targeted dynamically. The searches have focused on galaxies with known nuclear star clusters, but the nucleation fraction among galaxies in this stellar mass range is as high as $\sim 80-90\%$  \citep{georgievboker2014,sanchez-janssenetal2019}. Neumayer et al.\ (in preparation) do not see any differences between the nucleation fractions of red and blue galaxies. On the other hand, the nucleation fractions measured for blue galaxies in the $M_* = 10^9-10^{10}$~\msun\ range may be biased against galaxies with ongoing vigorous star formation. A case in point is the LMC, which has no readily-identified nuclear star cluster, although as discussed above its center is unknown. 

To explore these issues a bit more quantitatively, we rely on the nearby galaxy catalog of \citet{karachentsevetal2013}. This catalog includes $K-$band magnitudes and we make the simplifying assumption that all galaxies have $M_K/M_{\odot} \approx 1$. There are 21 galaxies with $1.5<D<5$~Mpc and $10^9<M_*/M_{\odot}<10^{10}$ by this definition. Of these 21, 13 have known and well-studied nuclear star clusters, while one galaxy has none \citep[NGC 55;][]{sethetal2006}. Most of the remaining seven galaxies have imaging with $HST$; it is a high priority to examine these for the presence of nuclei (Hoyer et al.\ in preparation). In terms of mass and size, we do not see obvious biases in the nuclear star cluster properties in this sample relative to \citet{georgievboker2014}, nor is there evidence for differences between red and blue galaxies \citep[see also][]{foordetal2017}. Finally it is worth noting again that dynamical measurements are particularly challenging in blue galaxies, due both to their complicated non axisymmetric kinematics and the shotnoise from individual young stars, which must be considered as more dynamical constraints become available.


\subsection{Dynamical Searches for IMBHs in the Milky Way}

Another special nucleus postulated to house an IMBH is the one within our own Galactic Center. The Milky Way may host a population of ``leftover'' IMBHs from past accretion of dwarf galaxies \citep[e.g.,][ and \S \ref{sec:seedmodels}]{rashkovmadau2014}. IMBHs have been invoked to explain several observational phenomena associated with the Galactic Center, although, to date, none of the evidence can be regarded as definitive. It is worth recalling that rather strict constraints exist from the small residual proper motion of Sgr A* perpendicular to the plane of the Galaxy: no dark object larger than $10^4$~\msun\ is permitted within $10^3-10^5$~AU from Sgr A$^*$ \citep{reidbrunthaler2004}.

The ``paradox of youth’’ of the stars within the central parsec of the Galactic Center (Ghez et al. 2003) has inspired mechanisms to shepherd stars into the Galactic Center with the aid of an IMBH \citep{hansenmilosavljevic2003}. The inward migration of an IMBH can also account for the origin of the kinematic distribution of these young stars \citep{yuetal2007}.   To this end, the bright infrared source IRS 13E, at a projected distance of only 0.13 pc from Sgr A$^*$, has been the subject of intense scrutiny.  High-spatial resolution observations by \citet{maillardetal2004} resolve the source into a compact group of several co-moving massive stars, prompting speculation that it constitutes the disrupted core of a young massive cluster in which an IMBH has coalesced by runaway growth \citep[e.g.,][]{portegieszwartetal2002}.  The proper motion measurements of \citet{schodeletal005} would require a large black hole mass of $>10^4$~\msun, which is difficult to reconcile with the absence of clear non-thermal radio and X-ray emission. A dark object of this mass scale, however, would satisfy the ionized gas kinematics recently reported by \citet{tsuboietal2017}. 

Gas kinematics can be notoriously tricky to interpret.  This challenge is well exemplified by the compact cloud CO-0.40-0.22, whose large line-of-sight velocity and large internal velocity dispersion ($\sim$100 km s$^{-1}$) prompted \citet{okaetal2016} to suggest that it experienced a gravitational kick from a dark $10^5$~\msun\ object within 60 pc of the Galactic Center.  The spectrum of its associated millimeter continuum and IR source, however, is more consistent with that of a protostellar disk instead of a scaled-down version of Sgr A$^*$ \citep{ravieetal2018}, and its detailed kinematics are more consistent with cloud-cloud collisions \citep{tanaka2018} or supernova-driven interactions \citep{yalinewichbeniamini2018}. Interest in this topic continues unabated \citep{takekawaetal2019,tsuboietal2019}.

\subsection{IMBH Demographics in Globular Clusters From Stellar Dynamics}
\label{sec:globdynamics}

Many papers have presented dynamical evidence for IMBHs in globular clusters, but as of this writing there are no systems for which such evidence is unambiguous. 

Perhaps the best-studied system is $\omega$ Cen. There are claims of a massive IMBH ($\sim 4-5 \times 10^4 M_{\odot}$) in this star cluster based on isotropic modeling of the radial velocity dispersion and surface brightness profiles \citep{noyolaetal2010,jalalietal2012,baumgardt2017}. However, the velocity dispersion signature of an IMBH is not found in the proper motions of central stars \citep{vandermareletal2010}, and no resolution of this observational issue has been given in the literature. On the modeling side, anisotropy and the presence of dark remnants could account for most or all of the IMBH signature found in other studies (e.g., \citealt{zocchietal2017,zocchietal2019,baumgardtetal2019}).
Some papers have highlighted the need for more sophisticated modeling of the velocities and proper motions of individual stars (see \S \ref{sec:pm}) rather than a binned dispersion profile in the context of $\omega$ Cen and other clusters \citep{lutzgendorfetal2012,baumgardtetal2019,mannetal2019}. Additional observations and modeling are clearly needed for $\omega$ Cen, but in any case, the contrasting results highlight the challenges of this work. These challenges are only magnified as one considers IMBHs of lower mass. 

Claims of dynamical evidence for massive black holes have been made in multiple other clusters, including for NGC 6388 and M54. The case of NGC 6388 is similar to $\omega$ Cen, with contrasting results rooted partially in conflicting observational results \citep{lutzgendorfetal2011,lanzonietal2013,lutzgendorfetal2015}. Multiple studies working with similar data sets have found tentative evidence for a $\sim 10^4 M_{\odot}$ IMBH in M54, with the same interpretation caveats as for $\omega$ Cen, and the additional complication that the cluster is embedded in the remnants of the Sagittarius dwarf galaxy \citep{ibataetal2009,baumgardt2017}. There is no accretion evidence for an IMBH in any of these clusters \citep[][\S \ref{sec:searchoff}]{tremouetal2018}. As a promising example of the dynamical limits possible using forefront instrumentation for nearby, dense globular clusters, \citet{kamannetal2016} show that any IMBH in NGC 6397 must be $\lesssim 600 M_{\odot}$ (Figure \ref{fig:existinglimits}), consistent with the radio limit for this cluster \citep{tremouetal2018}.

IMBH searches have been extended outside the Milky Way to the nearest massive galaxy with a large globular cluster population, M31. The best-studied globular cluster in M31 is G1. At this distance ($D=0.8$~Mpc), the observational complication is that the putative sphere of influence of an $\sim 2\times10^4 M_{\odot}$ IMBH is barely resolved with \emph{HST} spectroscopy and imaging. G1 was notable not solely for the contested interpretation of the central kinematics \citep{gebhardtetal2002,baumgardtetal2003,gebhardtetal2005}, which have the same issues already discussed, but also for the unique addition of X-ray and radio accretion evidence for a possible IMBH \citep[][and \S \ref{sec:searchnuclei}]{pooleyrappaport2006,ulvestadetal2007}. Unfortunately, this radio emission was not confirmed in deeper observations, and the multi-wavelength data are consistent with a standard low-mass X-ray binary \citep{millerjonesetal2012}. Hence G1 belongs to a similar category as $\omega$ Cen and M54, with debated dynamical evidence for an IMBH.

Several recent papers have taken a new dynamical tack: using the timing properties of millisecond pulsars in the cores of globular clusters to constrain the presence of an IMBH. \citet{kiziltanetal2017} modeled the properties of a subset of pulsars in the central regions of 47 Tuc, finding evidence that pulsar accelerations were best-explained by the presence of a central IMBH of mass $\sim 2300 M_{\odot}$. Subsequent analyses of similar pulsar data sets did not reproduce this result: they found no evidence for an IMBH, with a formal 99\% upper limit of $< 4000 M_{\odot}$ \citep{freireetal2017,abbateetal2018}. This non-detection is consistent with the results of several studies that modeled the detailed proper motions of individual stars in the core of 47 Tuc \citep{mclaughlinetal2006,mannetal2019}, and found $1\sigma$ upper limits in the range $< 1000$--1500 $M_{\odot}$, and also consistent with the $3\sigma$ radio upper limit of $< 1040 M_{\odot}$ \citep{tremouetal2018}. We conclude that there is no compelling evidence for an IMBH in 47 Tuc based on available observations and modeling. 

\citet{pereraetal2017} argue that a single pulsar in NGC 6624 has timing properties consistent with being in a very long period, eccentric, loosely bound orbit around an IMBH of mass $> 7500 M_{\odot}$. Other interpretations of these data are possible, and recent dynamical modeling of stars constrains an IMBH to be $\lesssim 1000 M_{\odot}$ \citep{baumgardtetal2019}. Future observations and modeling will help distinguish among these possibilities.

\begin{textbox}[b]
\subsection{IMBH Demographics in Globular Clusters From Microlensing}

There is also a promising alternative method for IMBH searches in globular clusters. Typically, the microlensing effects discussed are ``photometric", resulting in magnification of the background source by the lens. However, the optical depth for photometric microlensing by IMBHs is very low \citep{safonova2010},
and detections are unlikely for reasonable monitoring campaigns of Galactic globular clusters. \citet{kainsetal2016} instead suggest the possibility of using astrometric microlensing, which is much less sensitive to the angular separation of the source and lens than standard microlensing. The detectability of this astrometric microlensing signal is maximized for clusters close to the Sun that also have high background densities. This methodology was used to search for evidence of a central IMBH in M22 with \emph{HST} by \citet{kainsetal2018}. Owing to gaps in the astrometric  time series the resulting upper limit on an IMBH was not constraining compared to limits from other methods \citep{straderetal2012}, but future observations with the \emph{JWST} or the \emph{Wide Field Infrared Survey Telescope} (\emph{WFIRST}) should provide improved constraints to IMBH masses $\lesssim 10^{4} M_{\odot}$ in a subset of nearby clusters, including M22, M4, and 47 Tuc.

\end{textbox}

\subsection{Hypervelocity stars}

Hypervelocity stars are those with Galactocentric velocities in excess of the escape velocity at their present location. There is compelling evidence that some of the most extreme hypervelocity stars are due to interactions between stellar binaries and the supermassive black hole at the center of the Galaxy \citep{hills1988,brownetal2005,zhangetal2013,brownetal2018,Koposovetal2019}. The remainder, including some of the stars with less extreme velocities, likely have a wide range of origins, including close binaries disrupted by the death of one of the stars, dynamical encounters in star clusters, and even tidal material from accreted satellites (e.g., \citealt{hirschetal2005,peretsetal2012,abadietal2009,shenetal2018}).

If star clusters or Galactic satellites host IMBHs, then extreme hypervelocity stars whose kinematics exclude a Galactic Center origin could provide evidence for IMBHs. Perhaps the best example of this is for the hypervelocity star HVS3, which may have been ejected from the LMC \citep{edelmannetal2005} by an interaction with an IMBH \citep{gualandrisetal2007}. The star is at most $\sim 35$ Myr old, so the ejection must have been relatively recent, and from a young stellar population \citep{edelmannetal2005}. \citet{erkaletal2019} add \emph{Gaia} data to the analysis and show that an origin in the LMC is much more likely than from the Galactic Center \citep[see also][]{lennonetal2017}. They argue the relative velocity of HVS3 ($\sim 870$ km s$^{-1}$) could only have originated in an interaction with a $\sim 4 \times 10^3$--$10^4 M_{\odot}$ IMBH. This should intensify efforts to search for other evidence of an IMBH near the center of the LMC or in its young star clusters. We note that the confirmation of an IMBH in one of these clusters (which typically have $M_{\star} \lesssim 10^5 M_{\odot}$) would be remarkable.

Hypervelocity stars may also prove to be the most robust probe of IMBHs in the Galactic Center \citep{yutremaine2003,baumgardtetal2006}. The observed spectrum of ejection velocities appears to be inconsistent with theoretical expectations for a supermassive black hole--intermediate-mass black hole binary, but the current data leave much room for improvement \citep{sesanaetal2007}.  Even more powerful would be the eventual future detection of hypervelocity binary stars \citep{luetal2007,sesanaetal2009}. 

Future missions such as LSST and \emph{WFIRST} will improve the ability to search for hypervelocity stars as signposts to IMBHs in Galactic satellites or globular clusters, since they will be sensitive to more typical lower-mass stars. Because of the longer lifetimes of these stars, they may have more possible kinematic origins, and other information like chemical abundances and abundance ratios may be needed to interpret their origin.

\section{Searches for Accreting IMBHs in Galaxy Nuclei}
\label{sec:searchnuclei}

Even with next-generation facilities, dynamical measurements will only reach $\sim 10$~Mpc. Thus, to gain population statistics we must rely on accretion signatures to identify the presence of a black hole. We discuss the physics of accretion signatures at different wavelengths, and the samples that have resulted from searches so far. 

\subsection{Optical Spectroscopic Selection}

The two prototype low-mass active galactic nuclei (AGNs), NGC 4395 \citep{filippenkosargent1989,filippenkoho2003} and POX 52 \citep{kunthetal1987,barthetal2004}, were both originally identified based on optical spectroscopic signatures. In light of the apparent rarity of AGNs in late-type hosts \citep{hfs1997,ho2008}, large spectroscopic surveys are needed to tease out any significant statistical sample.  \citet{greeneho2004} performed the first systematic search for AGNs powered by low-mass black holes using SDSS DR1, producing 19 low-redshift ($z < 0.35$) broad-line AGNs with estimated black hole masses $\lesssim 10^6$~\msun, subsequently boosted to a sample of $\sim 200$ using SDSS DR4 \citep{greeneho2007a,greeneho2007b}. Although the AGN-based masses are uncertain (see later in this section), follow-up shows that the ensemble of these objects are powered by low-mass black holes. The host galaxies are sub$-L_*$, disky galaxies \citep{greeneetal2008,jiangetal2011}, with the low gas-phase metallicities expected at these masses \citep{ludwigetal2012}. Subsequent efforts have adopted variants of this strategy to enlarge and refine the broad-line sample, which to date stands at $\sim 500$ sources \citep{dongetal2012b,chilingarianetal2018,liuetal2018}. These objects can only be detected in SDSS when they radiate at close to their Eddington limits, and as such they are rare, comprising only $f_{\rm AGN} \approx 0.1\%$ of the local galaxy population with $M_*<10^{10}$~\msun.

In addition to selecting on broad-line properties, one can also pre-select low-mass galaxies and look for those whose emission lines classify them as AGN based on the ``BPT" diagram \citep{baldwinetal1981}, which classifies objects based on line ratios between strong inter-stellar medium lines. AGNs have long been known to separate in these diagrams due to their hard ionizing spectra \citep[e.g.,][]{hfs1997}. A number of groups have started with a stellar-mass selected sample, $M_* < 10^{10}$~\msun\ or less, to search for AGN signatures in dwarf galaxies \citep{barthetal2008,reinesetal2013,moranetal2014,sartorietal2015}. Moran et al.\ focus on 28 sources within 80 Mpc ($f_{\rm AGN} = 2.7\%$), while Reines et al.\ present 151 AGN candidates from a parent sample of $\sim 25,000$ emission-line galaxies ($f_{\rm AGN} \approx 0.5\%$).  Most AGN uncovered in this manner are narrow-line objects, but a subset have broad H$\alpha$ emission indicating  $M_{\rm BH} \approx 10^5-10^6$~\msun.  The late-type spiral RGG118 \citep{baldassareetal2017} even has a black hole as small as 50,000~\msun\ \citep{baldassareetal2015}.

For the local AGN selected from SDSS, there is substantial multi-wavelength follow-up that can inform further searches.  Multi-wavelength follow-up of the Barth et al.\ sample confirm their obscured AGN nature \citep{thorntonetal2009,hoodetal2017}. The AGN selected based on broad emission lines behave like AGN powered by more massive black holes. In the radio, a small number of objects have deep follow-up observations but few are radio loud \citep{greeneetal2006radio,wrobelho2006,wrobeletal2008}. More attention has been devoted to the X-rays \citep{greeneho2007c,desrochesetal2009,dongetal2012a,yuanetal2014,plotkinetal2016}, using \emph{Chandra} and \emph{XMM-Newton} observations of sufficient depth to perform detailed spectral and timing analysis \citep{moranetal2005,dewanganetal2008,thorntonetal2008,miniuttietal2009,aietal2011,kamizasaetal2012,jinetal2016}. The results from X-ray timing are particularly critical, as they help to lend confidence to and independently confirm the black hole mass estimates from broad H$\alpha$.  As a class, these objects are among the most rapidly variable extragalactic X-ray sources \citep{dewanganetal2008,miniuttietal2009,aietal2011,kamizasaetal2012}, pointing to low-mass black holes.  

In general, optically selected AGNs appear to be rare in dwarf galaxies. The AGN detection rate for low-mass galaxies using traditional optical tracers seems to hover around $\sim 1\%$ (Supplemental Table \ref{tab:fagn}). It so far has been prohibitive to correct for the major incompleteness in these samples stemming from star formation contamination, aperture dilution, and dust reddening \citep[e.g.,][]{greeneho2007a,trumpetal2015}.

\subsubsection{Reverberation Mapping}

For objects with broad emission lines, ``reverberation mapping'' yields information about the size scale of the broad-line region (BLR) by measuring the delay between the continuum and line light curve, emitted from the accretion disk and BLR respectively \citep{peterson2014}. Combining the BLR radius $r$ with the  line width $\Delta V$, yields a virial-like mass $M_{\rm BH} = f_{\rm vir} r (\Delta V)^2/G$, with $f_{\rm vir}$ the the virial constant. The low luminosities of low-mass AGNs suggests that their BLRs will be compact, and hence any attempt at reverberation mapping must have sufficiently high cadence to sample lags of less than a few days at most. We emphasize that the reverberation mapping-based masses are currently calibrated with the dynamical samples through  $f_{\rm vir}$. Because the structure and kinematics of the broad-line region are unknown, reverberation mapping yields a ``virial product'', which is currently scaled by $f_{\rm vir}$ such that the AGN samples obey the same \mbh-\sigmastar\ relations as inactive galaxies \citep[although see][]{pancoastetal2014}. This scaling has at least a factor of two ambiguity in it depending on what galaxy samples are used in the \mbh-\sigmastar\ calibration sample \citep[e.g.,][]{hokim2014}. 

The initial effort to monitor the prototype low-mass AGN NGC 4395 gave only a marginally useful constraint on the lag for H$\beta$ \citep{desrochesetal2006}, and preference has been given to the C {\small IV}~$\lambda 1549$ measurement of \citet{petersonetal2005}, which led to a black hole mass estimate of \mbh$= (3.6 \pm 1.1) \times 10^5$~\msun. This mass is consistent, within the considerable tolerance of the uncertainties, with the direct dynamical estimate of \mbh$= 4^{+8}_{-3} \times 10^5$~\msun\ by \citet{denbroketal2015}. \citet{wooetal2019} recently advocate a markedly lower value of \mbh$\approx 9100$~\msun, based on a short, 80-min lag detected from a narrow-band H$\alpha$ reverberation mapping campaign with rapid sampling. The published mass adopts a virial constant $f_{\rm vir} = 4.5$.  Using a value calibrated to lower-mass spiral galaxies \citep[$f_{\rm vir} = 3.2$][]{hokim2014}, the mass formally drops to \mbh$\approx 6500$~\msun. The difference in black hole mass between Woo and Peterson is mainly due to differences between the linewidth of C~{\small IV} and H$\beta$, while the two works recover consistent lags. Very few studies have done intercomparisons of multiple lines in the context of reverberation mapping, and certainly not in this mass and luminosity range \citep[e.g.,][]{parketal2017}.

Somewhat more massive but still in the neighborhood of $10^6$~\msun\ or less is UGC 6728, whose H$\beta$ lag of $\tau = 1.5 \pm 0.8$ days yields \mbh$= (5.2 \pm 2.9) \times 10^5$~\msun\ \citep[][scaled to our preferred $f_{\rm vir}=3.2$]{bentzetal2016}. The published lag for SDSS J114008.71+030711.4  \citep[GH08 from][]{greeneho2004} of $\tau = 1.5^{+4}_{-2}$~days is short and highly uncertain \citep[][]{rafteretal2011}, but it is likely less than 6 days, in which case \mbh$< 3.9 \times 10^5$~\msun. Lastly, we mention three Seyfert 1 nuclei with well-measured H$\beta$ lags hosted in late-type spiral galaxies, as summarized in \citet{hokim2014}: NGC 4051 with $\tau = 2.5 \pm 0.1$~days and log ($\mbh/\msun) = 6.11\pm0.04$; NGC 4253 (Mrk 766) with $\tau = 5.4^{0.2}_{-0.8}$~days and log ($\mbh/\msun) = 5.98 \pm 0.29$; and Mrk 202 with  $\tau = 3.5 \pm 0.1$~days and log ($\mbh/\msun) = 5.98\pm0.06$. 

Reverberation mapping also provides a relationship (the so-called ``radius-luminosity'' relation) between the AGN luminosity and the typical size of the broad-line region \citep[e.g.,][]{bentzetal2013}. Using the virial constant and the radius-luminosity relation, we can calculate ``single-epoch'' virial black hole masses for broad-line AGN. It is hard to know what the systematic uncertainties are on these masses, particularly at low black hole mass where the radius-luminosity relation is not directly measured.

\subsection{Multi-wavelength Searches and Confusion With Star Formation}

Searching for concrete evidence of AGNs in low-mass galaxies poses a set of unique challenges.  One of the major complications in applying the traditional optical BPT diagrams is that the AGNs become hopelessly intermingled with star-forming galaxies at low metallicity \citep{grovesetal2006,stasinskaetal2006,cannetal2019}. Along with a BPT selection, \citet{sartorietal2015} selected additional samples using both a He {\small II} $\lambda 4686$ diagnostic diagram and  mid-IR color cuts that have proven effective in selecting luminous AGNs \citep{jarrettetal2011,sternetal2012}. Distressingly, almost none of the samples identified by the three methods overlap. Why?  

As \citet{hainlineetal2016} emphasize, young starbursts in the low-metallicity environment of dwarf galaxies have red mid-IR colors that closely mimic those of AGNs.  This, unfortunately, calls into question the usage of mid-IR color to select AGNs in late-type, low-mass galaxies 
\citep{satyapaletal2014,marleauetal2017,kavirajetal2019}. The lower metallicity environment of dwarf galaxies is characterized by higher electron temperatures and higher levels of excitation for the ISM. In theory, there may also be a more top-heavy stellar initial mass function \citep[e.g.,][]{brommetal2002}. The preponderance of massive stars profoundly affects the heating and ionization of the gas.  High-mass X-ray binaries may be responsible for the ionizing photons for He {\small II}~$\lambda 4686$ \citep{schaereretal2019}.  If so, 
He~{\small II} ceases to be a useful AGN indicator in dwarf galaxies \citep{sartorietal2015,baretal2017}.   Massive O stars and Wolf-Rayet stars can generate sufficient extreme UV radiation to excite high-ionization lines such as [O {\small IV}] 25.89$\mu$m \citep{lutzetal1998,schaererstasinska1999}, which renders moot any attempt to use this line to select low-mass AGNs \citep{georgakakisetal2011}. [Ne {\small V}] 14.32$\mu$m, normally considered a robust AGN indicator because of its high ionization potential of 97.12 eV \citep{satyapaletal2007,satyapaletal2008,satyapeletal2009,gouldingetal2010} is not immune either, as it can be excited by fast-shocks produced by stellar winds from massive stars and supernovae \citep{contini1997,izotovetal2012}.

As a case in point, we draw attention to the local spheroidal NGC 185, whose nuclear optical line emission, though feeble, technically qualifies it as a ``Seyfert 2’’ galaxy \citep{hfs1997}.  However, the recent detailed spatially resolved optical and X-ray analysis of this object by \citet{vuceticetal2019} clearly shows that the excitation of the central nebula is due to supernova remnants. NGC 185 is a fake AGN. This caveat may well impact larger samples of AGN selected based on narrow emission lines.

Even broad emission lines are not sacrosanct.  While the presence of broad H$\alpha$ is typically regarded as ironclad evidence of an AGN, high-velocity gas can also be of stellar origin.  Wolf-Rayet galaxies, for example, often exhibit broad wings to the H$\alpha$ line \citep{mendezesteban1997}, and the optical spectra of some Type II supernovae bear an uncanny resemblance to Seyfert 1 nuclei \citep{filippenko1989}. \citet{baldassareetal2016} obtained multiple-epoch observations of the type 1 sources from \citet{reinesetal2013} with evidence of star formation in their narrow-line spectra and discovered that in most of them the broad H$\alpha$ line is transient over a baseline of several years, suggesting a supernova origin. Even when broad H$\alpha$ is persistent and too strong to be explained easily by supernovae, as is the case in some blue compact dwarfs \citep{izotovthuan2008,izotovetal2010}, no compelling, independent evidence for AGNs has yet surfaced.  Follow-up \emph{Chandra} observations of the metal-poor AGN candidates of \citet{izotovthuan2008} reveal that their X-ray emission is far weaker compared to their optical or mid-IR emission than expected for active galaxies \citep{simmondsetal2016}. Even sensitive X-ray and radio observations do not find compelling evidence for AGN in blue compact dwarfs \citep{latimeretal2019}. There is also the converse problem, that broadened lines with $\sim 200-400$~\kms\ may arise from the narrow-line region of an obscured AGN, with the width reflecting non-virial motions associated with the AGN. In such cases the linewidth may not have any relation to the black hole mass. Studies that push to ambitiously low linewidth may suffer this contamination \citep{chilingarianetal2018}.

Several of the AGN candidates in late-type galaxies originally identified through detection of  [O {\small IV}] 25.89~$\mu$m or [Ne {\small V}] 14.32~$\mu$m have since been followed up in X-rays \citep{gliozzietal2009,mcalpineetal2011,georgakakisetal2011,secrestetal2012,hebbaretal2019}, and in general the X-rays are found to be weaker than expected.  Diffuse, thermal emission is occasionally detected when the data are of sufficient quality.  Certainly one can appeal to absorption to explain the deficit of hard X-rays, but we cannot rule out the possibility that these galaxies actually lack AGNs.

\subsection{Pushing Down the Luminosity Function with X-ray or Radio Observations}

Given the observation of central black holes in the Local Group with very low Edddington ratios (including Sgr A$^{\star}$ and M31$^{\star}$), the minority population of highly accreting black holes identified by optical spectroscopy must be the tip of the iceberg. In addition to being faint, AGNs of low accretion rate have systematically lower ionization parameters \citep{ho2009}, and they may lack broad emission lines altogether \citep{elitzurho2009}.  X-ray observations provide a clean tool to overcome these problems in a wide range of circumstances.  Black hole accretion invariably generates X-ray emission \citep{brandtalexander2015}, and low accretion rates have the virtue of producing proportionately even more hard X-rays \citep{ho2008}. The resolving power of \emph{Chandra} and the low background of its ACIS instrument offer the ideal combination to detect faint compact sources in nearby galaxies, even with short exposures, and the excellent astrometric accuracy of the satellite can align the optical or near-IR nucleus of the galaxy to within 1'' or better.  The main source of confusion comes from X-ray binaries, but the degree of contamination can be estimated once the stellar mass and star formation rate are known \citep[e.g.,][]{milleretal2015}.

A number of studies have exploited this opportunity to evaluate the incidence of AGNs in nearby late-type galaxies, succeeding in identifying X-ray nuclei in star-forming \citep{ghoshetal2008,zhangetal2009,grieretal2011,sheetal2017a}, dwarf irregular \citep{lemonsetal2015}, and local Lyman-break analog \citep{jiangetal2011} galaxies. \citet{desrochesho2009} analyzed \emph{Chandra} images of 64 Scd-Sm spirals within 30 Mpc and discovered an X-ray core in 17 of them ($f_{\rm AGN} = 27\%$).  The Sc-Sm spirals in the sample of Zhang et al. (2009) yield a consistent result ($f_{\rm AGN} = 30\%$). \citet{sheetal2017} extended this effort to a more comprehensive census of more than 700 galaxies within 50 Mpc; among late-type, bulgeless spirals, the detection rate of X-ray cores is $f_{\rm AGN} = 21\%$.  Unlike the optically selected sources, these X-ray selected nuclei are all highly sub-Eddington, with median $L/L_{\rm bol} \approx 10^{-4}$. Archival work searching for X-rays in ultra-compact dwarfs has been far less conclusive \citep{pandyaetal2016}.

The above efforts, largely based on archival data, have been complemented by a series of experiments aimed at characterizing the incidence of X-ray nuclei using dedicated \emph{Chandra} surveys of nearby, lower mass early-type galaxies, focused on the Virgo cluster \citep{ghoshetal2008,galloetal2010}, the Fornax cluster \citep{leeetal2019}, and in the field \citep{milleretal2012,gallosesana2019}.  Among the $\sim 200$ early-type galaxies within 30 Mpc uniformly observed to date, \citet{milleretal2015} conclude that for galaxies with stellar mass $M_* < 10^{10}$~\msun, $f_{\rm AGN} > 20\%$. Their observations in this stellar mass range are sensitive to $L/L_{\rm bol} \approx 10^{-4}$, but in more massive galaxies, the same depths probe much lower $L/L_{\rm bol}$. By assuming that the Eddington ratio distribution can be modeled as a smooth function of $M_*$, Miller et al.\ bracket the occupation fraction to fall between $30\%$ and $100\%$ at $1 \sigma$ \citep[see also][]{gallosesana2019}. Work by \citet{airdetal2013} supports the idea that the Eddington ratio distribution varies only very mildly as a function of stellar mass. Finally, it is very encouraging that the active fractions uncovered from X-rays in both star-forming and quiescent low-mass galaxies are comparable. These X-ray results, combined with the dynamical ones above (\S \ref{sec:dynamics}), strongly suggest a high ($>50\%$) occupation fraction in $M_* = 10^9-10^{10}$~\msun\ galaxies.

Recently \citet{reinesetal2019} used deep and high-resolution radio imaging with the VLA to follow up 111 dwarf galaxies ($3 \times 10^7 < M_*/\msun < 3 \times 10^9$) with prior radio detections in FIRST. Of these 13 are likely to be powered by AGN, due to their point-like nature and luminosity relative to their star formation rates. Only one of these 13 was also identified by optical spectroscopy. These data cannot be used to measure the AGN fraction in a straightforward way because of the sample selection, but they do highlight the promise of even existing radio telescopes to unveil lower luminosity AGN populations. Intriguingly, many of the sources are found offset from their galaxy nucleus, perhaps consistent with predictions \citep{bellovaryetal2019,pfisteretal2019} that in low-mass galaxies the seed black holes may never settle at galaxy centers (\S \ref{sec:seedmodels}). Another possible origin for some of these sources may be fueling of wandering black holes (see also \S \ref{sec:searchoff}).

\subsection{Going Further with the Fundamental Plane of Radio Activity}

Combining radio emission with X-rays could be even more effective at probing AGNs with very low Eddington ratios. \citet{merlonietal2003} and \citet{falckeetal2004} found that both supermassive and stellar-mass black holes show fundamental similarities in their accretion flows in the radiatively inefficient low/hard state. Observationally, the X-ray luminosity ($L_X$; a product of the accretion rate and radiative efficiency) and the radio continuum luminosity ($L_R$; a measure of the jet power) scale with the mass of the black hole in a simple manner, such that a combination of these three quantities form a tolerably clean two-dimensional sequence (the fundamental plane) in three-dimensional space.

The physical cause of the fundamental plane is not necessarily straightforward, since the spectral energy distributions of accretion flows are expected to change systematically with black hole mass. There are no clear conclusions about whether most or all of the radio and X-ray emission is associated with a jet, or whether the X-ray emission might instead come from a corona that is separate from the jet, though still presumably linked to the accretion flow (see, e.g., the discussion in \citealt{plotkinetal2012}).  

A major challenge in using the fundamental plane comes from the large scatter. This considerable scatter must be at least partially intrinsic, based on a number of arguments. One is that among X-ray binaries, $L_R/L_X$ can vary by a factor of at least a few at fixed $L_X$ (e.g., \citealt{jonkeretal2012}). Another is that even in the newest modeling of the fundamental plane---with careful restriction to objects with high-quality dynamical masses and radio and X-ray data, and consideration of the measurement uncertainties---the scatter for determining black hole masses is still large at $\sim 1$ dex. The best-fit plane from \citet{gultekinetal2019} is: log $(M/10^8 M_{\odot})$ = $(1.09\pm0.10)$ log $(L_R/10^{38}$ erg s$^{-1}$) $(-0.59\pm0.16)$ log $(L_X/10^{40}$ erg s$^{-1}$)  + ($0.55\pm0.22$). We note that the use of previous fits based on careful sample selection and fitting (e.g., \citealt{plotkinetal2012}) would give similar results within the uncertainties in most cases.

Despite these uncertainties, many works have moved forward in a observational spirit to use the fundamental plane to constrain black hole masses in circumstances where more direct measurements are prohibitively difficult or impossible. 

Soon after the discovery of the fundamental plane, \citet{maccarone2004} pointed out that this would be a powerful tool to search for IMBHs in globular clusters (\S \ref{sec:searchoff}). The first high-profile use of the fundamental plane to show evidence for a modest mass nuclear black hole was in the starbursting galaxy Henize 2-10 \citep{reinesetal2011}. They used the radio and X-ray data to argue for the presence of a log $M/\msun = 6.3\pm1.1$ supermassive black hole. Follow-up very long baseline interferometry observations constrained the radio source size to be $<3 \times 1$ pc and confirmed a non-thermal origin \citep{reinesdeller2012}.  Higher-resolution data revealed multiple components to the central X-ray source, moving the fundamental plane mass estimate up to log $M/\msun \sim 7$ \citep{reinesetal2016}. Subsequent work showed that the X-ray spectrum of the nuclear source is more consistent with a supernova remnant than an AGN \citep{hebbaretal2019}, although the possible presence of hour-scale variability would not favor this scenario \citep{reinesetal2016,hebbaretal2019}.  There is also no evidence for AGN ionization in the galaxy center \citep{crescietal2017}. In the end, given the stellar mass of $10^{10}$~\msun\ \citep{nguyenetal2014}, the detection of a black hole may be of limited relevance for IMBH studies, but it does show that use of the fundamental plane is challenging in star-forming galaxies.


Another use of the fundamental plane to find evidence for a low-mass central black hole is in the low-mass galaxy NGC 404, where the plane yielded a mass estimate of log $M/\msun = 6.4\pm1.1$ \citep{nylandetal2012}. Newer observations may support an AGN interpretation of this system, although the evidence is not yet conclusive \citep{nylandetal2017}. On the other hand, dynamical studies of NGC 404 give a $3\sigma$ upper limit of \mbh$< 1.5\times10^5 M_{\odot}$ \citep{nguyenetal2017}. These measurements are formally consistent given the large scatter in the fundamental plane, but cannot be taken as a vote of confidence for the success of the fundamental plane near the IMBH regime either.

Going forward, for most reasonable central black hole mass functions, volume-limited fundamental plane surveys are almost certainly needed to effectively address the question of whether IMBHs exist, due to the usual Malmquist bias that a flux-limited survey is much more likely to detect massive black holes at larger distances. If IMBHs are less common than central black holes of higher masses (see \S \ref{sec:demographics}), then this bias would be exacerbated by the scatter in the fundamental plane, producing an Eddington-like bias in the IMBH candidate sample. 

Consider the use of the fundamental plane to find candidate IMBHs well-suited for dynamical follow-up in a nearly fixed distance sample, e.g., the \emph{Chandra} survey of Virgo galaxies by \citet{galloetal2008}. This survey found a large number of central X-ray sources with luminosities near their detection limit of $\sim 4 \times 10^{38}$~erg~s$^{-1}$. If a subset of these sources hosted IMBHs with masses in the range $10^4$--$10^5 M_{\odot}$ and they obeyed the fundamental plane, the 5 GHz radio continuum flux densities would typically be only a few $\mu$Jy. Such sensitivities are achievable with very long (10--20 hr) integrations on the Jansky Very Large Array, but the time investment for such a ``fishing expedition" would be challenging for more than a few galaxies. On the other hand, next generation facilities such as the next generation VLA (ngVLA), which has a factor of 10 higher sensitivity than the Jansky VLA, could carry out a survey of $\sim 100$ galaxies with a reasonable time investment, and with a resolution of $< 0.1"$, could pinpoint the location of any radio emission. Cross-matching with existing high-resolution \emph{HST} images could weed out possible non-nuclear contaminants such as star-forming regions, supernova remnants, or background galaxies.

\subsection{The Promise of Variability Selection}

All AGN vary, and this has been used as a successful selection tool \citep[e.g.,][]{sarajedinietal2006,macleodetal2011}. In the near future, a number of upcoming surveys will make variability selection very powerful for IMBH searches. 

Optical variability has been used to select low-mass black holes specifically \citep{morokumaetal2016}, but the promise of this technique has not been fully explored yet. \citet{heinisetal2016} find AGN in galaxies with stellar masses as low as $M_* \approx 10^{9.5}$~\msun\ using a variability selection.  \citet{baldassareetal2018,baldassareetal2019} use optical variability to find AGN candidates in galaxies with $M_*$ spanning $10^7-10^{10}$~\msun, a large fraction of which are not uncovered through optical emission-line selection. With upcoming surveys like LSST, this discovery space should grow.

AGN also vary in the radio, for diverse reasons both intrinsic (variations in the accretion rate; shocks in the relativistic jet) and extrinsic (scattering and/or magnification caused by interstellar plasma). Synoptic radio surveys have successfully used radio variability to select AGN (e.g., \citealt{mooleyetal2016}). Current and future radio continuum surveys such as the VLA Sky Survey and those with the Square Kilometer Array could identify candidate IMBHs as variable radio sources associated with low-mass galaxies.

IMBHs should also leave distinct signatures in their X-ray variability signals. Specifically, because the X-ray emission region is very compact compared to more massive black holes with AGN, the variability timescales should be short. This is clearly seen with NGC 4395 \citep{moranetal2005}. \citet{kamizasaetal2012} leveraged this idea to search the \emph{XMM-Newton} archive for low-mass black hole candidates. Those they found have a median black hole mass of $\sim 10^6$~\msun\ \citep{hokim2016}, but in principle surveys like eROSITA may further such a search. The X-ray excess variance, while promising as a search tool, may not track black hole mass anymore below $\lesssim 10^6$~\msun, where the variance-mass relation seems to flatten \citep{ludlametal2015,panetal2015}.

\subsection{Moving to Higher Redshift}

In addition to pushing to deeper limits for local samples, we could gain orthogonal demographic constraints by looking at high luminosity sources over much larger volumes; this work is in its infancy. So far, the effort to search for accreting black holes in dwarf galaxies beyond the local universe has focused on deep X-ray observations of well-studied extragalactic fields, finding candidates from $z < 0.5$ \citep{schrammetal2013,pardoetal2016,airdetal2018} out to $z \approx 2.4$ \citep{mezcuaetal2016,mezcuaetal2018}. We emphasize that these objects should be considered as candidates. In the case of the Mezcua objects in particular, many of the faint sources are proximate to more luminous objects, making the matching particularly challenging. The use of photometric redshifts with active galaxies also adds additional ambiguity. Finally, in these very high redshift sources, the implied Eddington ratios are substantially super-Eddington. Additional folow-up is needed. As high-redshift spectroscopic samples continue to grow \citep[e.g.,][]{takadaetal2014} we will continue to build reliable luminosity functions for lower-mass black holes at intermediate and high redshifts.

Studying the incidence of X-ray emission from low-mass host galaxies  ($M_* = 5 \times 10^9 - 2 \times 10^{10}$~\msun) out to $z \approx 1$, \citet{shietal2008} placed a strong lower limit of 12\% to the fraction of local low-mass galaxies harboring black holes.  This statistical result agrees well with the surveys of nearby galaxies summarized above. In the future, along with next-generation X-ray missions, deep spectroscopic surveys with \emph{JWST} and \emph{WFIRST} will certainly provide complementary samples of optically selected AGN at moderate redshift.

\section{Searches for Accreting Black Holes Outside of Galaxy Nuclei}
\label{sec:searchoff}

Perhaps the most promising tool to distinguish between seeding models comes from finding the black holes that are {\it not} in galaxy nuclei (\S \ref{sec:formation}). Thus far, dynamical searches in stellar clusters have been challenging to interpret (\S \ref{sec:dynamics}). However, some intriguing off-nuclear objects have surfaced due to their accretion signatures, which we review here.

\subsection{Ultra-Luminous X-ray Sources}

Ultra-luminous X-ray sources (ULXs) provided early impetus to think about black hole formation in stellar clusters \citep[e.g.,][]{ebisuzakietal2001}, but in recent years our understanding of these objects has evolved. There are many reviews of ultra-luminous X-ray sources (e.g., \citealt{kaaretetal2017}), and we provide only a brief discussion here of the work most directly relevant to IMBHs. 

Usually ULX samples are constructed from objects for which the Eddington limit is exceeded for a typical stellar mass black hole of $\sim 10 M_{\odot}$ (although the mass of the compact object may vary). ULXs were initially interpreted as strong candidates for IMBHs, largely from the simple argument that inferred isotropic luminosities $\gtrsim 10^{40}$ erg s$^{-1}$ imply accretors above $\sim 100 M_{\odot}$ (e.g., \citealt{colbertetal1999}).  By definition, ULXs are non-nuclear, to rule out quiescent or low/hard emission from central supermassive black holes.

Early surveys for ULXs were primarily statistical, since contamination from background AGN can dominate candidate ULX samples \citep{zolotukhinetal2016}. ULXs are overabundant in star-forming galaxies and indeed are primarily found near regions of recent star formation  themselves \citep{swartzetal2009}, consistent with ULXs being associated with young massive stars rather than old globular clusters or Population III remnants. ULXs are also much more common among metal-poor stellar populations than those of solar metallicity (per unit of star formation; \citealt{prestwichetal2013}).

X-ray spectra of ULXs show substantial variety, but overall are not congruent with the states observed for Galactic stellar-mass black holes scaled to the IMBH mass regime. Instead, some ULXs show distinct spectral states that can be physically interpreted as ``ultraluminous" states consistent with super-Eddington accretion onto stellar-mass black holes (see discussion in \citealt{kaaretetal2017}). 

Another breakthrough in the interpretation of ULXs was the observation of X-ray pulsations in some ULXs \citep{bachettietal2014,furstetal2016,israeletal2017},  proving that the accretors in these systems are neutron stars, likely with highly anisotropic accretion due to strong magnetic fields. Some ULXs mooted as IMBHs (e.g., M51 X-7; \citealt{earnshawetal2016}) have been proven to be neutron star ULXs \citep{rodriguezcastilloetal2019}.

Putting this  together, there is strong evidence that most ULXs are not IMBHs. There are a few important exceptions that we discuss below.

\subsubsection{HLX-1 and Friends}

Even though most typical ULXs (with $L_X \sim 10^{39}$--$10^{40}$ erg s$^{-1}$) are unlikely to be IMBHs, some rare sources have been observed with $L_X \gtrsim 10^{41}$ erg s$^{-1}$ that cannot be easily explained even within a super-Eddington paradigm for neutron stars or stellar-mass black holes. The clearest example is HLX-1 (a so-called ``hyperluminous" X-ray source), located in the disk galaxy ESO 243-49 ($D \sim 95$ Mpc), at a projected distance $\sim 4$ kpc from the galaxy's center \citep{farrelletal2009}.

The isotropic-equivalent X-ray luminosity of HLX-1 ranges from $L_X \sim 10^{40}$--$10^{42}$ erg s$^{-1}$. The association of HLX-1 with its host, and hence its extreme luminosity, have been confirmed via optical spectroscopy \citep{wiersemaetal2010}. Unlike typical ULXs, the source shows spectral behavior more similar to standard accretion disks than super-Eddington accretion, including typical high-thermal and low-hard states and state transitions (e.g., \citealt{servillatetal2011}).

Modeling of the optical and X-ray emission through the state changes are consistent with a black hole mass of a few $\times 10^4 M_{\odot}$, with a fair degree of uncertainty \citep{davisetal2011,godetetal2012,straubetal2014}. Radio emission associated with the state changes gives similar mass estimates, in the range $\mbh \sim 10^4$--$10^5 M_{\odot}$ \citep{webbetal2012}. The fundamental plane of X-ray/radio emission allows a mass as high as $\sim 3\times 10^6 M_{\odot}$ \citep{csehetal2015} and hence is of limited utility. 

An intriguing puzzle in HLX-1 is the nature of the X-ray luminosity variations and state changes. Since X-ray monitoring of the source began in 2008, the state changes appeared nearly periodic at intervals of $\sim 1$ yr, leading to the idea that they were associated with the orbital period of a tidally captured companion star on an eccentric orbit \citep{lasotaetal2011}. Challenging this simple scenario, only a few years later the state change interval began elongating unpredictably, perhaps consistent with the unstable, tidally affected orbit of a compact donor such as a white dwarf \citep{godetetal2014}. Other models, in which the luminosity variations do not directly reflect the orbital period of a donor star, have also been proposed (e.g., \citealt{soriaetal2017}).
Future state changes, or a lack thereof, can provide additional constraints on these models.

Observations of the surrounding stellar population give additional insight into HLX-1. While the modeling of the data is not conclusive due to optical emission associated with HLX-1 itself, the photometry is most consistent with a relatively massive (few $\sim \times 10^6 M_{\odot}$) star cluster dominated by intermediate to old stellar population \citep{soriaetal2017}. No compelling evidence for a recent major merger has been observed around the galaxy \citep{webbetal2017}, but nevertheless a self-consistent scenario is that HLX-1 represents a central massive black hole on the low-mass end of the mass distribution of nucleated $\sim 10^9$--$10^{10} M_{\star}$ galaxies whose parent galaxy was accreted and tidally stripped, leaving only the bare nuclear star cluster \citep[e.g.,][]{mapellietal2013}. Star formation, perhaps due to the merger, could have increased the capture rate for a central black hole to acquire a companion star on which it is currently feeding. In this scenario HLX-1 does not represent a unique formation channel for IMBHs, but it is one of the best candidates we have for a (previously central) $\sim 10^4$~\msun\ black hole. HLX-1 and similar systems could help inform our understanding of low-mass central black holes in galaxies. 

Besides HLX-1, no other very luminous ULXs are as convincing as IMBH candidates. \citet{pashametal2014} argue that M82 X-1 (which can reach $L_X \sim 10^{41}$ erg s$^{-1}$) contains a $\sim 430\pm100 M_{\odot}$ IMBH via an extrapolation of a stellar-mass black hole scaling relation for X-ray quasi-periodic oscillations. \citet{brightmanetal2016} model X-ray data of the source over a wide range of energies and prefer a model of super-Eddington accretion onto a stellar-mass black hole, although depending on the spin, the black hole mass might be as high as  $\sim 100 M_{\odot}$. The interpretation of these data are not settled, and may require future X-ray missions; confusion with the bright ULX pulsar M82 X-2 is a challenge for low-resolution observations. M82 is a starbursting dwarf only $\sim 3.5$ Mpc distant, and a confirmation of a luminous IMBH at such a distance would suggest a high space density of IMBHs.

Another source of recent interest is an off-nuclear ULX in the Seyfert galaxy NGC 5252 at $\sim 100$ Mpc \citep{kimetal2015}. This ULX has an associated optical and radio source that has been studied at high resolution with the \emph{HST} and very long baseline interferometry. Overall, these data are consistent with a massive black hole in the $\sim 10^5$--$10^6 M_{\odot}$ regime whose stars have been stripped, though a somewhat lower mass in the IMBH range cannot be excluded \citep{mezcuaetal20185252}. 

An even less settled case is that of NGC 2276-3c, an $L_X = 2 \times 10^{40}$ erg s$^{-1}$ ULX in a face-on starbursting spiral at $\sim 33$ Mpc. The source was associated with a large (100s of pc) radio nebula in Very Large Array (VLA) radio imaging \citep{mezcuaetal2013}. Follow-up quasi-simultaneous X-ray and very long baseline radio data were used to find unresolved hard X-ray emission and detect a compact pc-scale radio jet at modest significance \citep{mezcuaetal2015}. A compact, steady jet is consistent with an AGN in the low/hard state, and the fundamental plane was then used to estimate a mass of $\sim 5\times 10^{4} M_{\odot}$ (all such mass estimates have uncertainties of at least 1 dex). However, \citet{yangetal2017} independently reduced the radio data and did not confirm the high-resolution radio detection. Furthermore, there is no claimed optical counterpart to the ULX, which might be expected in a scenario in which the IMBH is associated with a star cluster or stripped nucleus. The large number of ULXs found in NGC 2276 \citep{wolteretal2015} suggests that NGC 2276-3c is truly associated with the galaxy. Nonetheless, given the galaxy's high star formation rate, the lack of an optical counterpart, and uncertain compact jet detection, it seems more likely that NGC 2276-3c is a super-Eddington stellar-mass black hole or a neutron star ULX than an IMBH. In this scenario the extended radio emission might well be associated with NGC 2276-3c, as radio nebulae due to ULXs are not uncommon (see the discussion in \citealt{urquhartetal2018}).

\subsection{Fundamental Plane Searches in Globular Clusters}

Accretion constraints on the presence of IMBHs in globular clusters rest on the assumption that, in a manner analogous to that of low-luminosity AGN, such IMBHs will accrete a fraction of the gas within their sphere of influence, resulting in detectable radio or X-ray  emission from the accretion flow or jet \citep{maccarone2004}. To convert a measurement or upper limit in radio or X-ray luminosity to a corresponding IMBH mass or limit requires assumptions about the accretion rate and radiative efficiency of the accretion process. 

In a globular cluster, the winds of evolved stars represent a source of low-velocity gas that is continually replenished: no long-term accumulation of gas is needed to produce an observable level of accretion. Assuming that the gas is ionized, the expected electron density is $n_e \sim 0.05-0.5$ cm$^{-3}$ \citep{pfahletal2001}. The predicted level of ionized gas was first observed in the core of the globular cluster 47 Tuc by \citet{freireetal2001}, using the radial distribution of the dispersion measure of millisecond pulsars. This single observation has had little follow-up in subsequent years, since the method requires (a) a large population of millisecond pulsars, and (b) a low foreground of ionized material. This combination is, at present, still only satisfied for 47 Tuc. Nonetheless, \citet{abbateetal2018} revisit the measurement for this well-studied cluster with updated pulsar parameters, finding an even larger gas density than in the original paper ($n_e = 0.23\pm0.05$ cm$^{-3}$, compared to $\sim 0.07$ cm$^{-3}$). This measurement is consistent with theory and supports the basic underpinning of the accretion-based mass constraints: globular clusters should have some amount of gas that is available to be accreted by an IMBH. Other routes to fueling at higher rates are also available, such as winds or tidally stripped material from binary companion stars acquired dynamically \citep{macleodetal2016b}.

Since the X-ray luminosity ($L_X$) is observed to be a non-negligible fraction of the bolometric luminosity for known low-luminosity AGN, it is the most directly accessible tracer of the accretion rate. Given the expected gas density, IMBHs with $10^3$--$10^4 M_{\odot}$ accreting at the Bondi rate with high ($\epsilon \sim 0.05$--0.1) radiative efficiency would be X-ray sources with $L_X \sim 10^{34}$--10$^{36}$ erg s$^{-1}$ at the centers of globular clusters. X-ray sources are observed at this $L_X$, but are identified as individual X-ray binaries rather than IMBHs. Hence it is clear that if IMBHs are present in Galactic globular clusters, they accrete at levels below that of the Bondi rate or with lower radiative efficiency. 

\citet{maccarone2004} first pointed out that, based on the fundamental plane, radio continuum observations put the most stringent accretion constraints on IMBHs accreting at low radiative efficiency, at least for typical observational sensitivities of radio and X-ray data. Assuming radiatively inefficient accretion and the sub-Bondi accretion rates observed for nearby low-luminosity AGN (e.g., \citealt{pellegrini2005}), many papers have used the fundamental plane and radio upper limits to set stringent constraints on the presence of accreting IMBHs in globular clusters \citep{maccaroneetal2005,maccaroneetal2008,csehetal2010,lukong2011,straderetal2012}.

These observational efforts culminated in \citet{tremouetal2018}, which uses a similar formalism but a much larger sample of deep radio continuum imaging of 50 Galactic globular clusters from the Jansky VLA and the Australia Telescope Compact Array. No emission consistent with an IMBH was observed for any cluster or for a cluster stack, including clusters with dynamical IMBH claims, such as $\omega$ Cen, M54, and NGC 6388 (see above). For the clusters with  more sensitive VLA data, using the assumed formalism, the median $3\sigma$ upper limit to an IMBH is $\lesssim 1100 M_{\odot}$, corresponding to very low accretion rate upper limits of $\lesssim$ few $\times 10^{-11} M_{\odot}$ yr$^{-1}$. 

There are plausible mechanisms that could temporarily remove all gas from the immediate vicinity of the IMBH, rendering it invisible in radio continuum emission. Hence the lack of radio emission alone does not definitively prove the absence of an IMBH in a particular cluster. On the other hand, scatter in the accretion rate and efficiency is expected in both directions, so if IMBHs were common, at least some obvious sources would be observed. They are not.

It is reasonable to use the full sample of objects with radio constraints to set limits on the occupation fraction. The \citet{tremouetal2018} work is sensitive to $\gtrsim 1000 \, M_{\odot}$ IMBHs and searches 
$\gtrsim 10^5 M_{\odot}$ globular clusters. Assuming the conservative accretion luminosities outlined by \citet{tremouetal2018}, their non-detections set a $3\sigma$ upper limit of $10-15\%$ on the fraction of such massive globular clusters that could host $\sim 1000$~\msun\ black holes just based on Poisson statistics. The modest size of the Galactic globular cluster system overall makes it challenging to substantially improve this limit on the occupation fraction with future data, except to push to fainter emission levels and hence lower (or more conservative) mass limits. Deep X-ray observations can also provide complementary constraints if the observing times are sufficiently long \citep{grindlayetal2001,haggardetal2013}.

Perhaps the best way to improve the limit, especially if IMBHs closer to $10^4 M_{\odot}$ are being considered, is with observations of extragalactic globular clusters in nearby galaxies with next-generation radio continuum telescopes: the large number of clusters (hundreds to thousands per galaxy) should allow the detection of the high accretion tail of the distribution even if the occupation fraction is modest \citep{wrobeletal2016,wrobeletal2019}. 

\section{IMBH Searches with Transient Phenomena}
\label{sec:searchupcoming}

Thus far we have focused on search techniques specialized for either galaxy nuclei that can extend downward to the upper-end of the IMBH regime or stellar-cluster focused work aiming to identify sources in the $10^3$~\msun\ realm. However, tidal disruptions and gravitational waves have the potential to work across these boundaries, and potentially find wandering black holes in lower-mass stellar systems, should they exist.

\subsection{Tidal Disruption Events}

Tidal disruption events (TDEs) are the electromagntic signature that may result if a star passes within its tidal radius of a black hole \citep[e.g.,][]{rees1988}. TDEs are powered by accretion onto massive black holes, but their rates and environments will provide an independent probe of the space density of IMBHs, since the rates depend on stellar (rather than gas) dynamics. In principle it is possible to derive \mbh\ from modeling of the TDE light curve itself \citep{lodatoetal2009,guillochonramirezruiz2013,mockleretal2019} as the emission from these events depends on both the mass and radius of the star and the mass of the black hole \citep[e.g.,][]{law-smithetal2017}. The TDE literature merits its own review; we focus here only on aspects that may bear directly on IMBH demographics. 

\citet{weversetal2017} present black hole masses for a sample of optically selected TDEs\footnotetext{\citet{weversetal2019} play a similar game for X-ray selected samples, but in those cases identifying a reliable optical counterpart is more complicated, so we focus on the optical samples here.} using stellar velocity dispersion measurements.  We also note an X-ray--detected transient that is a likely TDE with an \mbh-\sigmastar-based mass estimate of $1.3-6.3 \times 10^5$~\msun\ \citep{maksymetal2013}. If the \msigma\ relation can be extrapolated below \sigmastar$\sim 100$~\kms\ (we provide direct support for this in \S \ref{sec:scaling}) then it is very likely some $<10^6$~\msun\ black holes have already produced detectable TDEs.

\citet{vanvelzen2018} goes a step further and claims that a constant black hole occupation fraction from stellar masses $M_* \approx 5 \times 10^9$~\msun\ to  $3 \times 10^{10}$~\msun\ is required to reproduce the observed TDE rate as a function of mass. This result highlights the promise of TDE observations to independently constrain the occupation fraction, and is consistent with our inferences from dynamics (\S \ref{sec:dynamics}) and X-ray observations of local galaxies (\S \ref{sec:searchnuclei}) above. That said, to fully utilize the promise of TDEs, some key assumptions  need to be explicitly checked. We do not know how TDE rates in low-mass galaxies may depend on star formation rate or galaxy structure, although we know that TDE rates in more massive galaxies are a function of galaxy properties \citep[e.g.,][]{arcavietal2014}. 
Another open question is whether the TDE emission properties depend systematically on \mbh\ in a way that biases the mass distributions of samples as a function of their selection \citep[e.g.,][]{strubbequataert2009}. Finally, for white dwarf disruptions, there is an interesting literature comparing the emission signatures to those of Type Ia supernovae \citep{rosswogetal2009,macleodetal2016,anninosetal2018}. These events are particularly exciting since they could be accompanied by the gravitational wave signature of an extreme mass-ratio inspiral (\S \ref{sec:searchupcoming}).

TDE rates are so uncertain that in theory, all AGN activity in low-mass galaxies may be powered by TDEs, and most black hole mass density at low-mass may be built up through tidal capture and TDEs \citep{milosavljevicetal2006,macleodetal2016,stoneetal2017,zubovas2019}. Zubovas in particular claims that the AGN fraction matches TDE expectations at low mass by assuming (a) rates from \citet{stonemetzger2016} and (b) unity occupation fraction. It should be possible to use multi-epoch observations of active nuclei in low-mass galaxies to test whether they fade, as would be expected for TDEs, but at present our time baselines of $\sim 10$~yr are too short to be discriminating \citep[e.g.,][]{baldassareetal2016,baldassareetal2018}. On the positive side, \citet{jonkeretal2019} present some hopeful evidence for persistent (5-10 year) X-ray emission from optically identified TDEs, suggesting that eROSITA may find as many as 1000s of TDEs powered by low-mass black holes.

\subsubsection{Peculiar explosions and possible off-nuclear TDEs}

In theory there may be numerous off-nuclear IMBHs living in globular clusters, or wandering with only a small number of tightly bound stars \citep{olearyloeb2012,fragioneetal2018}, which may give rise to off-nuclear TDEs observed as peculiar transients. For instance, there is a class of rapid blue transients \citep[e.g.,][]{droutetal2014,vinkoetal2015,tanakaetal2016,pursiainenetal2018,perleyetal2019} that are not easily explained as standard classes of supernovae. These do not occur in galaxy centers, nor do they quite look like TDEs. Various groups have suggested that the object presented by \citet{vinkoetal2015} may be powered by the tidal disruption of a white dwarf \citep[e.g.,][]{law-smithetal2017}. By contrast, \citet{marguttietal2019} argue that AT2018cow \citep{perleyetal2019} has too much circumstellar material to be explained by an IMBH TDE \citep[although see also][]{kuinetal2019}. 

Another class of TDE candidate is associated with off-nuclear sources that show transient X-ray emission, often ascribed to a large temporary increase in the accretion rate of the putative IMBH associated with a TDE. Perhaps the most compelling of these is 3XMM J215022.4--055108, which is an X-ray source associated with a $z \sim 0.06$ lenticular galaxy \citep{linetal2018}. This source shows a TDE-like X-ray light curve measured over 10 years, with a luminosity  of at least $10^{43}$ erg s$^{-1}$. The X-ray source matches the position of a compact star cluster (possibly a stripped nucleus) with an old stellar population and a mass of $\sim 10^7 M_{\odot}$. Fitting the X-ray spectra suggests a black hole mass of $5 \times 10^4-10^5 M_{\odot}$, depending on the unknown spin of the black hole. The position of the source on a diagram of luminosity and temperature for a standard thin accretion disk is very similar to that of HLX-1, suggesting a similar mass.

Other similar sources have been discovered (e.g., \citealt{linetal2016,linetal2017}), although the evidence for a TDE-like decay (rather than typical AGN activity) is less compelling due to a smaller number of deep observations over time, and generally the masses of the associated black holes  are more in the regime of ``normal" low-mass central black holes (masses $10^5-10^6 M_{\odot}$) rather than in the IMBH regime. If TDEs associated with low-mass central black holes in stripped nuclei are as common as suggested by these recent works, then future sensitive X-ray missions such as \emph{Athena} should confirm such sources in substantial quantities, with rates up to $\sim 100$ yr$^{-1}$ possible \citep{linetal2018,cassanoetal2018}. 

While these TDE-like sources decay over timescales of years, another type of source is worth a quick mention: Irwin et al.~(2016) found extremely brief (100s of seconds), luminous ($L_X \sim 10^{40}$--$10^{41}$ erg s$^{-1}$) X-ray flares from star clusters associated with two nearby early-type galaxies. One of the clusters appears to be a typical massive globular cluster, while the other has properties consistent with being a stripped nucleus. These X-ray sources show properties most similar to flares from young pulsars with strong magnetic fields---sources not expected to exist in old stellar populations. If the flares are Eddington-limited, then they could represent accretion onto IMBHs with masses of $\sim 100 - 1000 M_{\odot}$. So far, these results are only suggestive, and other evidence (such as radio emission from the sources in their lower state) would be necessary to provide stronger evidence for an IMBH classification. 

Very likely, these peculiar objects have diverse origins, and only with next-generation observations, along with detailed modeling, will we be in a position to find the most likely TDE candidates among all the rich stellar death phenomenology. 

\subsection{Gravitational Waves}
\label{sec:gravwaves}

Given the range of mass and mass ratio that we can hope to detect in the upcoming decades, we summarize some of the expected discoveries related to IMBHs from LIGO and \emph{LISA}.

\subsubsection{IMBH-IMBH Mergers at High Redshift} 

\emph{LISA} will be sensitive to black holes mergers with mass ratios $q \sim 0.1-1$ for \mbh$\sim 10^4-10^5$~\msun\  out to very high redshift ($z \approx 20$). The black hole masses and redshifts can be measured from high S/N gravitational waveforms \citep{amaro-seoaneetal2017}. These measurements will provide insight into the seeding mechanisms and fueling rates \citep[e.g.,][]{sesanaetal2011}. However, in detail the rates will depend not only on the seeding mechanism(s) at play, but also on the accretion history and the dynamics driving the mergers of the black holes \citep[e.g.,][]{sesanaetal2007,kleinetal2016}. In practice it will not be trivial to disentangle these effects. Observations in the X-ray and optical/IR with missions like \emph{JWST} and \emph{Lynx} that should detect the actively accreting $10^5-10^6$~\msun\ black holes at similar epochs would be very complementary \citep[e.g.,][\S \ref{sec:future}]{haimanetal2019}.

\subsubsection{Low Redshift Constraints from Intermediate and Extreme Mass-ratio Inspirals} 

If there are IMBHs floating around in more massive halos, then occasionally they should merge with the primary supermassive black hole in an intermediate-mass ratio inspiral event \citep[e.g.,][]{holley-bockelmanetal2010}. In addition, an IMBH embedded in a star cluster (nuclear or otherwise) could merge with a stellar-mass black hole in an extreme mass-ratio inspiral with even lower mass ratios \citep[e.g.,][]{gairetal2010}. Again, the rates of both events are heavily dependent on a number of unknown factors, including the spin of the black hole, the dynamics of the surrounding stellar cluster, and the number density of IMBHs \citep[e.g.,][]{amaro-seoaneetal2015,fragioneetal2018,berryetal2019}, which hopefully can be partially determined through complementary electromagnetic observations.

\subsubsection{Gravitational Runaway Constraints at Low Redshift} 

As alluded to in \S \ref{sec:formation} above, if the gravitational runaway channel operates to make seeds, then it should be operating at present in young and forming star clusters. It should be possible to catch the early stages of this process with LIGO detections \citep{kovetzetal2018,antoninietal2019}. Current limits on the merger of $\sim 100$~\msun\ black holes are not yet constraining \citep{abbateetal2018}, but should become so as LIGO progresses. The one caveat relates to the possible subset of runaway processes involving stars (as opposed to black holes) that may require low metallicity. If so, this channel also may operate only at high redshift. We have argued, based on the similarity of observed properties of globular clusters as a function of metallicity, that metallicity dependence is in some tension with observations.  

\section{IMBH Candidate Tables} 

We summarize the prior three sections in three tables containing all of the credible IMBH candidates in the literature to date. For clarity, we separate measurements in nuclei (Table \ref{tab:allnuclei}) from off-nuclear candidates (Table \ref{tab:offnuclei}) and constraining limits (Table \ref{tab:limnuclei}). To be included in the table, an object must have a published black hole mass or limit that falls below $10^6$~\msun. Furthermore, the mass must be based on a primary or secondary black hole mass determination method. We include \mbh\ estimates based on stellar and gas dynamics, reverberation mapping, and scaling from the \mbh-\sigmastar\ relation in the specific case of TDE events that seem to have reliable determinations of the host galaxies. We also include a small number of measurements based on modeling of the X-ray emission or on the radio/X-ray fundamental plane. We do not include single-epoch virial black hole mass estimates for AGN (but see references in \S \ref{sec:searchnuclei}). For the credible candidates, we also include other representative published measurements or limits, which in many cases are contradictory. However, we do not include off-nuclear sources that have \emph{only} limits with no claimed detections. 

The tables neatly summarize the state of the field. We have high confidence that nuclear black holes extend downward to $\sim 10^5$~\msun. There is tantalizing evidence for objects below this limit in galaxy nuclei (NGC 205 and upper limits), but not definitive evidence to date. HLX-1 and objects like it provide additional circumstantial evidence for black holes with masses in the $10^4-10^5$~\msun\ range. There is no compelling evidence yet for any object with \mbh$\sim 10^3$~\msun, and it is likely that many of the candidates listed in the table will not be confirmed as IMBHs.

While all the search techniques described above have their own challenges, there are a few themes worth drawing out that are special roadblocks when searching for IMBHs. Angular resolution is currently the limiting factor for dynamical methods, and enhances accretion searches by eliminating more contamination from stellar sources and enables one to better tease out the signal from low-level accretion \citep[e.g.,][]{dickeyetal2019}. 

However, even the order-of-magnitude improvements in angular resolution coming soon will not completely eliminate this confusion. Indeed, nearly all techniques suffer contamination from stars. Even at high Eddington ratio, IMBHs cannot be uniquely distinguished on the basis of their luminosity, and for a short period of time luminous supernovae can masquerade as accreting black holes \citep{filippenko1989,baldassareetal2015}. We have already hit a confusion floor with X-ray observations since going deeper than $\sim 10^{38}$~erg~s$^{-1}$ simply yields large samples of low-mass X-ray binaries. In principle incorporating radio observations could help. However, in nearly all accreting objects the X-ray and radio emission are linked in some manner, leading the fundamental plane to ``return" a seemingly reasonable mass even in cases where its use turns out later to have been not justified. For example, early discussions of ULXs argued that the fundamental plane might support their identification as IMBHs (e.g., \citealt{miller2005,kaaretetal2009}), while at least some of these sources turned out to be pulsars. Even dynamical masses in nuclei for sources with \mbh$<10^4$~\msun\ will suffer from confusion with clusters of compact objects, and measurements in globular clusters are very challenging due to this confusion. 

\section{Scaling Relations}
\label{sec:scaling}

Scaling relations between \mbh\ and macroscopic galaxy properties are useful as a tool to estimate \mbh\ for exciting objects \citep[e.g.,][]{weversetal2017}, or to calculate the black hole mass density in the universe \citep[e.g.,][]{marconietal2004}. These relations may also encode the evolutionary history of black holes. In the case of IMBHs, scaling relations at low mass might differ based on seeding mechanisms \citep[e.g.,][]{volonterietal2008,vanwassenhoveetal2010}, although accretion processes at early times may also wash out these signatures \citep[e.g.,][]{volonterignedin2009}.

\begin{figure*}
\vbox{ 
\vskip 0mm
\hskip -10mm
\includegraphics[width=0.95\textwidth]{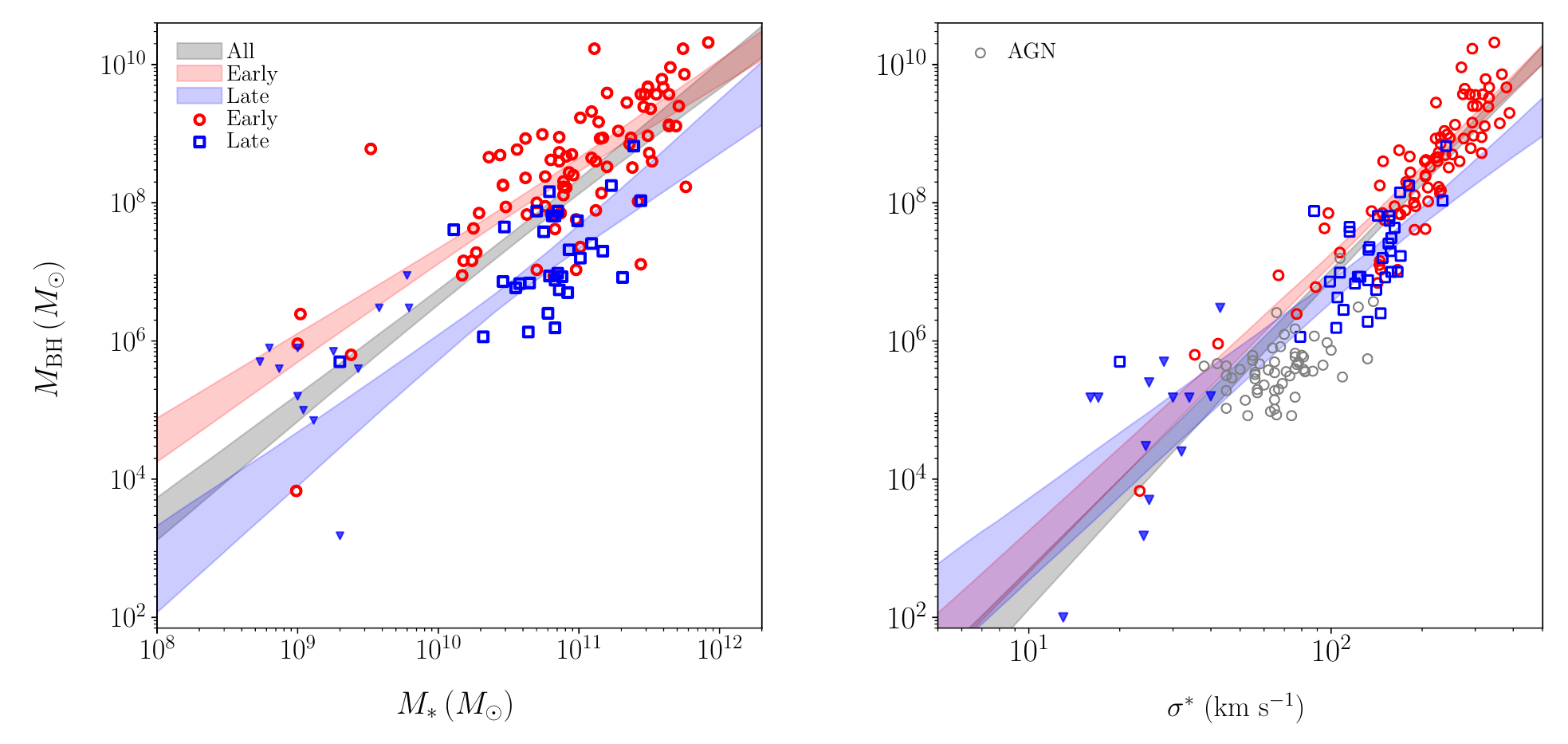}
}
\vskip -0mm
\caption{{\it Left}: Relationship between \mbh\ and $M_*$ for dynamical early-type (red open circles), late-type (blue open squares), and dynamical upper limits (blue triangles). We show fits to the early and late-type galaxies (red and blue shaded regions) and the full sample (grey). It is clear that the late-type galaxies have a comparable slope, but lower normalization, compared to the early-type galaxies. We also see hints for increased scatter at low mass, in part perhaps due to non-unity occupation fraction at low mass. 
{\it Right}: Same as {\it left}, but for \mbh\ versus \sigmastar. Here we also include the sample of active galactic nuclei from \citet[][grey dots]{xiaoetal2011}. By construction through $f_{\rm vir}$, the AGN points are consistent with the late-type galaxies. }
\label{fig:MbhMstar}
\end{figure*}

\subsection{$M_{\rm BH}-\sigma_*$}
\label{sec:mbhsigma}

We consider the low-mass end of the \mbh-\sigmastar\ relation, based on dynamical masses from \citet{kormendyho2013}, supplemented with more recent work \citep{greeneetal2016,sagliaetal2016,krajnovicetal2018,thateretal2019}, particularly at low masses \citep{denbroketal2015,nguyenetal2018,nguyenetal2019}, although our results do not change if we focus exclusively on Kormendy \& Ho augmented by low-mass objects. Crucially, we explore the importance of including upper limits \citep{bokeretal1999,barthetal2009,neumayerwalcher2012,nguyenetal2017}. There has been considerable literature on the scatter in scaling relations as a function of galaxy properties \citep{hopkinsetal2007,kormendyho2013,sagliaetal2016}. We do not attempt to address these issues, since the measurement uncertainties are still very large in the low-mass systems that concern us here. Finally, since we need it for our black hole mass function determinations below, we also split the sample into early (elliptical and S0) and late-type (spiral) galaxies, using the Hubble Types from \citet{sagliaetal2016} where available and the individual papers for all other galaxies (Supplemental Tables \ref{tab:scaleappendix}, \ref{tab:scalekh1}, and \ref{tab:scalekh2}). 

As we will show, including the constraining upper limits on low-mass galaxies \citep{barthetal2009,neumayerwalcher2012,delorenzietal2013} will make a real difference in the fitted slopes of the relations. We note that Neumayer and Walcher provide both a ''best'' and ''maximum'' allowed black hole mass, using the best and minimum $M/L$ from population synthesis models. We use their maximum value as the upper limits in the fits. 

Assuming that $\log (M_{\rm BH}/\msun) = \alpha + \beta \, \log (\sigma^*/160~{\rm km s^{-1}}) + \epsilon$, where $\epsilon$ is the intrinsic scatter, we present our fits in Supplemental Table \ref{tab:scaling} and Figure \ref{fig:MbhMstar}. Our results are broadly consistent with prior work. However, we are able to explore some additional issues due to the inclusion of low-mass \mbh\ measurements and limits. First of all, when only detections are included in the fitting, the slope we fit to the late-type galaxies alone is very shallow, likely due to the bias in \mbh\ measurements towards the most massive black holes at a given galaxy property \citep[see also][]{batcheldoretal2010,pacuccietal2018}. Similar flattening or breaks have been reported based on AGN-based \mbh\ values \citep[e.g.,][]{greeneho2006sig,martinnavarromezcua2018}, perhaps suffering from a similar bias. When limits are included, the fit to late and early-type galaxies become much more similar. 
In either case, the limits mitigate the bias seen in the detections, and we see no evidence for a change in \mbh-\sigmastar\ relations at low \sigmastar\ \citep[as also concluded by][]{barthetal2005,neumayerwalcher2012}. 

Early work raised the exciting prospect that the shape and scatter in the \mbh-\sigmastar\ relation for low \mbh\ objects might depend on the seeding mechanism \citep{volonterietal2008,volonterinatarajan2009}. Certainly, if all seeds were made at or above $10^5$~\msun, then the scaling relations would flatten at low mass. As of now, we do not see evidence for flattening at $10^5$~\msun, but rather NGC 205 and the published upper limits argue for a broad distribution of \mbh\ at the low-mass end, particularly given the additional evidence for a high occupation fraction in this stellar mass range. Nominally, the observed \mbh-\sigmastar\ relation disfavors heavy seed models that make exclusively \mbh$\gtrsim 10^5$~\msun.

\subsection{$M_{\rm BH}-M_*$}
\label{sec:mbhmstar}

Unfortunately, there are not \sigmastar\ functions measured for low-mass galaxies, and bulge fractions are no longer meaningful in these galaxies either \citep[e.g.,][]{macarthuretal2003}, although see \citet{schutteetal2019}. For these reasons, we revisit the correlation between $M_{\rm BH}$ and $M_*$ from dynamical studies including black holes with \mbh$<10^6$~\msun. We will then use this relation to estimate the black hole mass function in that same mass range.

Recently, \citet{reinesvolonteri2015} presented a fit to the $M_{\rm BH}-M_*$ relation using the dynamical sample of \citet{kormendyho2013}. We add additional galaxies that have been published since then as above \citep{greeneetal2016,sagliaetal2016,krajnovicetal2018,thateretal2019}, including low-mass black holes \citep{denbroketal2015,nguyenetal2018,nguyenetal2019}. 

As for \citet{kormendyho2013}, we measure $K-$band magnitudes and $B-V$ colors for the new galaxies. We use this single color and the fitting functions from \citet{belletal2003} to calculate $M_*$ for all samples, so that all $M_*$ estimates share a common IMF and stellar population assumptions. The one exception is the low-mass galaxies ($M_*<10^{10}$~\msun), where we use the stellar masses from the dynamical papers. \citet{nguyenetal2019} gives a more detailed comparison between different color-$M/L$ relations.  All black hole mass measurements and stellar masses used in this fit are tabulated in Supplemental Tables (Tables \ref{tab:scaleappendix}, \ref{tab:scalekh1}, \& \ref{tab:scalekh2}).

In Figure \ref{fig:MbhMstar} (Supplemental Table \ref{tab:scaling}), we present the updated $M_{\rm BH}-M_*$ scaling relation. One thing that is immediately apparent is that, when plotted against $M_*$, the dearth of dynamical black hole mass measurements for $M_* < 3 \times 10^{10}$~\msun\ is striking. A high priority for understanding black hole demographics at low $M_*$ is to study the supermassive black hole demographics in Milky-Way--mass galaxies \citep{krajnovicetal2018,nguyenetal2019b}.

Focusing on the IMBH regime, again we find that without upper limits, the fit to late-type galaxies returns a very shallow relation, because the measured \mbh\ values are biased high. When we include the limits, however, the slope of the relation becomes consistent between the red and blue populations. On the other hand, the overall normalization for the late-type galaxies is considerably lower than for the early-type galaxies \citep[in agreement with ][]{greeneetal2010,reinesvolonteri2015,greeneetal2016,laskeretal2016}. This difference has been interpreted to mean that black hole mass does not correlate with disk mass \citep[e.g.,][]{kormendyrichstone1995}, although the different trends have also been used to constrain the relationship between star formation and black hole growth \citep{caplaretal2015}. We note that our fit to the full sample including limits has a lower normalization and slightly steeper slope than the Reines \& Volonteri fit, while it is very similar to that presented recently by \citet{gallosesana2019}.

\subsection{Using AGN to Probe Scaling Relations}

We overplot the sample of low-mass AGNs from \citet{xiaoetal2011} in the $M_{\rm BH}-\sigma_*$ plane (Figure \ref{fig:MbhMstar}; excluding those with uncertain broad lines). We adjust the single-epoch virial masses to an $f_{\rm vir}$ calibrated with late-type galaxies, as advocated by \citet{hokim2014}. By construction, the AGNs align with the late-type galaxy fit. We again see no evidence in the AGN sample for any flattening or offset at low mass, given the caveat that we do not know the absolute black hole masses of these objects. Given the large uncertainty in the single-epoch masses we cannot make any additional statements about scatter from this sample.

More generally, we would urge extreme caution in using AGN signatures to infer scaling relations, particularly in this low-mass regime. Even the nearby AGN NGC 4395 ($D \approx 4$~Mpc) has published reverberation mapping masses that differ by more than an order of magnitude \citep{petersonetal2005,edrietal2012,wooetal2019}. There are legitimate reasons that we do not yet know to fully interpret the reverberation masses. One problem is the $f_{\rm vir}$ uncertainty. Beyond that, in the case of NGC 4395, the disagreement in black hole mass is mostly attributable to different line widths between the UV resonance line C{\small IV} and the optical recombination line H$\beta$. Much more work is needed to effectively harness reverberation mapping for AGN with low-mass black holes.

\begin{textbox}[b]
\subsection{Feedback}

One plausible explanation of black hole-galaxy scaling relations more generally is that there is a feedback loop between black hole growth and star formation, such that black holes are able to remove or heat gas in galaxies when they reach a critical mass relative to the galaxy potential \citep[e.g.,][]{silkrees1998}. For low-mass galaxies in particular, AGN feedback has largely been ignored. However, recent theoretical interest and observational evidence have brought the possibility of AGN feedback in dwarf galaxies to the fore. 

\citet{silk2017} argues on theoretical grounds that AGN feedback in low-mass galaxies may be important to solve small-scale challenges for $\Lambda$CDM \citep[see also][]{dashyanetal2018}. \citet{pennyetal2018} present evidence for AGN activity in 10\% of LMC-mass quenched galaxies. They interpret misaligned kinematics as signs of AGN-driven feedback. \citet{dickeyetal2019} examine 20 of the rare $M_* \approx 10^9$~\msun\ galaxies that are both isolated and non-star--forming, and find evidence for AGN activity in 16 of them \citep[see also][]{bradfordetal2018}. Dickey et al.\ conclude that AGN activity can ``self-quench'' even relatively isolated low-mass galaxies that otherwise are blue \citep{gehaetal2012}. \citet{nylandetal2017} also identify possible evidence for feedback in the jet in NGC 404. In short, there is now intriuging evidence that even dwarf galaxies with low-mass black holes may suffer the impacts of AGN feedback. We do not yet know whether these episodes are crucial in setting scaling relations in this regime.

\end{textbox}

\subsection{Scaling With Nuclear Star Clusters}

\begin{figure*}
\vbox{ 
\vskip 0mm
\hskip +2mm
\includegraphics[width=0.8\textwidth]{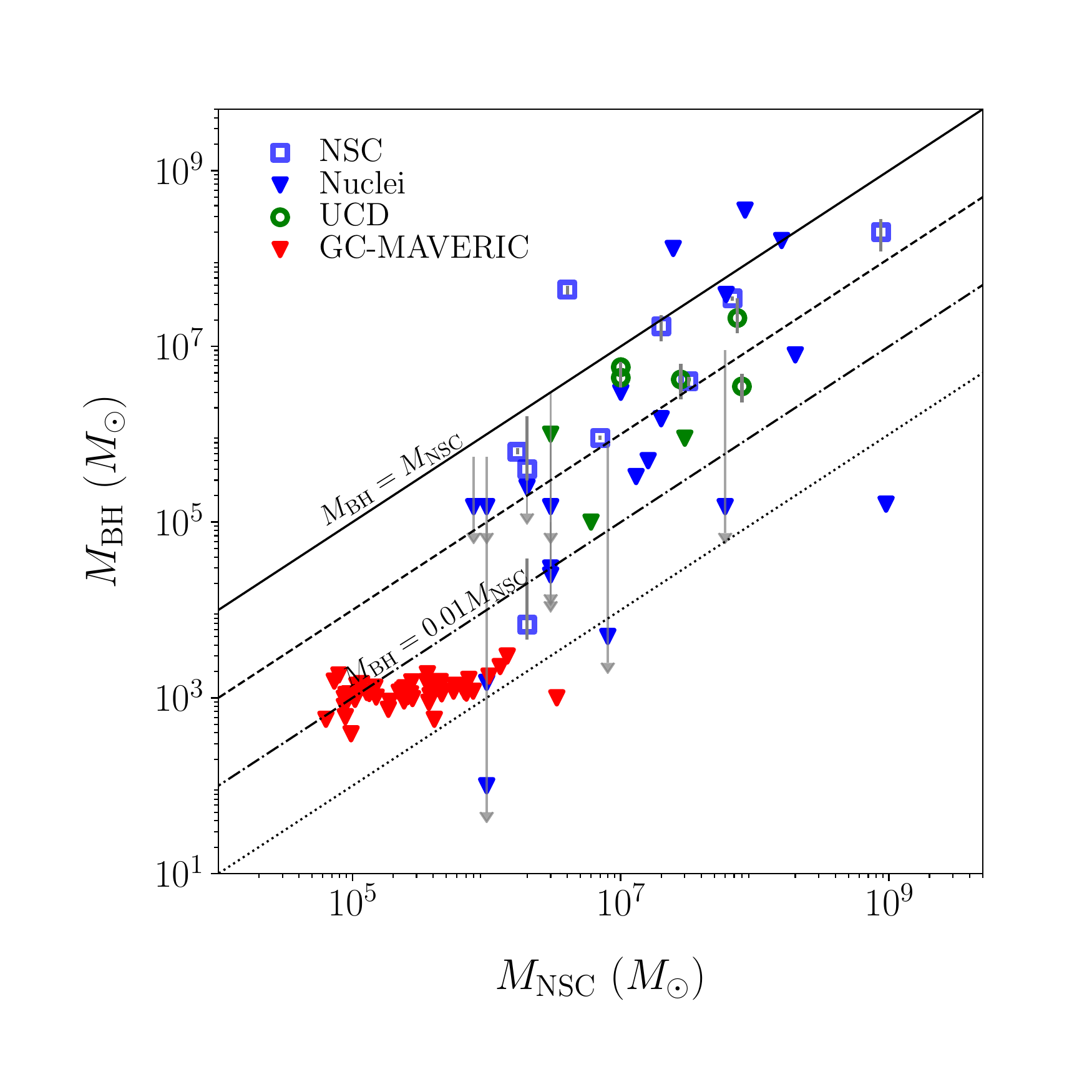}
}
\vskip -0mm
\caption{
The ratio of black hole to cluster mass, including globular clusters from \citet[][ red]{tremouetal2018}, dynamical black hole mass measurements with known nuclear star clusters from the literature \citep[blue; building on ][]{sethetal2008,nguyenetal2018}, limits. For the limits from Neumayer et al., we show their ``best'' measurement as a symbol and the ``maximum'' allowed black hole mass as the top of the arrow. We further include ultra-compact dwarfs with \mbh\ masses or upper limits published (green). The median ratio of \mbh\ to nuclear star cluster mass for objects with black holes detected is $\sim 25\%$, but with a factor of two scatter. The globular clusters are consistent with a $0.1\%$ mass fraction for a small fraction of the Milky Way globular cluster system.}
\label{fig:MncMbh}
\end{figure*}

In addition to scaling relations between galaxy properties and black hole mass, we consider possible scaling relations between the central black hole and the surrounding nucear star cluster. We observe a high incidence of black holes in nuclear star clusters. As we will argue below (\S \ref{sec:demographics}), a fraction $>50\%$ of $10^9-10^{10}$~\msun\ galaxies harbor black holes, and nearly all such galaxies harbor nuclear star clusters. Nuclear star clusters do not seem to replace black holes as the ``central compact object'' \citep{ferrareseetal2006}. Rather the two appear to coexist often at low galaxy mass \citep{sethetal2008}. In contrast, most higher-mass galaxies contain supermassive black holes \citep{gultekinetal2011} but show a very low incidence of nuclear star clusters \citep[e.g.,][]{grahamspitler2009}. Likely the growing black hole contributes to the demise of the nuclear star cluster \citep{antoninietal2019}. 

It is not obvious what (if any) causal relationship exists between black holes and nuclear star clusters. We have already discussed the possibility that black holes form via gravitational runaway processes in stellar clusters (\S \ref{sec:formation}), and the dearth of concrete observational evidence to date for black holes with \mbh$>1000$~\msun\ in globular clusters \citep[][\S \ref{sec:searchoff}]{tremouetal2018}. Nuclear star clusters are different from globular clusters in four key ways: they have higher stellar densities (and thus higher interaction rates), they have longer relaxation times, they have deeper potential wells (and hence higher escape velocities), and they have well-documented multi-age populations \citep[e.g.,][]{kacharovetal2018}. While some theoretical work postulated a higher likelihood of gravitational runaway in clusters with the highest \sigmastar\ \citep{millerdavies2012}, more recent work suggests that massive black holes likely can only form via rapid processes in nuclear star clusters \citep{breenetal2013,stoneetal2017}. 

Instead, the high coincidence of black holes and nuclear star clusters may suggest that globular clusters with IMBHs are more successful at surviving and migrating to a galaxy center, where through continued gas accretion they can become a nuclear star cluster. Or, the black hole may form early via other means (\S \ref{sec:formation}) and then sit at the center of a cluster that grows through mergers and accretion to be a present-day nuclear star cluster.

Going beyond occupation fractions, the mass fraction of the cluster bound up in the black hole may constrain models. We extend the work of \citet{sethetal2008}, \citet{georgievetal2016}, and \citet{nguyenetal2018} based on the increased number of dynamical black hole masses available now in the literature (Figure \ref{fig:MncMbh} and Supplemental Table \ref{tab:extranuclei}). Our full sample includes upper limits on \mbh\ in globular clusters from \citet{tremouetal2018}, ultra-compact dwarfs with dynamical black hole mass measurements or upper limits \citep{sethetal2014,ahnetal2017,ahnetal2018,afanasievetal2018}, and dynamical black holes in low-mass galaxies (\S \ref{sec:dynamics}), drawing nuclear star cluster masses from those papers. For the black hole masses and limits in \citet{krajnovicetal2018} and \citet{pechettietal2017}, we calculate the nuclear star cluster masses using color and luminosity information from \citet{coteetal2006} and the relations of \citet{belletal2003}. Finally, we include NGC 1023 and NGC 3384, two early-type galaxies that \citet{laueretal2005} identify as containing stellar nuclei, with the caveat that in some cases it can be challenging to determine whether these point sources are stellar or nonthermal in nature \citep[e.g.,][]{ravindranathetal2001}.

Roughly $\sim 0.1\%$ mass fractions are predicted from gravitational runaway scenarios (\S \ref{sec:formation}). The observed mass fractions in nuclear star clusters far exceed these limits. We see from Figure \ref{fig:MncMbh} that in most cases \mbh\ comprises a much higher fraction of the nuclear star cluster mass: the median value of \mbh/$M_{\rm NSC}$ for the objects with black hole detections is $\sim 0.25$, with more than a factor of two scatter. There is even more scatter when the limits are considered. This scatter likely reflects both a lack of black holes in some clusters and intrinsic scatter in the growth histories of both constituents. Interestingly, the largest outliers with low ratios of black hole to nuclear star cluster mass are found in the low-mass and late-type spiral galaxies. 

Given the wide range of \mbh/$M_{\rm NSC}$ that we observe, it seems possible that stochastic late-time fueling and/or merging plays a substantial role in the growth of each component \citep[e.g.,][]{naimanetal2015}. Ultra-compact dwarfs provide an interesting testing ground for how much black hole growth occurs via later-time accretion versus initial formation. Since ultra-compact dwarfs were stripped, likely quite early in some cases \citep{pfefferetal2014}, we would expect less late-time growth, leaving their \mbh/$M_{\rm NSC}$ ratios closer to their value at formation, while there may be more scatter in the population due to different stripping times. So far, the ultra-compact dwarfs do not obviously segregate in this plane, but the constraint will be more interesting as their number with \mbh\ constraints continues to grow.

\section{Demographics}
\label{sec:demographics}

\subsection{Existing limits on the occupation fraction}

\begin{figure}
\vbox{ 
\vskip 0mm
\hskip +0mm
\includegraphics[width=0.85\textwidth]{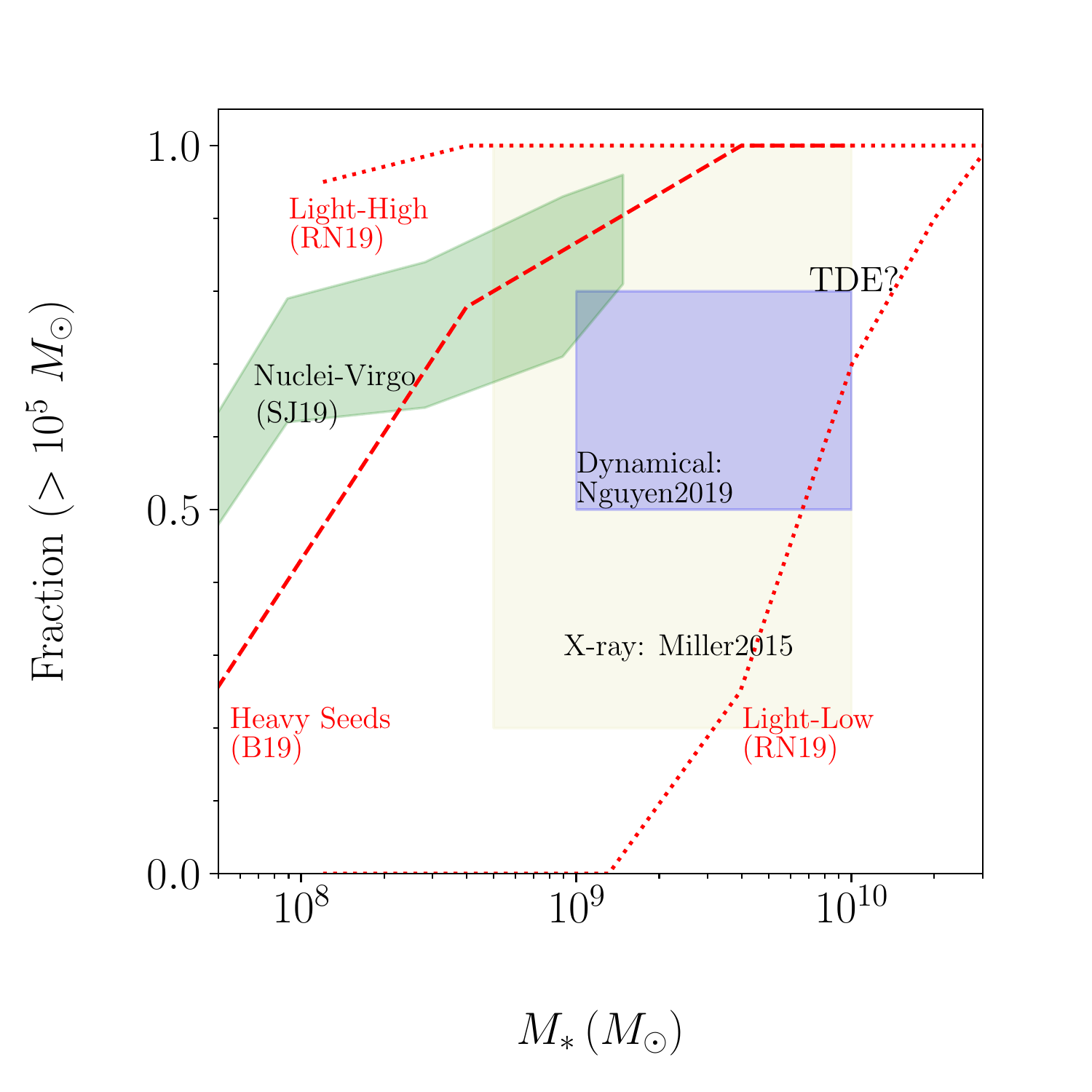}
}
\vskip -0mm
\caption{
Multiple constraints are converging towards a relatively high value of occupation fraction $>50\%$ of black holes with \mbh$\gtrsim 10^5$~\msun\ in galaxies with stellar masses $10^9<M_*/M_{\odot}<10^{10}$. Observational constraints from \citet[][yellow box]{milleretal2015} are  from X-ray observations of low-mass red galaxies. Dynamical constraints are summarized in \citet[][blue box]{nguyenetal2019} and suggest occupation fractions $>50\%$. Likewise, the TDE mass function suggests a high occupation fraction \citep{vanvelzen2018} even below $10^{10}$~\msun. Intriguingly, the nuclear star cluster occupation fraction is consistent with existing black hole limits as well, as exemplified by the green region derived from Virgo by \citet{sanchez-janssenetal2019}. For reference, we include predictions from three models (red). We include \citet[][dotted]{ricartenatarajan2018} predictions for black holes more massive than $3 \times 10^5$~\msun. Their Population III models span the full range shown here, with the Light-Low case corresponding to their power-law accretion mode while the Light-High case is their main-sequence accretion mode. Their direct collapse seeds fall in between these two limiting cases. We also show the limits on occupation fraction from \citet[][dashed]{bellovaryetal2019}, which is effectively a heavy seeding model.}
\label{fig:OccFrac}
\end{figure}

After \citet{gebhardtetal2001} published an upper limit of $<1500$~\msun\ on any putative black hole in the nuclear star cluster in M33, the community assumed that massive black holes were rare in low-mass, late-type galaxies with little or no bulge component. In fact, black holes do not appear to be rare, at least in galaxies with $M_* <10^{10}$~\msun. 

Recall that there are ten galaxies within 4 Mpc and $10^9<M_*/\msun<10^{10}$~\msun\ with published dynamical masses or limits (\S \ref{sec:dynamics}). Of these, five are detections. From the dynamical measurements, we infer an occupation fraction $> 50\%$ (Figure \ref{fig:OccFrac}, blue region). If the upper limit in M33 of \mbh$<1500$~\msun\ is taken at face value, it would suggest that the occupation fraction is not unity for \mbh=$10^3-10^6$~\msun\ black holes in galaxies with $M_* \approx 10^9$~\msun. However, better statistics and more limits are needed. 

In addition to dynamical constraints, X-ray surveys of local galaxies also allow us to make a concrete measurement of the occupation fraction. Just based on the number of detected X-ray sources in galaxies with $M_*=10^9-10^{10}$~\msun, \citet{milleretal2015} and \citet{sheetal2017} both find a lower limit of 20\% on the occupation fraction. Miller et al.\ argue for a likely occupation fraction of $\sim 70\%$ based on jointly modeling the $L_{\rm X}/M_*$ relation and the occupation fraction. Taking $\sim 20\%$ as the lower limit, we find that sensitive X-ray surveys of local galaxies already point towards relatively high occupation fractions, in agreement with the dynamics (Figure \ref{fig:OccFrac}, yellow box).

As yet, we see no concrete evidence for differences in mass distribution between red and blue low-mass galaxies, but given the small numbers involved this is mostly just an assumption (which also appears to hold for the X-ray detection fractions described above \S \ref{sec:searchnuclei}). Since blue galaxies are far more common than red at these masses, any difference could have significant ramifications for the black hole mass density.

The growing field of tidal disruptions will soon start to put competitive constraints on the occupation fraction if we can understand the rates (\S \ref{sec:searchupcoming}). \citet{vanvelzen2018} makes a first attempt. He finds that the observed luminosity function of tidal disruption events can only be explained if the occupation fraction is basically flat from LMC to Milky Way mass galaxies. While this result is hard to turn into an exact occupation fraction, it certainly argues that a high fraction of galaxies in this mass range host an IMBH.

As a point of comparison, we also plot the fraction of Virgo galaxies with $M_*<10^9$~\msun\ containing nuclear star clusters \citep{sanchez-janssenetal2019}. It is interesting to see that if most nuclear star clusters host a central black hole, then the black hole occupation fraction would be within the limits derived by other methods \citep[see also][]{sethetal2008,foordetal2017}. Below, when we calculate limits on the black hole mass function, we will use the measured nuclear star cluster fraction as one estimate of the black hole occupation fraction. We also show predictions from recent models for the occupation fraction (Figure\ \ref{fig:OccFrac}). The \citet{ricartenatarajan2018} models with Population III seeds span nearly all possible occupation fractions, depending on their fueling model (red dotted lines in Figure\ \ref{fig:OccFrac}), while the direct collapse models lie in between, consistent with \citet[][red dashed line]{bellovaryetal2019}. 

In summary, at least $\gtrsim 50\%$ of galaxies with $M_* \approx 10^9-10^{10}$~\msun\ host a massive black hole with \mbh$\sim 10^4-10^6$~\msun.

\subsection{Inferred Black Hole Mass Functions at Low Mass}

\begin{figure}
\vbox{ 
\vskip 0mm
\hskip +0mm
\includegraphics[width=0.95\textwidth]{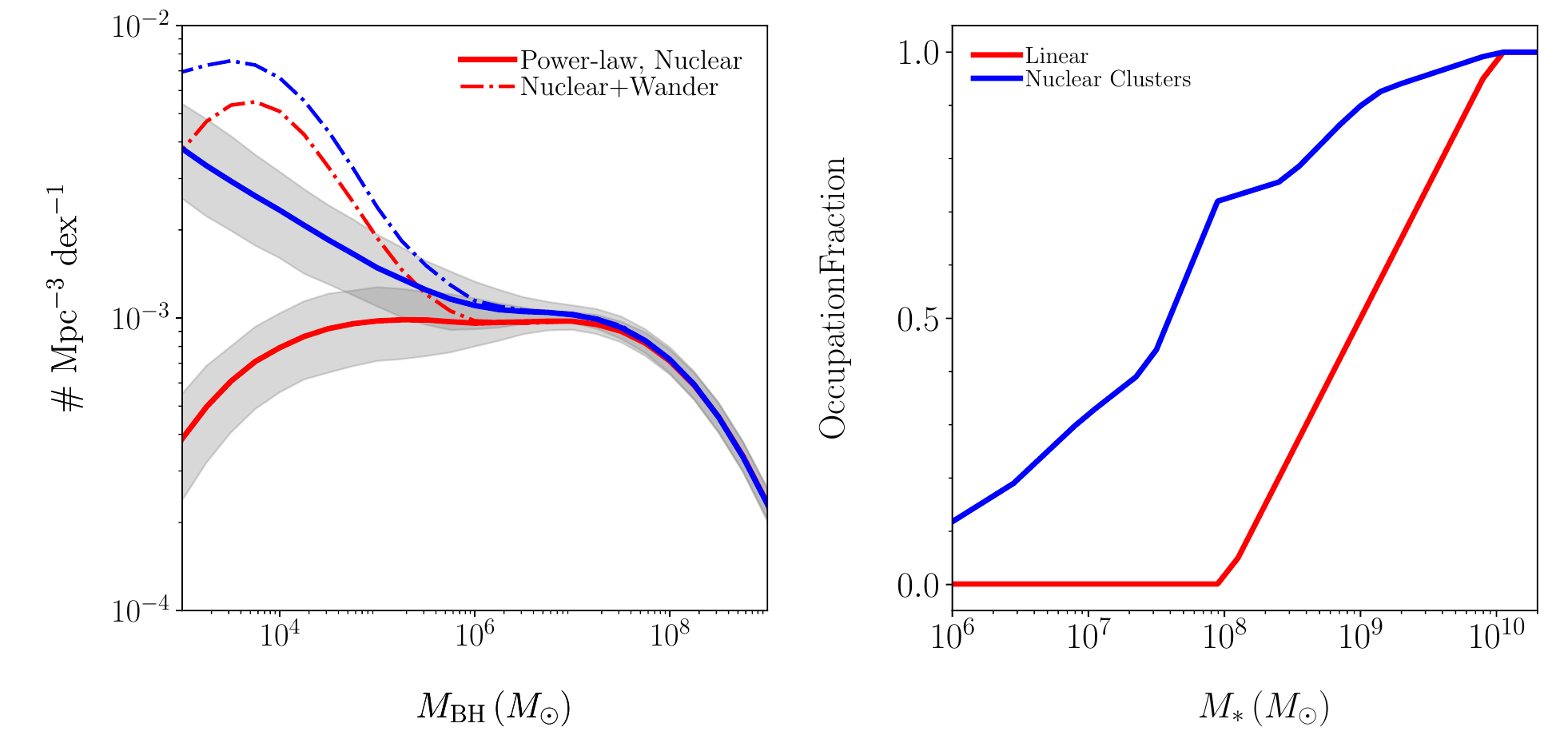}
}
\vskip -0mm
\caption{
{\it Left}: Inferred black hole mass function based on the GAMA stellar mass function \citep{wrightetal2017} and the $M_{\rm BH}-M_*$ relation from \S \ref{sec:scaling}. We take two cartoon occupation fractions as illustrated on the {\it right}, with the optimistic ``nuclear cluster'' model (blue) assuming that every nuclear star cluster houses an IMBH and the nucleation fraction coming from \citet{sanchez-janssenetal2019}, while the more pessimistic line (red) is just meant to be consistent with existing observations. In the solid lines, we show the black hole mass density from nuclear sources alone, assuming a single power-law relation between \mbh\ and $M_*$. We additionally add to the default power-law model a wandering black hole component with a number density tied to ultra compact dwarfs (dash-dot).
\label{fig:Massfunction}}
\end{figure}

Predictions for the rates of events like tidal disruptions \citep[e.g.,][]{stonemetzger2016} or extreme mass-ratio inspirals \citep[e.g.,][]{gairetal2010} depend on the black hole mass function into the IMBH mass range. However, there are few observational constraints on the \mbh\ mass function below $\sim 10^6$~\msun\ \citep[e.g.,][]{marconietal2004,greeneho2007b}. We now have the ingredients needed to calculate a range of possible black hole mass functions down to $\sim 10^4$~\msun, albeit with significant uncertainty. 

Using the \mbh-$M_*$ relation derived in \S \ref{sec:scaling} above, we convert the observed galaxy mass function into a black hole mass function \citep[e.g.,][]{marconietal2004}. The mass function of \citet{wrightetal2017} extends to $M_* \, \approx 10^6$~\msun, and thus allows us to explore the ramifications of a non-zero occupation fraction to very low stellar mass. The galaxy stellar mass is converted into black hole mass density using the \mbh-$M_*$ relation that we fit in \S \ref{sec:scaling}, including intrinsic scatter. We consider the red and blue galaxies separately, since they have quite different scaling zeropoints \citep[see also][]{shankaretal2016}, and for this we take guidance from the red and blue galaxy luminosity functions presented by \citet{blantonmoustakas2009}. We thus input a galaxy mass function that is dominated by blue galaxies below, and red galaxies above, $M_* = 10^{10}$~\msun. 

Below $M_* \approx 10^{10}$~\msun, the occupation fraction may deviate from unity. We adopt two different forms for the occupation fraction, both of which are consistent with the current data and models (Figure \ref{fig:Massfunction}). As a pessimistic case, we assume that the occupation fraction drops linearly from unity at $10^{10}$ to zero at $3 \times 10^7$~\msun. As an optimistic limit, we take the fraction of galaxies with nuclear star clusters and assume that every nuclear star cluster harbors a massive black hole \citep[derived in Virgo by][]{sanchez-janssenetal2019}. In the default mass function, there are only nuclear black holes and a single power-law relation between stellar and black hole mass. The resulting range of possible mass functions are shown in Figure \ref{fig:Massfunction} (Supplemental Table \ref{tab:massfunc}). There is quite a range of possible \mbh\ densities for \mbh$<10^6$~\msun. We note again that we have assumed comparable occupation fractions for red and blue galaxies, which is a systematic uncertainty that must be tested.

We also explore what a contribution from wandering black holes might look like  (\S \ref{sec:formation}). These black holes may reveal themselves in our own galaxy through dynamical signatures, and they may be the sites of extreme mass-ratio inspirals or white dwarf TDEs as well. We tie the number of wandering black holes to the number of ultra-compact dwarfs, as described in \S \ref{sec:seedmodels}. In short, we take the number of ultra-compact dwarfs as a proxy for the number of disrupted satellites and then assume a 10--50\% occupation fraction for them. We are not adopting a cosmologically motivated mass spectrum of satellites, nor do we have an empirical way to assign black hole masses to this population. The ultra-compact dwarfs with known black holes orbit more massive hosts and likely had more massive progenitors than those we consider here \citep[][]{sethetal2014,voggeletal2019}. Thus, we simply take a typical black hole mass of $3 \times 10^3$~\msun, chosen to fall below current detection limits, and a log-normal width of 0.5 dex. These are arbitrary, but at least qualitatively demonstrate how much higher the black hole mass density might be for relatively conservative estimates on wandering populations. 


There are a number of systematic uncertainties here, including the shape of the $M_{\rm BH}-M_*$ relation and all the challenges inherent in calculating stellar masses \citep{conroy2013}. \citet{gallosesana2019} nicely demonstrate the systematic effect of different scaling relation fits \citep[see also][]{shankaretal2016}.  Furthermore, we do not know the functional form of the occupation fraction with stellar mass. Because the stellar mass function rises steeply, even a small occupation fraction in low-mass galaxies dramatically changes the black hole mass density at low mass. Nevertheless, the mass functions we present here may be used to bracket the expected event rates of TDEs and extreme mass-ratio inspirals.

\section{The Future}

\begin{figure}
\vbox{ 
\vskip 0mm
\hskip +1mm
\includegraphics[width=0.90\textwidth]{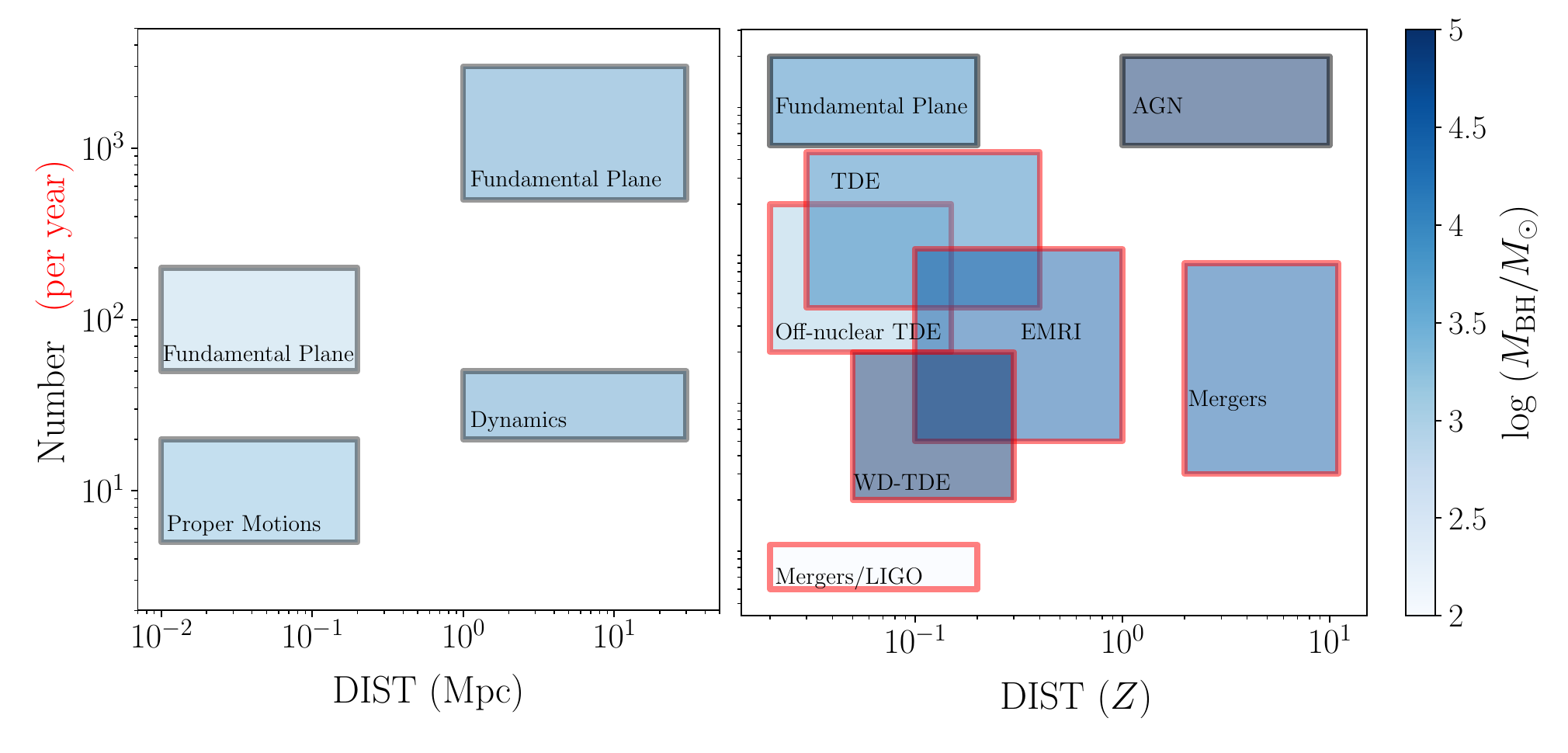}
}
\vskip -0mm
\caption{
{\it Left}: Volume and number of targets to be reached by next-generation facilities. The color-bar reflects the black hole mass range (from 100-$10^5$~\msun, with darker color for more massive black holes, see scale bar). The red outline signals that we tabulate a rate per year rather than a number. 
{\it Right}: Same as left, but for more distant black holes. 
\label{fig:Future}}
\end{figure}

In the coming two decades, we hope to break into the elusive $\sim 100-10^4$~\msun\ regime with robust dynamical constraints from stars, gas, and gravitational wave detections. We provide a short synopsis of the most promising ongoing and upcoming experiments, along with the distance distributions and available number of objects that they can each explore (Figure\ \ref{fig:Future}). We provide a full explanation for the numbers used in the figure in Supplementary Materials (I).  If we are to take full advantage of these next-generation opportunities, we must begin to lay the groundwork now to be maximally ready to exploit upcoming surveys. We present here a list of our high priority items.

Next-generation extremely large telescopes and ALMA at full capacity will have stunning spatial resolution, but we will only be able to find putatitive $<10^5$~\msun\ black holes in $<10^9$~\msun\ galaxies if we know where to look. We must use all the tools at our disposal, from astrometry with \emph{Gaia} to gas kinematics to determine centers for Local Group dwarfs \citep{vandermarelkallivayalil2014}. Moving out to larger radius, a complete sample of nuclear star clusters within 5 Mpc would be immensely useful to plan for next-generation dynamical searches. Likewise a survey of the molecular gas content of $<10^9$~\msun\ galaxies in the South to identify good ALMA targets is needed. Finally, it is a high priority to start mining \emph{Gaia} for potential wandering black hole candidates. 

We have assumed throughout this article that red and blue low-mass galaxies have similar occupation fractions, which is consistent with the current X-ray and dynamical results. Even a larger sample of constraining limits for more nuclear star clusters \citep{neumayerwalcher2012}, particularly including all the available red galaxies within 10 Mpc, could potentially shed light on this important issue before many more detections are in reach. Because red and blue galaxies have very different number density at low mass, it is important to know if there are differences in their occupation fractions.

We will not lose our reliance on accretion signatures, even as dynamical capabilities grow. We urgently need theoretical predictions for the spectral energy distributions and presence or absence of broad lines in black holes of $10^3-10^5$~\msun\ black holes \citep[e.g.,][]{cannetal2019} to search effectively for them with upcoming sensitive instruments like \emph{JWST} and/or \emph{Lynx}. ``Standard'' AGN search techniques have identified targets with inferred masses pushing towards $10^4$~\msun\ \citep[e.g.,][]{baldassareetal2015}, but we may be unable to identify the majority of sources at this black hole mass because their accretion signatures do not match our expectations \citep[][]{cannetal2019}. This question is important not only for selection reasons, but also because the ionizing spectrum of low-mass black holes is needed to think about the impact of IMBHs on reionization \citep[e.g.,][]{madauhaardt2015}. 

There are some hints that the spectral energy distributions of accreting black holes may be changing in interesting ways as we push to low mass. Most concretely, AGN with inferred low black hole mass and high Eddington ratio show a very pronounced soft X-ray excess, that may even be the tail of emission from the accretion disk itself \citep[e.g.,][]{doneetal2012,jinetal2012,yuanetal2014}, as would be expected in the low-mass regime. Such objects are also highly variable \citep{kamizasaetal2012}. Along with this, the ratio of optical to X-ray emission also shows interesting behavior among the low-mass black holes \citep[e.g.,][]{dongetal2012a,plotkinetal2016}, but these changes may be tied more closely to the high Eddington ratios rather than low masses of this sample. On the flip side, \citet{ludwigetal2012} did not find the ratio of He {\small II}/H$\beta$ to depend on black hole mass, as one would expect if the big blue bump moves to higher temperature at lower masses. At the low accretion rate end, there is also need for theoretical guidance in terms of the expected radio and X-ray emission. We would like to see calculations like those of \citet{ressleretal2019}, built for the Galactic Center, applied to the globular cluster context to directly predict accretion rates based on the angular momentum of the stellar winds.

\section{Summary}

By reviewing the observational literature, we draw the following conclusions about IMBH populations:

\begin{itemize}
    \item  There are clear and compelling cases for black holes in the mass range of $10^5-10^6$~\msun\ in the nuclei of low-mass galaxies. Moving downward into the true IMBH regime, there are tantalizing hints from NGC 205 and some constraining limits for lower-mass black holes, while HLX-1 and a handful of similar objects provide additional circumstantial evidence that $\sim 10^4$~\msun\ black holes exist. There is not even circumstantial evidence yet for $10^3$~\msun\ black holes.
    
    \item Excepting HLX-1, most ultra-luminous X-ray sources are not powered by IMBHs.

    \item Among $M_* \, > \,10^5$~\msun\ globular clusters in the Milky Way, observations limit the occupation fraction of $\sim 10^3$~\msun\ black holes to 10-15\%. In contrast, both dynamical measurements and X-ray observations in $10^9-10^{10}$~\msun\ galaxies argue for $>50\%$ occupation fractions of black holes in these galaxies.
    
    \item The most promising avenues to find $\sim 100-10^3$~\msun\ black holes are either to identify stellar binaries, which are expected to be common or to detect them in gravitational radiation. Ground-based gravitational wave limits on the low end of this mass regime are continuing to improve.
   
    \item We do not see evidence for a change in the \mbh-\sigmastar\ relation for \sigmastar$<100$~\kms. However, the existing limits suggest there is a broad range of black hole masses in these galaxies just below our detection limit, particularly since observations suggest that a large fraction of these galaxies do host black holes with \mbh$\, \gtrsim \, 10^4$~\msun.
    
    \item Folding together the empirically allowed range of occupation fractions with the observed correlation between stellar mass and black hole mass, we are able to bracket the allowed range of black hole number density down to $\sim 10^4$~\msun. More dynamical observations of black holes in $<10^{10}$~\msun\ galaxies, which will be facilitated by ALMA at full capacity along with extremely large telescopes in the coming decade, are needed to determine the scaling relations and occupation fractions for both red and blue galaxies.

\end{itemize}

By putting together theoretical predictions of occupation fractions with galaxy luminosity functions, we evaluate the current constrains on seeding mechanisms:

\begin{itemize}
\item Early seeding mechanisms---direct collapse type models or Population III stars---will be hard to distinguish using local observations alone, because varied accretion histories can wash out the early differences in these models. However, early-time luminosity functions and gravitational wave observations at high redshift could help.

    \item Theoretical predictions are quite divided on whether direct collapse scenarios can produce sufficient black holes. If direct collapse models do operate, the typical formation mass must be $<10^5$~\msun.
    
    \item From both a theoretical and observational perspective, gravitational runaway is unlikely to take off in typical globular clusters, but the most massive star clusters may still host such events.  
\end{itemize}

We highlight a few concrete conclusions related to next-generation constraints here:

\begin{itemize}
    \item Due to mass segregation, the total number of stars within the sphere of influence of a $\sim 1000$~\msun\ black hole may be insufficient to place stringent dynamical constraints on objects in this mass range even with extremely large telescopes, while $\sim 3000$~\msun\ black holes should be measurable.
    
    \item To unlock the potential of the fundamental plane for understanding black hole demographics, next generation radio telescopes will be crucial to detect low-mass, low-accretion--rate black holes in dwarf galaxies.
    
    \item As the number of ultra-compact dwarfs with black holes grow, these may help illuminate the relative importance of birth and accretion for black holes in nuclear clusters, since their accretion histories were likely truncated by falling into a bigger halo.
    
    \item Nuclear tidal disruption events are on the cusp of providing important clues to black hole demographics, but we must understand the rates as a function of galactic environment.
    
    \item It should be possible to identify (or rule out) wandering black holes in our galaxy by searching for compact and low-mass star clusters with \emph{Gaia}. Complementary off-nuclear tidal disruption events offer a promising avenue to getting large-scale statistics of such objects, again provided we can understand the rates and emission mechanisms. 
    
\end{itemize}

We are very excited about future prospects for finding true IMBHs in the coming decades, between improved sensitivity and frequency range for gravitational wave detection, the leap in angular resolution and sensitivity afforded by next-generation optical, radio, and X-ray telescopes, and the continuing reach of time-domain surveys.

\section{Acknowledgements}

We thank Gillian Bellovary, Ruben Sanchez-Janssen, and Angelo Ricarte for sharing their data. We thank Ruancun Li for performing bespoke 2MASS fitting and providing luminosities and colors for more than 40 galaxies. We thank Joan Wrobel for catching an error in an earlier version of this manuscript. We are grateful to Suvi Gezari, Zoltan Haiman, Morgan MacLeod, Nadine Neumayer, Anil Seth, Sjoert van Velzen, Marta Volonteri for useful discussions. LH is supported by the National Science Foundation of China (11721303) and the National Key R\&D Program of China (2016YFA0400702). JS acknowledges support from the Packard Foundation and from National Science Foundation grant AST-1514763. JEG acknowledges support from National Science Foundation grants AST-1713828 and AST-1815417.



\begin{table}[h]
\caption{Seeds to Local Predictions\label{tab:predict}}
\begin{center}
  \begin{tabular}{lccccccc}
\hline
 Model & Redshift & Tooth & $F_{\rm occ}$ & $F_{\rm occ}$ &
 \# Mpc$^{-3}$ & $M_{\rm BH}$ &  \# Mpc$^{-3}$ \\ 
    {} & {} & Fairies & $10^8$ & $10^9$ & Nuclei & Today &  Wander \\
\hline
Direct collapse & $z>10$ & UV background & $0.2-0.4$ & $0.4-0.8$ & $0.1-0.15$ & $10^4-10^6$ & $0.1-0.3$ \\
&  & pristine gas & & & & & \\
Population III & $z>15$ & Super-Eddington & $0.2-1.0$ & $0.5-1.0$ & $0.1-0.4$ & $10^4-10^6$ & $0.1-0.3$ \\
&  & accretion & & & & & \\
Fast  & All & BH retention & $0.1-0.7$ & $0.1-1.0$ & $0.02-0.25$ & $10^3-10^5$ & $>0.3$ \\
Runaway &  & high stellar density & & & & & \\
Slow & All & BH retention & $0.1-0.7$ & $0.1-1.0$ & $0.02-0.25$ & $10^3$ & $>0.3$ \\
Runaway &  & high $\sigma_*$ & & & & & \\
\hline
  \end{tabular}
  \end{center}
\begin{tabnote}
  {We show the range of occupation fractions and implied number densities for each seed formation channel. The direct collapse numbers are based on \citet{bellovaryetal2019} and the Population III numbers are based on \citet{ricartenatarajan2018}. For the gravitational runaway channel, we have relied on a number of sources described in \S 2.3.}
\end{tabnote}
\end{table}

\vspace{-1cm}

\begin{table}
  \caption{Detections in Nuclei\label{tab:allnuclei}}
  \begin{center}
  \begin{tabular}{llclcccl}
\hline
    Object & Dist & Method & $M_{\rm BH}$ & $M_{\rm cluster}$ & $M_*$ & $\sigma_*$ & Reference \\
\hline
NGC205  &  0.82 & Dynamics &   $(6.8_{-2.2}^{+32}) \times 10^3$    & $1.8 \times 10^6$ &  $9.7 \times 10^8$  & 40.0   & 1 \\
NGC5102 &  3.2  & Dynamics &   $(9.1_{-0.51}^{+0.61}) \times 10^5$  & $7 \times 10^6$ & $5.9 \times 10^9$  &  42.3   & 1 \\
NGC5206 &  3.5  & Dynamics &   $(6.3_{-0.61}^{+0.69}) \times 10^5$   & $1.7 \times 10^6$ &  $2.4 \times 10^9$  & 35.4   & 1 \\
\hline
NGC4395 &  4.4  & Dynamics &   $(4.0_{-3}^{+8}) \times 10^5$    & $2 \times 10^6$  &  $2 \times 10^9$ & $<30$   & 2 \\
NGC4395 &  4.4  & RM   &   $(3.6 \pm 1.1) \times 10^5$          & $2 \times 10^6$ &  $2 \times 10^9$  & $<30$  & 3 \\
NGC4395 &  4.4  & RM   &   $(9.1^{+1.5}_{-1.6}) \times 10^3$                    & $2 \times 10^6$ &  $2 \times 10^9$  &  $<30$  & 4 \\
UGC 6728 &  27   & RM    & $(5.2 \pm 2.9) \times 10^5$            & \nodata  & $7.5 \times 10^9$ & 52  & 5 \\          
\hline
iPTF–16fnl &  66.6 & $M_{\rm BH}-\sigma^*$ & $3.2 \times 10^5$ & \nodata & $9.7 \times 10^{9}$ & 55  & 6, 7 \\
ASASSN-14ae & 200 & $M_{\rm BH}-\sigma^*$ & $2.6 \times 10^5$ & \nodata & $1.4 \times 10^{10}$ & 55  & 6, 8 \\
PTF-09axc   & 536 & $M_{\rm BH}-\sigma^*$ & $4.8 \times 10^5$ & \nodata & $2.9 \times 10^{10}$ & 60  & 6, 9 \\
PS1-10jh    & 822 & $M_{\rm BH}-\sigma^*$ & $7.1 \times 10^5$ & \nodata & $1.5 \times 10^{10}$ & 65   & 6, 10 \\
PTF-09djl   & 900 & $M_{\rm BH}-\sigma^*$ & $6.6 \times 10^5$ & \nodata & $3.0 \times 10^{10}$ & 64  &  6, 10 \\
WINGS J1348 & 265 & $M_{\rm BH}-\sigma^*$ & $5.1-5.6 \times 10^5$ & \nodata & $2.5 \times 10^8$  & $<50$ & 11 \\
\hline
  \end{tabular}
  \end{center}
\begin{tabnote}
{Column (1): Object name. Column (2): Distance (Mpc). Column (3): Black hole mass determination method.
  Column (4): Black hole mass ($M_{\odot}$). Column (5): Star cluster mass, when relevant ($M_{\odot}$).
  Column (6): Stellar mass when galaxy host ($M_{\odot}$). Column (7): Stellar velocity dispersion, cluster or galaxy (km~s$^{-1}$).
  Column (8): (1) \citet{nguyenetal2019}; (2) \citet{denbroketal2015}; (3) \citet{petersonetal2005}; (4) \citet{wooetal2019}; (5) \citet{bentzetal2016};
  (6) \citet{weversetal2017}; (7) \citet{blagorodnovaetal2017}; (8) \citet{holoienetal2014} ; (9) \citet{arcavietal2014}; (10) \citet{gezarietal2012}; (11) \citet{maksymetal2013,maksymetal2014}. }
\end{tabnote}
\end{table}

\newpage

\begin{table}
\caption{IMBH Candidates Outside Nuclei\label{tab:offnuclei}}
\begin{center}
  \begin{tabular}{ccccccc}
\hline
Object & Dist & Method & $M_{\rm BH}$ & $M_{\rm cluster}$ & $\sigma^*$ & Ref \\
\hline
47 Tuc             & 0.005 & Dynamics   &  2300                                  & $7.8 \times 10^5$ & 12.2   & 1 \\
47 Tuc             & 0.005 & Dynamics   & $< 1650$                               & $7.8 \times 10^5$ & 12.2    & 2 \\ 
47 Tuc             & 0.005 & FP              & $< 1040$                          & $7.8 \times 10^5$ & 12.2   & 3 \\
$\omega$ Cen       & 0.005  & Dynamics   & $(4.7 \pm1.0)  \times 10^4$           & $3.6 \times 10^6$ & 17.6    & 4 \\
$\omega$ Cen       & 0.005  & Dynamics   & $<1.8 \times 10^4$                    & $3.6 \times 10^6$ & 17.6   & 5 \\
$\omega$ Cen       & 0.005  & FP         &  $<1000$                              & $3.6 \times 10^6$ & 17.6   & 3 \\
M62                & 0.007  & Dynamics   &  $ (2000 \pm 1000) $                  & $7.1 \times 10^5$ & 15.2    & 6 \\
M62                & 0.007  & FP         &  $ < 1130$                            & $7.1 \times 10^5$ & 15.2   & 3 \\
NGC6624            & 0.008  &   Dynamics  &  $>7500$                             & $7.3 \times 10^4 $  & 6.1      & 7 \\
NGC6624            & 0.008  &   Dynamics  &  $< 650$                             & $7.3 \times 10^4 $  & 6.1      & 8 \\
NGC6624            & 0.008  &   Dynamics  &  $< 1000$                            & $7.3 \times 10^4 $  & 6.1      & 9 \\
NGC6624            & 0.008  &   FP        &  $<1550$                             & $7.3 \times 10^4 $  & 6.1   & 3 \\
M15                 & 0.010  &   Dynamics &  $1700-3200$                        & $4.5 \times 10^5 $  & 12.9  & 10 \\
M15                 & 0.010  &   Dynamics &  $<2000$                             & $4.5 \times 10^5 $  & 12.9  & 11 \\
M15                 & 0.010  &   FP       &  $<1530$                             & $4.5 \times 10^5 $  & 12.9  & 3 \\
NGC6388            & 0.011  & Dynamics    & $(2.8 \pm 0.4) \times 10^4$          & $1.1 \times 10^6$   & 13.1  & 12 \\
NGC6388            & 0.011  & Dynamics    & $< 2000$                             & $1.1 \times 10^6$   & 13.1 & 13 \\
NGC6388            & 0.011  & FP          & $< 1770$                             & $1.1 \times 10^6$   & 13.1 & 3 \\
NGC1904            & 0.013  & Dynamics    &  $(3000 \pm 1000$)                   & $1.7 \times 10^5$   & 8.7    &  6 \\
M54                & 0.024  &   Dynamics  &  9400                                & $1.4 \times 10^6$  & 20.2   & 14  \\
M54                & 0.024  &   Dynamics  &  $1.1 \times 10^4$                   & $1.4 \times 10^6$  & 20.2   & 15  \\
M54                & 0.024  &   FP        &  $<3000$                             & $1.4 \times 10^6 $ & 20.2   & 3 \\
M31/G1             & 0.780  &  Dynamics   &  $(1.8\pm0.5) \times 10^4$           & $3.1 \times 10^6$  & 25.8   & 16 \\
M31/G1             & 0.780  &  Dynamics   &  $\sim 0$                            & $3.1 \times 10^6$  & 25.8 & 11 \\
M31/G1             & 0.780  &  FP         &  $<9700$                             & $3.1 \times 10^6$  & 25.8     & 17 \\
\hline
HLX-1 & 95 & X-ray spectra & $\sim 10^4$--$2 \times 10^5$ & $2 \times 10^5$--$3 \times 10^6$ &  \nodata  & 18 \\
HLX-1 & 95 & State Changes & $\sim 10^4-10^5$ & $2 \times 10^5$--$3 \times 10^6$ & \nodata              & 19 \\
HLX-1 & 95 & FP & $< 3 \times 10^6$ & $2 \times 10^5$--$3 \times 10^6$ & \nodata                        & 20 \\
M82 X-1 & 3.5 & X-ray QPOs & $430\pm100$ & \nodata & \nodata                                            & 21 \\
M82 X-1 & 3.5 & X-ray spectra & $<100$ & \nodata & \nodata                                              &  22 \\
NGC5252-ULX1 & 100 & FP & 3000--$2 \times 10^6$ & \nodata & \nodata                                     &  23 \\
\hline
  \end{tabular}
  \end{center}
\begin{tabnote}
{Column (1): Object name. Column (2): Distance (Mpc). Column (3): Black hole mass determination method.
  Column (4): Black hole mass ($M_{\odot}$). When limits are listed,
  an effort has been made to quote a $3 \sigma$ limit. Column (5): Star cluster mass, when relevant ($M_{\odot}$).
  Column (6): Stellar velocity dispersion, cluster or galaxy (km~s$^{-1}$).
  Column (7): (1) \citet{kiziltanetal2017}; (2) \citet{mannetal2019}; \citet{tremouetal2018}; (4) \citet{noyolaetal2010}; (5) \citet{vandermarelanderson2010}; 
  (6) \citet{lutzgendorfetal2013}; (7) \citet{pereraetal2017}; (8) \citet{gielesetal2018}; (9) \citet{baumgardtetal2019};
  (10) \citet{gerssenetal2002}; (11) \citet{baumgardtetal2003}; (12) \citet{lutzgendorfetal2015}; (13) \citet{lanzonietal2013};
  (14) \citet{ibataetal2009}; (15) \citet{baumgardt2017}; (16) \citet{gebhardtetal2005}; (17) \citet{millerjonesetal2012}; (18) \citet{straubetal2014};
  (19) \citet{webbetal2012}; (20) \citet{csehetal2015}; (21) \citet{pashametal2014}; (22) \citet{brightmanetal2016}; (23) \citet{mezcuaetal20185252}  
}
\end{tabnote}
\end{table}

\newpage

\begin{table}
\caption{Constraining Upper Limits\label{tab:limnuclei}}
\begin{center}
  \begin{tabular}{llllcccc}
\hline
Object & Dist & Method & $M_{\rm BH}$ & $M_{\rm cluster}$ & $M_*$ & $\sigma_*$ & Ref. \\
\hline
Fornax  & 0.135 & Dynamics & $<1 \times 10^5$ & \nodata        & $2 \times 10^7$ & 10  & 1 \\
UMi     & 0.076 & Dynamics & $<3 \times 10^4$  & \nodata        & $4.5 \times 10^5$ & \nodata   & 2 \\
IC342   & 3.9   & Dynamics & $<3.3 \times 10^5$ & $1.3 \times 10^7$ & \nodata  & 33 & 3 \\
NGC300 &  2.2  & Dynamics & $<1 \times 10^5$                  & $1 \times 10^6$ & $2.0 \times 10^9$ &  13   & 4 \\
NGC404 &  3.1  & Dynamics & $<1.5 \times 10^5$              & $1 \times 10^7$  & $8.5 \times 10^8$   & 40  & 5 \\
NGC428  &  16.1 & Dynamics & $<7 \times 10^4$                 & $3 \times 10^6$ & $6.3 \times 10^9$ & 24   & 4 \\
NGC1042 &  18.2 & Dynamics & $<3.0 \times 10^6$                & $3 \times 10^6$ & $1.0 \times 10^{10}$ & 32  & 4 \\
NGC1493 &  11.4 & Dynamics & $<8.0 \times 10^5$                & $2 \times 10^6$ & $4.0 \times 10^{9}$ & 25  & 4 \\  
NGC2139 &  23.6 & Dynamics & $<4.0 \times 10^5$                & $8 \times 10^5$ & $1.3 \times 10^{10}$ &  17  & 4 \\  
NGC3423 &  14.6 & Dynamics & $<7.0 \times 10^5$                & $3 \times 10^6$ & $7.9 \times 10^{9}$ &  30  & 4 \\
NGC3621 &  6.6  & Dynamics & $<3.0 \times 10^6$               & $1 \times 10^7$ & $1.6 \times 10^{10}$ &  43  & 6 \\
NGC4244 &  4.4 & Dynamics & $<5.0 \times 10^5$                & $1.6 \times 10^7$ & $2.0 \times 10^9$ & 28  & 7 \\
NGC7418 &  18.4 & Dynamics & $<9.0 \times 10^6$                & $6 \times 10^7$ & $1.6 \times 10^{10}$ & 34   & 4 \\ 
NGC7424 &  10.9 & Dynamics & $<4.0 \times 10^5$                & $1 \times 10^6$ & $4.0 \times 10^{9}$ & 16   & 4 \\ 
NGC7793 &  3.3  & Dynamics & $<8.0 \times 10^5$                & $8 \times 10^6$ & $4.0 \times 10^{9}$ & 25   & 4 \\
\hline
  \end{tabular}
  \end{center}
\begin{tabnote}
{Column (1): Object name. Column (2): Distance (Mpc). Column (3): Black hole mass determination method.
  Column (4): Black hole mass limit, $3 \sigma$ when possible ($M_{\odot}$). For the Neumayer \& Walcher (2012) sample, we use
  their maximum black hole masses. Column (5): Star cluster mass, when relevant ($M_{\odot}$).
  Column (6): Stellar mass when galaxy host ($M_{\odot}$). Column (7): Stellar velocity dispersion, cluster or galaxy (km~s$^{-1}$).
  Column (8): (1) \citet{jardelgebhardt2012}; (2) \citet{loraetal2009}; (3) \citet{bokeretal1999}; (4) \citet{neumayerwalcher2012}; (5) \citet{nguyenetal2017};
  (6) \citet{barthetal2009}; (7) \citet{delorenzietal2013}.}
\end{tabnote}
\end{table}

\clearpage

\newpage

\section{Supplementary Material 1: The Future}
\label{sec:future}


\subsection{Next-generation Opportunities}


We include here a full description of all the numbers displayed in Figure \ref{fig:Future}. Starting in the Local Volume with dynamical signatures, for the densest and most massive globular clusters, we expect to reach limits of a few $\times 10^3$~\msun\ using the proper motions of stars with extremely large telescopes \citep[e.g.,][]{greeneetal2019}, although as mentioned in \S \ref{sec:dynamics} above, the sensitivity will be a function of the number of stars in the sphere of influence of the black hole. Moving out to the Local Volume, assuming stellar velocity dispersions of $10-30$~\kms\ for the stars, extremely large telescopes and ALMA at full capabilities should be able to detect or put robust limits on $10^4$~\msun\ black holes out to 5-10 Mpc, although systematics in the modeling and in degeneracy with clusters of lower-mass compact objects will be limiting factors at this mass. A conservative number of targets would be the $\sim 20-40$ galaxies with masses less than the LMC in the Local Volume.

Beyond the relatively modest number of galaxies and star clusters accessible for dynamical measurements, we will have to rely on accretion signatures to determine the presence of black holes. The proposed next-generation Very Large Array (ngVLA) instrument offers the largest discovery space \citep{murphyetal2018}. With the sensitivity of the ngVLA, it will become possible to detect or rule out IMBHs at $1000$~\msun\ in all massive globular clusters in the Milky Way, as well as Local Volume galaxy nuclei and thousands of extragalactic globular clusters \citep{wrobeletal2019}, even given the large scatter in the Fundamental Plane and the uncertain gas density and angular momentum content of that gas.  That said, as emphasized in \S \ref{sec:searchnuclei}, there are many sources of confusion from stellar processes when attempting to use radiation to find IMBHs. Multi-epoch and multi-wavelength follow-up will be needed. Complementary time-domain surveys like the Zwicky Transient Factory \citep[ZTF;][]{bellmetal2019} and LSST \citep{ivezicetal2008}, with next-generation X-ray missions like eROSITA and \emph{Athena}, will begin to uncover yet more off-nuclear TDE candidates, optimistically at rates up to $\sim 100$ yr$^{-1}$ \citep{linetal2018,cassanoetal2018}.

Our current best constraints on the occupation fraction from radiation come from X-ray observations of galaxies within 30 Mpc \citep{milleretal2015}, and while the number of targets may increase by factors of a few \citep[e.g.,][]{gallosesana2019}, the X-ray searches cannot go any deeper than current limits because they hit a confusion limit with X-ray binaries at luminosities below $\sim 10^{38} $erg~s$^{-1}$ \citep{sivakoffetal2007,galloetal2008,ghoshetal2008}. Going forward, radio studies with the ngVLA will be the most promising way to improve the existing constraints from Miller et al.\ As shown by \citet{plotkinreines2018}, with relatively modest exposures times of an hour, one could systematically study $10^5$~\msun\ black holes accreting at $\sim 10^{-5}$ of their Eddington limits out to Gpc distances, and $10^4$~\msun\ black holes out to the Virgo cluster. The black hole fundamental plane could be calibrated within 5 Mpc using dynamical measurements from extremely large telescopes, and then extended to map the occupation fraction of $10^4$~\msun\ black holes out to 20 Mpc. Again, we note that deep and high-resolution radio imaging alone will be insufficient; we will need multi-wavelength information to determine the true nature of these radio sources (\S \ref{sec:searchnuclei}).

As advanced LIGO, VIRGO, and other gravitational wave experiments continue, the constraints on 100~\msun\ black hole mergers (which appear as bursts in the LIGO band) should get more stringent, putting yet more concrete constraints on that population. Unfortunately, those limits constrain a product of the number density and the merger rate, and we see no immediate prospect for independent number density measurements of $\sim 100$~\msun\ black holes.

Moving beyond the Local Volume, we increasingly will rely on accretion at high Eddington ratio and gravitational wave signatures to find IMBHs. Ongoing and upcoming time-domain surveys have the prospect to uncover unprecedented numbers of tidal disruption events \citep[e.g.,][]{vanvelzenetal2011}. Although at present only a very small fraction of the events have black hole mass primaries with \mbh$<10^6$~\msun\ \citep{maksymetal2014,weversetal2017,mockleretal2019}, the increased depth of LSST will improve the situation \citep[although see][]{bricmangomboc2019}.

A particularly exciting prospect is white dwarf TDEs, which happen uniquely around  $<10^5$~\msun\ black holes, and may be accompanied by a gravitational wave signal. As argued by \citet{eracleousetal2019}, based on depth, we can expect $\sim 10$ events per year in the ZTF+LSST era \citep[see also][]{hungetal2018}, although we should note that the rates are highly uncertain, as is our ability to uniquely identify such events from electromagnetic signatures alone \citep[e.g.,][]{macleodetal2016}.

The EMRI rates will include not only the white dwarf-black hole mergers, but also neutron star and stellar-mass black hole mergers with the central black hole. These rates are also uncertain, as they depend on both the unknown black hole mass function (\S \ref{sec:demographics}) and the unknown stellar populations in the vicinity of low-mass black holes \citep[see][and references therein]{berryetal2019}. However, the ``optimistic'' number may be hundreds of events per year \citep{gairetal2010} with the pessimistic one being a couple of events per year. 

\emph{LISA} will also be sensitive to black hole-black hole mergers, and here the event rate will depend on the seed formation mechanism, as well as on the early accretion history of the seeds. \citet{ricartenatarajan2018} present example rates for their different seeding and accretion prescriptions, used here.

Finally, first \emph{JWST} and then a \emph{Lynx-}like mission in the X-ray, should each be sensitive to $\sim 10^5$~\mbh\ black holes even at $z \approx 10$ \citep[e.g.,][]{barrowetal2018,haimanetal2019}. The challenge will be identifying such accreting low-mass black holes, which will require extensive multi-wavelength data. Nevertheless, we again use the predictions of \citet{ricartenatarajan2018} to estimate the numbers of accreting black holes that will be detectable by upcoming X-ray missions.

\section{Supplement II: Additional Tables}

\begin{table}
\caption{AGN Fractions\label{tab:fagn}}
\begin{center}
  \begin{tabular}{ccccc}
\hline
Sample & Selection & $N_{\rm tot}$ & $F_{\rm AGN}$ & $L_{\rm Bol, lim}$ \\
\hline
\citet{dongetal2012a}	& BLR &  451,000 &    0.0007 &	42.0$^a$ \\
\citet{moranetal2014}	& BPT &	 1040   &      0.03   &	39.0 \\
\citet{reinesetal2013}	& BPT &	25,000  &      0.005  &	39.3 \\
\hline
\citet{desrochesho2009} & X-ray & 64 &     0.27   & 39.3 \\
\citet{zhangetal2009}   & X-ray &  93    &     0.30   & 39.0 \\
\citet{sheetal2017a}    & X-ray &  212   &     0.25   & 39.0 \\
\citet{pardoetal2016}	& X-ray &  605   &     0.01   & 41.0 \\
\citet{milleretal2015}	& X-ray &  104   &     0.20   &	39.3 \\
\hline
  \end{tabular}
  \end{center}
\begin{tabnote}
{(a) In the case of this sample, the number represents total number of galaxies searched, without any cut on morphology or stellar mass, thus these numbers cannot be compared with others in this table.
  (1) Sample (see \S 4). (2) Type of selection. Broad-line region (BLR) means that a black hole mass cut was applied. BPT used the strong emission lines. X-ray was in all cases a search in \emph{Chandra} data for X-ray point soures. (3) Total number of objects with $M_*<10^{10} \, M_{\rm \odot}$ searched. (4) Fraction of objects found to harbor black holes. (5) Approximate detection threshold, converted to a bolometric luminosity assuming a bolometric correction of 10 for both the X-ray and optical continuum.}
\end{tabnote}
\end{table}

\begin{table}
\caption{Stellar Masses, \sigmastar\ and Black Hole Masses for Scaling Relations (New) \label{tab:scaleappendix}}
\begin{center}
  \begin{tabular}{llcccccccccc}
\hline
Galaxy & Distance & $\sigma_*$ & $\Delta \sigma_*$ & $K$ & $B-V$ & $M_*$  & $M_{\rm BH}$ & $\Delta M_{\rm BH, low}$ & $\Delta M_{\rm BH, high}$ & HT & Ref. \\
\hline
ESO558-G009 & 102 &  170.0  & 20.0 & \nodata & \nodata  &  \nodata  & 1.7e+07 & 1.0e+06 & 1.0e+06 & S & 1\\ 
IC4296 & 49 &  322.1  & 16.3 & 7.37 & 1.00  &  11.7  & 1.3e+09 & 2.2e+08 & 2.2e+08 & E & 2\\ 
J0437+2456 & 66 &  110.0  & 13.0 & \nodata & \nodata  &  \nodata  & 2.8e+06 & 2.0e+05 & 2.0e+05 & S & 1\\ 
Mrk1029 & 124 &  132.0  & 15.0 & \nodata & \nodata  &  \nodata  & 1.9e+06 & 6.0e+05 & 6.0e+05 & S & 1\\ 
NGC0307 & 52 &  204.2  & 3.3 & 9.60 & 0.97  &  10.9  & 4.0e+08 & 5.5e+07 & 5.5e+07 & E & 2\\ 
NGC0584 & 19 &  189.0  & 5.0 & 7.31 & 0.95  &  10.9  & 1.3e+08 & 5.0e+07 & 5.0e+07 & E & 3 \\ 
NGC1320 & 49 &  141.0  & 16.0 & 9.41 & 0.86  &  10.9  & 5.5e+06 & 2.5e+06 & 2.5e+06 & S & 2 \\ 
NGC1398 & 24 &  233.9  & 3.8 & 6.51 & 0.97  &  11.4  & 1.1e+08 & 2.1e+07 & 2.1e+07 & S & 2\\ 
NGC1600 & 64 &  293.0  & 30.0 & 7.83 & 1.04  &  11.7  & 1.7e+10 & 1.5e+09 & 1.5e+09 & E & 4 \\ 
NGC2784 & 9 &  188.0  & 8.0 & 6.36 & 1.15  &  10.7  & 1.0e+08 & 6.0e+07 & 6.0e+07 & E & 3 \\ 
NGC2974 & 21 &  227.0  & 11.0 & 7.57 & 0.99  &  10.9  & 1.7e+08 & 3.6e+07 & 3.6e+07 & E & 2 \\ 
NGC3031 & 3 &  142.9  & 6.9 & 3.88 & 1.04  &  10.8  & 6.5e+07 & 2.0e+07 & 2.0e+07 & S & 2\\ 
NGC3079 & 15 &  145.9  & 7.1 & 7.13 & 0.81  &  10.8  & 2.5e+06 & 2.8e+05 & 2.8e+05 & S & 2\\ 
NGC3414 & 25 &  205.1  & 9.9 & 7.72 & 0.87  &  11.0  & 2.5e+08 & 4.1e+07 & 4.1e+07 & E & 2\\ 
NGC3627 & 10 &  122.5  & 0.6 & 5.89 & 0.83  &  10.9  & 8.5e+06 & 9.4e+05 & 9.4e+05 & S & 2\\ 
NGC3640 & 26 &  176.0  & 8.0 & 7.43 & 0.93  &  11.1  & 7.7e+07 & 5.0e+07 & 5.0e+07 & E & 3\\ 
NGC3923 & 20 &  222.3  & 10.2 & 6.41 & 1.07  &  11.3  & 2.8e+09 & 7.5e+08 & 7.5e+08 & E & 2 \\ 
NGC4151 & 20 &  156.0  & 7.9 & 7.48 & 0.60  &  10.8  & 6.5e+07 & 1.1e+07 & 1.1e+07 & S & 2\\ 
NGC4281 & 24 &  227.0  & 11.0 & 7.91 & 0.96  &  10.9  & 5.4e+08 & 8.0e+07 & 8.0e+07 & E & 3 \\ 
NGC4339 & 16 &  95.0  & -9.0 & 8.50 & 0.89  &  10.2  & 4.3e+07 & 3.0e+07 & 3.0e+07 & E & 5\\ 
NGC4371 & 16 &  142.6  & 1.6 & 7.65 & 0.99  &  10.7  & 7.0e+06 & 1.2e+06 & 1.2e+06 & S & 2\\ 
NGC4434 & 22 &  98.0  & -9.0 & 9.16 & 0.94  &  10.3  & 7.0e+07 & 2.5e+07 & 2.5e+07 & E & 5 \\ 
NGC4501 & 16 &  157.4  & 2.9 & 6.26 & 0.89  &  11.2  & 2.0e+07 & 3.7e+06 & 3.7e+06 & S & 2\\ 
NGC4552 & 15 &  251.8  & 12.2 & 6.70 & 1.00  &  10.9  & 5.0e+08 & 5.9e+07 & 5.9e+07 & E & 2\\ 
NGC4570 & 17 &  170.0  & 8.0 & 7.73 & 0.99  &  10.6  & 6.8e+07 & 2.0e+07 & 2.0e+07 & E & 3 \\ 
NGC4578 & 16 &  107.0  & -9.0 & 8.49 & 0.87  &  10.3  & 1.9e+07 & 1.8e+07 & 1.8e+07 & E & 2 \\ 
NGC4621 & 18 &  224.9  & 10.9 & 6.64 & 0.97  &  11.1  & 4.0e+08 & 7.9e+07 & 7.9e+07 & E & 2\\ 
NGC4699 & 18 &  181.1  & 4.2 & 6.42 & 0.91  &  11.2  & 1.8e+08 & 2.1e+07 & 2.1e+07 & S & 2 \\ 
NGC4762 & 22 &  134.0  & -9.0 & 7.34 & 0.86  &  11.0  & 2.3e+07 & 7.5e+06 & 7.5e+06 & S & 5 \\ 
NGC5018 & 40 &  209.4  & 3.4 & 7.63 & 0.98  &  11.4  & 1.0e+08 & 1.9e+07 & 1.9e+07 & E & 2\\ 
NGC5419 & 56 &  367.3  & 9.3 & 7.55 & 1.07  &  11.8  & 7.2e+09 & 2.4e+09 & 2.4e+09 & E & 2 \\ 
NGC5495 & 93 &  166.0  & 20.0 & \nodata & \nodata  &  \nodata  & 1.0e+07 & 1.0e+06 & 1.0e+06 & S & 1\\ 
NGC5765b & 113 &  162.0  & 19.0 & \nodata & \nodata  &  \nodata  & 4.4e+07 & 5.0e+06 & 5.0e+06 & S & 1\\ 
NGC5813 & 32 &  230.1  & 11.1 & 7.30 & 1.02  &  11.4  & 7.1e+08 & 9.5e+07 & 9.5e+07 & E & 2 \\ 
NGC5846 & 24 &  237.1  & 12.0 & 6.94 & 1.03  &  11.3  & 1.1e+09 & 1.5e+08 & 1.5e+08 & E & 1\\
NGC7049 & 29 &  245.0  & 8.0 & 7.08 & 1.03  &  11.4  & 3.2e+08 & 8.0e+07 & 8.0e+07 & E & 3 \\ 
\hline
  \end{tabular}
  \end{center}
\begin{tabnote}
{First group of galaxies are added to \citet{kormendyho2013}, while the rest are taken from that paper. 
(1) Galaxy. (2) Distance (Mpc). (3) \sigmastar\ (\kms). (4) $K-$band magnitude. (5) $B-V$ color (mag).
(6) Stellar mass derived using \citet{belletal2003}. (7) Black hole mass (\msun). (8) Lower $1 \sigma$ limit on black hole mass (\msun). 
(9) Upper $1 \sigma$ limit on black hole mass (\msun). (10) Crude Hubble type. (11) Reference: (1) \citet{greeneetal2016};
(2) \citet{sagliaetal2016}; (3) \citet{thateretal2019}; (4) \citet{thomasetal2016}; (5) \citet{krajnovicetal2018}.}
\end{tabnote}
\end{table}

\begin{table}
\caption{Stellar Masses, \sigmastar\ and Black Hole Masses for Scaling Relations (KHI) \label{tab:scalekh1}}
\begin{center}
  \begin{tabular}{llcccccccccc}
\hline
Galaxy & Distance & $\sigma_*$ & $\Delta \sigma_*$ & $K$ & $B-V$ & $M_*$  & $M_{\rm BH}$ & $\Delta M_{\rm BH, low}$ & $\Delta M_{\rm BH, high}$ & HT \\
\hline
M32 & 0 &  77.0  & 3.0 & 5.10 & 0.90  &  9.0  & 2.4e+06 & 1.4e+06 & 3.5e+06 & E \\
NGC1332 & 22 &  328.0  & 9.0 & 7.05 & 0.93  &  11.1  & 1.5e+09 & 1.3e+09 & 1.7e+09 & E \\  
NGC1374 & 19 &  167.0  & 3.0 & 8.16 & 0.91  &  10.6  & 5.9e+08 & 5.4e+08 & 6.5e+08 & E \\  
NGC1399 & 20 &  315.0  & 3.0 & 6.31 & 0.95  &  11.4  & 8.8e+08 & 4.4e+08 & 1.8e+09 & E \\  
NGC1407 & 29 &  276.0  & 2.0 & 6.46 & 0.97  &  11.6  & 4.6e+09 & 4.2e+09 & 5.4e+09 & E \\  
NGC1550 & 52 &  270.0  & 10.0 & 8.77 & 0.96  &  11.2  & 3.9e+09 & 3.2e+09 & 4.5e+09 & E \\ 
NGC3091 & 53 &  297.0  & 12.0 & 8.09 & 0.96  &  11.5  & 3.7e+09 & 3.2e+09 & 3.8e+09 & E \\  
NGC3377 & 11 &  145.0  & 7.0 & 7.16 & 0.83  &  10.5  & 1.8e+08 & 8.5e+07 & 2.7e+08 & E \\  
NGC3379 & 10 &  206.0  & 10.0 & 6.27 & 0.94  &  10.8  & 4.2e+08 & 3.1e+08 & 5.2e+08 & E \\ 
NGC3608 & 22 &  182.0  & 9.0 & 7.62 & 0.92  &  10.9  & 4.6e+08 & 3.7e+08 & 5.6e+08 & E  \\ 
NGC3842 & 92 &  270.0  & 27.0 & 8.84 & 0.94  &  11.6  & 9.1e+09 & 6.3e+09 & 1.1e+10 & E  \\ 
NGC4291 & 26 &  242.0  & 12.0 & 8.42 & 0.93  &  10.7  & 9.8e+08 & 6.7e+08 & 1.3e+09 & E  \\ 
NGC4374 & 18 &  296.0  & 14.0 & 5.75 & 0.94  &  11.5  & 9.2e+08 & 8.4e+08 & 1.0e+09 & E  \\ 
NGC4472 & 16 &  300.0  & 7.0 & 4.97 & 0.94  &  11.7  & 2.5e+09 & 2.4e+09 & 3.1e+09 & E  \\ 
NGC4473 & 15 &  190.0  & 9.0 & 7.16 & 0.94  &  10.8  & 9.0e+07 & 4.5e+07 & 1.4e+08 & E  \\ 
M87 & 16 &  324.0  & 35.0 & 5.27 & 0.94  &  11.6  & 6.2e+09 & 5.8e+09 & 6.5e+09 & E  \\ 
NGC4486B & 16 &  185.0  & 9.0 & 10.39 & 0.99  &  9.5  & 6.0e+08 & 4.0e+08 & 9.0e+08 & E  \\ 
NGC4649 & 16 &  380.0  & 19.0 & 5.49 & 0.95  &  11.5  & 4.7e+09 & 3.7e+09 & 5.8e+09 & E  \\ 
NGC4697 & 12 &  177.0  & 8.0 & 6.37 & 0.88  &  10.9  & 2.0e+08 & 1.5e+08 & 2.5e+08 & E  \\ 
NGC4751 & 32 &  355.0  & 14.0 & 8.24 & 0.98  &  11.0  & 1.7e+09 & 1.5e+09 & 1.8e+09 & E  \\ 
NGC4889 & 102 &  347.0  & 5.0 & 8.41 & 1.03  &  11.9  & 2.1e+10 & 4.9e+09 & 3.7e+10 & E  \\ 
NGC5077 & 38 &  222.0  & 11.0 & 8.22 & 0.99  &  11.2  & 8.6e+08 & 4.1e+08 & 1.3e+09 & E  \\ 
NGC5328 & 64 &  333.0  & 2.0 & 8.49 & 1.00  &  11.5  & 4.8e+09 & 2.8e+09 & 5.6e+09 & E  \\ 
NGC5516 & 55 &  328.0  & 11.0 & 8.31 & 0.99  &  11.4  & 3.7e+09 & 2.6e+09 & 3.8e+09 & E  \\ 
NGC5576 & 25 &  183.0  & 9.0 & 7.83 & 0.86  &  10.9  & 2.7e+08 & 1.9e+08 & 3.4e+08 & E  \\ 
NGC5845 & 25 &  239.0  & 11.0 & 9.11 & 0.97  &  10.4  & 4.9e+08 & 3.3e+08 & 6.4e+08 & E  \\ 
NGC6086 & 138 &  318.0  & 2.0 & 9.97 & 0.96  &  11.5  & 3.7e+09 & 2.6e+09 & 5.5e+09 & E  \\ 
NGC6861 & 28 &  389.0  & 3.0 & 7.71 & 0.96  &  11.1  & 2.1e+09 & 2.0e+09 & 2.7e+09 & E  \\ 
NGC7619 & 53 &  292.0  & 5.0 & 8.03 & 0.97  &  11.5  & 2.3e+09 & 2.2e+09 & 3.4e+09 & E  \\ 
NGC7768 & 116 &  257.0  & 26.0 & 9.34 & 0.91  &  11.6  & 1.3e+09 & 9.3e+08 & 1.8e+09 & E  \\ 
IC1459 & 28 &  331.0  & 5.0 & 6.81 & 0.97  &  11.5  & 2.5e+09 & 2.3e+09 & 3.0e+09 & E  \\ 
NGC1316 & 20 &  226.0  & 9.0 & 5.32 & 0.87  &  11.8  & 1.7e+08 & 1.4e+08 & 2.0e+08 & E  \\ 
NGC2960 & 67 &  166.0  & 16.0 & 9.78 & 0.88  &  11.0  & 1.1e+07 & 1.0e+07 & 1.1e+07 & E  \\ 
NGC4382 & 17 &  182.0  & 5.0 & 5.76 & 0.86  &  11.4  & 1.3e+07 & 1.0e+07 & 2.2e+08 & E  \\ 
NGC5128 & 3 &  150.0  & 7.0 & 3.49 & 0.90  &  11.0  & 5.7e+07 & 4.6e+07 & 6.7e+07 & E  \\ 
NGC2778 & 23 &  175.0  & 8.0 & 9.51 & 0.91  &  10.2  & 1.4e+07 & 1.0e+07 & 2.9e+07 & E  \\ 
NGC3607 & 22 &  229.0  & 11.0 & 6.99 & 0.91  &  11.2  & 1.4e+08 & 9.0e+07 & 1.8e+08 & E  \\ 
NGC4261 & 32 &  315.0  & 15.0 & 6.94 & 0.97  &  11.5  & 5.3e+08 & 4.2e+08 & 6.4e+08 & E  \\ 
NGC4459 & 16 &  167.0  & 8.0 & 7.15 & 0.91  &  10.8  & 7.0e+07 & 5.6e+07 & 8.3e+07 & E  \\ 
NGC7052 & 70 &  266.0  & 13.0 & 8.57 & 0.86  &  11.5  & 4.0e+08 & 2.4e+08 & 6.7e+08 & E  \\ 
A1836BCG & 152 &  288.0  & 14.0 & 9.99 & 1.04  &  11.6  & 3.7e+09 & 3.2e+09 & 4.2e+09 & E  \\ 
A3565BCG & 49 &  322.0  & 16.0 & 7.50 & 0.96  &  11.6  & 1.3e+09 & 1.1e+09 & 1.5e+09 & E  \\
\hline
  \end{tabular}
  \end{center}
\begin{tabnote}
{First group of galaxies are added to \citet{kormendyho2013}, while the rest are taken from that paper. 
(1) Galaxy. (2) Distance (Mpc). (3) \sigmastar\ (\kms). (4) $K-$band magnitude. (5) $B-V$ color (mag).
(6) Stellar mass derived using \citet{belletal2003}. (7) Black hole mass (\msun). (8) Lower $1 \sigma$ limit on black hole mass (\msun). 
(9) Upper $1 \sigma$ limit on black hole mass (\msun). (10) Crude Hubble type.}
\end{tabnote}
\end{table}

\begin{table}
\caption{Stellar Masses, \sigmastar\ and Black Hole Masses for Scaling Relations (KHII) \label{tab:scalekh2}}
\begin{center}
  \begin{tabular}{llcccccccccc}
\hline
Galaxy & Distance & $\sigma_*$ & $\Delta \sigma_*$ & $K$ & $B-V$ & $M_*$  & $M_{\rm BH}$ & $\Delta M_{\rm BH, low}$ & $\Delta M_{\rm BH, high}$ & HT \\
\hline
NGC524 & 24 &  247.0  & 12.0 & 7.16 & 0.98  &  11.2  & 8.7e+08 & 8.2e+08 & 9.6e+08 & S0  \\ 
NGC821 & 23 &  209.0  & 10.0 & 7.71 & 0.89  &  10.9  & 1.6e+08 & 9.2e+07 & 2.4e+08 & S0  \\ 
NGC1023 & 10 &  205.0  & 10.0 & 6.24 & 0.95  &  10.8  & 4.1e+07 & 3.7e+07 & 4.6e+07 & S0  \\ 
NGC1194 & 58 &  148.0  & 24.0 & 9.76 & 0.89  &  10.9  & 7.1e+07 & 6.8e+07 & 7.4e+07 & S0  \\ 
NGC1277 & 73 &  333.0  & 17.0 & 9.81 & 0.98  &  11.1  & 1.7e+10 & 1.4e+10 & 2.0e+10 & S0  \\ 
NGC2549 & 12 &  145.0  & 7.0 & 8.05 & 0.91  &  10.2  & 1.4e+07 & 3.1e+06 & 1.6e+07 & S0  \\ 
NGC3115 & 9 &  230.0  & 11.0 & 5.88 & 0.93  &  10.9  & 9.0e+08 & 6.2e+08 & 9.5e+08 & S0  \\ 
NGC3245 & 21 &  205.0  & 10.0 & 7.86 & 0.89  &  10.8  & 2.4e+08 & 1.6e+08 & 2.7e+08 & S0  \\ 
NGC3585 & 20 &  213.0  & 11.0 & 6.70 & 0.91  &  11.2  & 3.3e+08 & 2.7e+08 & 4.7e+08 & S0  \\ 
NGC3998 & 14 &  275.0  & 7.0 & 7.37 & 0.94  &  10.6  & 8.4e+08 & 7.8e+08 & 9.2e+08 & S0  \\ 
NGC4026 & 13 &  180.0  & 9.0 & 7.58 & 0.90  &  10.5  & 1.8e+08 & 1.4e+08 & 2.4e+08 & S0  \\ 
NGC4342 & 22 &  225.0  & 11.0 & 9.02 & 0.93  &  10.4  & 4.5e+08 & 3.0e+08 & 7.2e+08 & S0  \\ 
NGC4526 & 16 &  222.0  & 11.0 & 6.47 & 0.94  &  11.1  & 4.5e+08 & 3.5e+08 & 5.9e+08 & S0  \\ 
NGC4564 & 15 &  162.0  & 8.0 & 7.94 & 0.90  &  10.5  & 8.8e+07 & 6.4e+07 & 1.1e+08 & S0  \\ 
NGC7457 & 12 &  67.0  & 3.0 & 8.18 & 0.84  &  10.2  & 9.0e+06 & 3.6e+06 & 1.4e+07 & S0  \\ 
NGC3384 & 11 &  146.0  & 7.0 & 6.75 & 0.91  &  10.7  & 1.1e+07 & 5.9e+06 & 1.6e+07 & S0  \\ 
NGC3945 & 19 &  192.0  & 10.0 & 7.53 & 0.93  &  10.8  & 8.8e+06 & 1.0e+06 & 2.6e+07 & S0  \\ 
NGC3998 & 14 &  275.0  & 7.0 & 7.37 & 0.94  &  10.6  & 2.3e+08 & 1.4e+08 & 3.3e+08 & S0  \\ 
NGC4596 & 16 &  136.0  & 6.0 & 7.46 & 0.92  &  10.7  & 7.7e+07 & 4.4e+07 & 1.1e+08 & S0  \\ 
M31 & 0 &  169.0  & 8.0 & 0.57 & 0.86  &  10.8  & 1.4e+08 & 1.1e+08 & 2.3e+08 & S \\ 
M81 & 3 &  143.0  & 7.0 & 3.83 & 0.88  &  10.8  & 6.5e+07 & 5.0e+07 & 9.0e+07 & S \\ 
NGC4258 & 7 &  115.0  & 10.0 & 5.46 & 0.68  &  10.8  & 3.8e+07 & 3.7e+07 & 3.8e+07 & S \\ 
NGC4594 & 9 &  240.0  & 12.0 & 4.62 & 0.93  &  11.4  & 6.6e+08 & 6.2e+08 & 7.0e+08 & S \\ 
Circinus & 2 &  79.0  & 3.0 & 4.71 & 0.41  &  10.3  & 1.1e+06 & 9.4e+05 & 1.3e+06 & S \\ 
NGC1068 & 15 &  151.0  & 7.0 & 5.79 & 0.71  &  11.3  & 8.4e+06 & 8.0e+06 & 8.8e+06 & S \\ 
NGC1300 & 21 &  88.0  & 3.0 & 7.56 & 0.65  &  10.9  & 7.6e+07 & 3.9e+07 & 1.5e+08 & S \\ 
NGC2273 & 29 &  125.0  & 9.0 & 8.48 & 0.83  &  10.8  & 8.6e+06 & 8.2e+06 & 9.1e+06 & S \\ 
NGC2748 & 23 &  115.0  & 5.0 & 8.72 & 0.71  &  10.5  & 4.4e+07 & 2.6e+07 & 6.2e+07 & S \\ 
NGC2787 & 7 &  189.0  & 9.0 & 7.26 & 0.94  &  10.1  & 4.1e+07 & 3.6e+07 & 4.5e+07 & S \\ 
NGC3227 & 23 &  133.0  & 12.0 & 7.64 & 0.80  &  10.9  & 2.1e+07 & 9.8e+06 & 2.8e+07 & S \\ 
NGC3368 & 10 &  125.0  & 6.0 & 6.32 & 0.84  &  10.8  & 7.7e+06 & 6.1e+06 & 9.2e+06 & S \\ 
NGC3393 & 49 &  148.0  & 10.0 & 9.06 & 0.81  &  11.0  & 1.6e+07 & 5.8e+06 & 2.6e+07 & S \\ 
NGC3489 & 12 &  113.0  & 4.0 & 7.37 & 0.81  &  10.5  & 5.9e+06 & 5.1e+06 & 6.8e+06 & S \\ 
NGC4388 & 16 &  99.0  & 10.0 & 8.00 & 0.71  &  10.5  & 7.3e+06 & 7.1e+06 & 7.5e+06 & S \\ 
NGC4736 & 5 &  120.0  & 6.0 & 5.11 & 0.73  &  10.6  & 6.8e+06 & 5.2e+06 & 8.3e+06 & S \\ 
NGC4826 & 7 &  104.0  & 3.0 & 5.33 & 0.80  &  10.8  & 1.6e+06 & 1.2e+06 & 2.0e+06 & S \\ 
NGC4945 & 3 &  134.0  & 20.0 & 4.44 & 1.20  &  10.6  & 1.4e+06 & 8.7e+05 & 2.0e+06 & S \\ 
NGC7582 & 22 &  156.0  & 19.0 & 7.32 & 0.74  &  11.0  & 5.5e+07 & 4.6e+07 & 6.8e+07 & S \\ 
IC2560 & 37 &  141.0  & 10.0 & 8.69 & 0.89  &  10.9  & 5.0e+06 & 1.0e+06 & 5.7e+06 & S \\ 
UGC3789 & 49 &  107.0  & 12.0 & 9.51 & 0.86  &  10.9  & 9.6e+06 & 8.1e+06 & 1.1e+07 & S \\ 
NGC4596 & 16 &  136.0  & 6.0 & 7.46 & 0.92  &  10.7  & 7.7e+07 & 4.4e+07 & 1.1e+08 & S \\ 
\hline
  \end{tabular}
  \end{center}
\begin{tabnote}
{First group of galaxies are added to \citet{kormendyho2013}, while the rest are taken from that paper. 
(1) Galaxy. (2) Distance (Mpc). (3) \sigmastar\ (\kms). (4) $K-$band magnitude. (5) $B-V$ color (mag).
(6) Stellar mass derived using \citet{belletal2003}. (7) Black hole mass (\msun). (8) Lower $1 \sigma$ limit on black hole mass (\msun). 
(9) Upper $1 \sigma$ limit on black hole mass (\msun). (10) Crude Hubble type.}
\end{tabnote}
\end{table}

\newpage

\begin{table}
  \caption{Scaling Relation Fits \label{tab:scaling}}
  \begin{center}
  \begin{tabular}{clccc}
\hline
Fit & Sample & $\alpha$ & $\beta$ & $\epsilon$ \\
\hline
$M_{\rm BH}-\sigma_*$  & All, no limits & $7.88 \pm 0.05$ & $4.34 \pm 0.24$ & $0.53 \pm 0.04$ \\
$M_{\rm BH}-\sigma_*$  & All, limits & $7.87 \pm 0.06$ & $4.55 \pm 0.23$ & $0.55 \pm 0.04$ \\ 
$M_{\rm BH}-\sigma_*$  & Early & $8.03 \pm 0.06$ & $4.24 \pm 0.25$ & $0.43 \pm 0.04$ \\
$M_{\rm BH}-\sigma_*$  & Late, no limits & $7.40 \pm 0.10$ & $2.54 \pm 0.50$ & $0.50 \pm 0.07$ \\
$M_{\rm BH}-\sigma_*$  & Late, limits &  $7.44 \pm 0.12$ & $3.61 \pm 0.50$ & $0.58 \pm 0.09$ \\
\hline
$M_{\rm BH}-M_*$  & All, no limits & $7.56 \pm 0.09$ & $1.39 \pm 0.13$ & $0.79 \pm 0.05$ \\
$M_{\rm BH}-M_*$  & All, limits & $7.43 \pm 0.09$ & $1.61 \pm 0.12$ & $0.81 \pm 0.06$ \\ 
$M_{\rm BH}-M_*$  & Early & $7.89 \pm 0.09$ & $1.33 \pm 0.12$ & $0.65 \pm 0.05$ \\
$M_{\rm BH}-M_*$  & Late, no limits & $6.94 \pm 0.13$ & $0.98 \pm 0.27$ & $0.60 \pm 0.08$ \\
$M_{\rm BH}-M_*$  & Late, limits & $6.70 \pm 0.13$ & $1.61 \pm 0.24$ & $0.65 \pm 0.09$ \\ 
\hline
  \end{tabular}
  \end{center}
\begin{tabnote}
  {(1) Fit being presented. We fit log (\mbh) = $\alpha \, + \beta \, {\rm log} (\sigma^*/\sigma_0)+ \epsilon$ where $\sigma_0 = 160$~\kms; log (\mbh)$= \alpha + \beta \, {\rm log} (M_*/M_0) + \epsilon$ for $M_0 = 3 \times 10^{10}$~\msun. (2) Sample. (3) Intercept ($\alpha$). (4) Slope ($\beta$). (5) Intrinsic scatter ($\epsilon$).}
\end{tabnote}
\end{table}

\clearpage

\begin{table}
  \caption{Black Hole Mass Functions \label{tab:massfunc}}
  \begin{tabular}{ccccccc}
\hline
Log ($M_{\rm BH}$) & Linear & Low & High &  NC & Low & High  \\
{} & $\# \, \rm{Mpc}^{-3} \rm{dex}^{-1}$ & $\# \, \rm{Mpc}^{-3} \rm{dex}^{-1}$ & $\# \, \rm{Mpc}^{-3} \rm{dex}^{-1}$ & $\# \, \rm{Mpc}^{-3} \rm{dex}^{-1}$ & $\# \, \rm{Mpc}^{-3} \rm{dex}^{-1}$ & $\# \, \rm{Mpc}^{-3} \rm{dex}^{-1}$ \\
\hline
4.0 & $-$3.10 & $-$3.25 & $-$2.98 & $-$3.10 & $-$3.25 & $-$2.98 \\
4.25 & $-$3.06 & $-$3.21 & $-$2.94 & $-$3.06 & $-$3.21 & $-$2.94 \\
4.5 & $-$3.04 & $-$3.19 & $-$2.92 & $-$3.04 & $-$3.19 & $-$2.92 \\
4.75 & $-$3.02 & $-$3.16 & $-$2.91 & $-$3.02 & $-$3.16 & $-$2.91 \\
5.0 & $-$3.01 & $-$3.15 & $-$2.89 & $-$3.01 & $-$3.15 & $-$2.89 \\
5.25 & $-$3.01 & $-$3.14 & $-$2.90 & $-$3.01 & $-$3.14 & $-$2.90 \\
5.5 & $-$3.01 & $-$3.13 & $-$2.91 & $-$3.01 & $-$3.13 & $-$2.91 \\
5.75 & $-$3.01 & $-$3.12 & $-$2.92 & $-$3.01 & $-$3.12 & $-$2.92 \\
6.0 & $-$3.02 & $-$3.10 & $-$2.93 & $-$3.02 & $-$3.10 & $-$2.93 \\
6.25 & $-$3.01 & $-$3.08 & $-$2.95 & $-$3.01 & $-$3.08 & $-$2.95 \\
6.5 & $-$3.01 & $-$3.05 & $-$2.96 & $-$3.01 & $-$3.05 & $-$2.96 \\
6.75 & $-$3.01 & $-$3.04 & $-$2.98 & $-$3.01 & $-$3.04 & $-$2.98 \\
7.0 & $-$3.01 & $-$3.04 & $-$2.98 & $-$3.01 & $-$3.04 & $-$2.98 \\
7.25 & $-$3.02 & $-$3.05 & $-$2.99 & $-$3.02 & $-$3.05 & $-$2.99 \\
7.5 & $-$3.04 & $-$3.08 & $-$3.01 & $-$3.04 & $-$3.08 & $-$3.01 \\
7.75 & $-$3.09 & $-$3.13 & $-$3.05 & $-$3.09 & $-$3.13 & $-$3.05 \\
8.0 & $-$3.15 & $-$3.19 & $-$3.11 & $-$3.15 & $-$3.19 & $-$3.11 \\
8.25 & $-$3.23 & $-$3.28 & $-$3.19 & $-$3.23 & $-$3.28 & $-$3.19 \\
8.5 & $-$3.34 & $-$3.39 & $-$3.29 & $-$3.34 & $-$3.39 & $-$3.29 \\
8.75 & $-$3.47 & $-$3.52 & $-$3.42 & $-$3.47 & $-$3.52 & $-$3.42 \\
9.0 & $-$3.63 & $-$3.69 & $-$3.58 & $-$3.63 & $-$3.69 & $-$3.58 \\
9.25 & $-$3.82 & $-$3.88 & $-$3.77 & $-$3.82 & $-$3.88 & $-$3.77 \\
9.5 & $-$4.04 & $-$4.10 & $-$3.99 & $-$4.04 & $-$4.10 & $-$3.99 \\
9.75 & $-$4.29 & $-$4.35 & $-$4.23 & $-$4.29 & $-$4.35 & $-$4.23 \\
10.0 & $-$4.57 & $-$4.65 & $-$4.50 & $-$4.57 & $-$4.65 & $-$4.50 \\
10.25 & $-$4.88 & $-$4.97 & $-$4.80 & $-$4.88 & $-$4.97 & $-$4.80 \\
10.5 & $-$5.23 & $-$5.33 & $-$5.13 & $-$5.23 & $-$5.33 & $-$5.13 \\
10.75 & $-$5.62 & $-$5.78 & $-$5.48 & $-$5.62 & $-$5.78 & $-$5.48 \\
\hline
  \end{tabular}
\begin{tabnote}
  {(1) Black hole mass bin. (2) Number density of black holes, assuming the more pessimistic ``linear'' occupation fraction. (3) Lower $68\%$ confidence limit on number density. (4) Upper $68\%$ limit on number density. (5) Number density of black holes, assuming that every nuclear star cluster hosts a black hole. (6) Lower $68\%$ confidence limit on number density. (7) Upper $68\%$ limit on number density.}
\end{tabnote}
\end{table}

\newpage

\begin{table}
    \caption{Other Nuclei \label{tab:extranuclei}}
\begin{center}
  \begin{tabular}{lcccc}
\hline
Object & Dist & Log ($M_{\rm NSC}$) & Log ($M_{\rm BH}$) & Ref \\
\hline
NGC1023  &  11.7  &  6.6  &  $7.6^{0.05}_{0.05}$  &   1,2 \\
NGC3384  &  11.6  &  7.3  &  $7.2^{0.12}_{0.17}$  &   1,2 \\ 
NGC4458  &  16.5  &  8.9  &  $8.3^{0.15}_{0.22}$  &   3,4 \\ 
NGC4578  &  16.2  &  7.8  &  $7.5^{0.02}_{0.02}$  &   5,4 \\ 
MW  &  0.008  &  7.5  &  $6.63^{0.05}_{0.05}$  & 6 \\ 
\hline
IC342  &  1.9  &  7.1  &  $<$5.5  &   7  \\ 
NGC4474  &  15.6  &  7.3  &  $<$6.2  &   5,4 \\ 
NGC4551  &  16.1  &  8.3  &  $<$6.9  &   5,4 \\ 
NGC4379  &  16.5  &  7.9  &  $<$8.6  &   3,4 \\ 
NGC4387  &  16.5  &  7.8  &  $<$7.6  &   3,4 \\ 
NGC4474  &  16.5  &  7.4  &  $<$7.7  &   3,4 \\ 
NGC4612  &  16.5  &  7.4  &  $<$8.1  &   3,4 \\ 
NGC4623  &  16.5  &  8.2  &  $<$8.2  &   3,4 \\ 
\hline
M59-UCD3  &  14.9  &  7.4  &  $6.6^{0.18}_{0.23}$  &  8 \\ 
M59cO  &  16.5  &  7.0  &  $6.8^{0.06}_{0.08}$  &  9 \\ 
M60-UCD1  &  16.5  &  7.9  &  $7.3^{0.22}_{0.18}$  &  10 \\ 
VUCD3  &  16.5  &  7.0  &  $6.6^{0.08}_{0.11}$  &  9 \\ 
F-UCD3  &  20.9  &  7.9  &  $6.5^{0.15}_{0.18}$  &  11 \\ 
\hline
NGC4546-UCD1  &  13.1  &  7.5  &  $<$6.0  &   12 \\ 
CenA-UCD330  &  3.8  &  6.8  &  $<$5.0  &  13 \\ 
CenA-UCD320  &  3.8  &  6.5  &  $<$6.0  &  13 \\ 
\hline
  \end{tabular}
  \end{center}
\begin{tabnote}
{Column (1): Object name (galaxy or UCD). Column (2): Distance (Mpc).  
Column (3): Mass of the nuclear star cluster ($M_{\odot}$). Errors are dominated by
systematics in $M/L$ determinations and are likely 0.2-0.3 dex. Column (4): BH mass.
Column (5): Reference for $M_{\rm BH}$ and $M_{\rm NSC}$. (1) \citet{sagliaetal2016};
(2) \citet{laueretal2005}; (3) \citet{pechettietal2017}; (4) \citet{coteetal2006};
(5) \citet{krajnovicetal2018}; (6) \citet{schodeletal2007}; (7) \citet{bokeretal1999}; 
(8) \citet{ahnetal2018}; (9) \citet{ahnetal2017};  (10)  \citet{sethetal2014};  
(11) \citet{afanasievetal2018}; (12) \citet{norriskannappan2011}; (13) \citet{voggeletal2019}.}
\end{tabnote}
\end{table}

\end{document}